\documentclass[a4paper,11pt]{article}
\pdfoutput=1
\usepackage{jcappub}
\usepackage{aas_macros}
\usepackage{placeins}
\usepackage[usenames,dvipsnames,svgnames]{xcolor}
\usepackage{bm}
\usepackage{bbold}
\usepackage{comment}  

\newif\ifshow
\showfalse  

\ifshow
  \includecomment{hide}
\else
  \excludecomment{hide}
\fi

\let\L\relax
\DeclareMathOperator{\L}{\mathcal{L}}
\DeclareMathOperator{\deltaD}{\delta^\textsc{d}}
\DeclareMathOperator{\order}{\mathcal{O}}
\DeclareMathOperator{\cov}{Cov}
\DeclareMathOperator{\corr}{Corr}
\renewcommand{\d}{\mathrm{d}}
\newcommand{\deltaK}{\delta^\textsc{k}}
\newcommand{\deltaW}{\delta_W}
\newcommand{\vk}{{\bm k}}
\newcommand{\vq}{{\bm q}}
\newcommand{\vx}{{\bm x}}
\newcommand{\vs}{{\bm s}}
\newcommand{\vv}{{\bm v}}

\newcommand{\vtheta}{{\bm \theta}}
\newcommand{\vrp}{{\bm r_p}}
\newcommand{\uvk}{{\hat{\vk}}}

\newcommand{\uvs}{{\hat{\vs}}}
\newcommand{\uvn}{{\hat{\bm n}}}
\newcommand{\uvv}{{\hat{\bm v}}}
\newcommand{\vl}{{\bm\ell}}
\newcommand{\uvtheta}{\hat{\vtheta}}
\newcommand{\uvl}{\hat{\vl}}
\newcommand{\nbar}{\bar{n}}
\newcommand{\W}{\mathcal{W}}
\renewcommand{\S}{\mathcal{S}}
\newcommand{\Q}{\mathcal{Q}}

\newcommand{\QW}[1][]{\Q_{\W #1}}
\newcommand{\QS}[1][]{\Q_{\S #1}}
\newcommand{\QX}[1][]{\Q_{\times #1}}
\newcommand{\Qone}[1][]{\Q_{\mathbb{1} #1}}
\newcommand{\lmax}{\ell_\mathrm{max}}
\newcommand{\setpart}{\mathcal{P}}
\newcommand{\data}{\mathrm{data}}
\newcommand{\rand}{\mathrm{rand}}
\newcommand{\shot}{\mathrm{shot}}
\newcommand{\disc}{\mathrm{disc}}
\newcommand{\conn}{\mathrm{conn}}
\newcommand{\diag}{\mathrm{diag}}
\newcommand{\mock}{\mathrm{mock}}
\newcommand{\smooth}{\mathrm{sm}}
\renewcommand{\SS}{\mathrm{SS}}
\newcommand{\Mpc}{\mathrm{Mpc}}
\newcommand{\Gpc}{\mathrm{Gpc}}
\newcommand{\Mpch}{{\,\Mpc/h}}
\newcommand{\Gpch}{{\,\Gpc/h}}
\newcommand{\hMpc}{{\,h/\Mpc}}
\newcommand{\mean}[1]{\ensuremath{\left\langle{#1}\right\rangle}}
\newcommand{\ensavg}[1]{\bigl\langle #1 \bigr\rangle}

\newcommand{\wj}[6]{
\begin{pmatrix} 
    #1 & #2 & #3 \\
    #4 & #5 & #6
\end{pmatrix}}
\newcommand{\HL}[1]{\textcolor{blue}{#1}} 

\newcommand{\sukhdeep}[1]{\textcolor{red}{SS: #1}}

\title{Disconnected Covariance of 2-point Functions in Large-Scale Structure}

\author[a,b]{Yin Li,}
\author[a]{Sukhdeep Singh,}
\author[a]{Byeonghee Yu,}
\author[a]{Yu Feng,}
\author[a,c]{\& Uro{\v s} Seljak}

\affiliation[a]{Berkeley Center for Cosmological Physics, Department of
Physics, \& Lawrence Berkeley National Laboratory, University of California,
Berkeley, CA 94720, USA}
\affiliation[b]{Kavli Institute for the Physics and Mathematics of the Universe
(WPI), UTIAS, The University of Tokyo, Chiba 277--8583, Japan}
\affiliation[c]{Department of Astronomy, University of California, Berkeley, CA
94720, USA}
\emailAdd{yin.li@berkeley.edu}
\emailAdd{sukhdeep1@berkeley.edu}
\emailAdd{bhyu@berkeley.edu}
\emailAdd{yfeng1@berkeley.edu}
\emailAdd{useljak@berkeley.edu}

\abstract{
Optimal analyses using the 2-point functions of large-scale structure probes
require accurate covariance matrices.
A covariance matrix of the 2-point function comprises the disconnected part and
the connected part.
While the connected covariance only becomes important on small scales, the
disconnected covariance is dominant on large scales, where the survey window
has a significant impact.
In this work, we develop an analytical method to compute the disconnected
covariance, accounting for the window effect.
Derived under the flat-sky approximation, our formalism is applicable to wide
surveys by swapping in the curved-sky window functions.
Our method works for both the power spectrum and the correlation function, and
applies to the covariances of various probes including the multipoles and the
wedges of 3D clustering, the angular and the projected statistics of clustering
and shear, as well as the cross covariances between different probes.
We verify the analytic covariance against the sample covariance from the galaxy
mock simulations in two test cases: (1) the power spectrum multipole
covariance, and (2) the joint covariance of the projected correlation function
and the correlation function multipoles.
Our method achieve good agreement with the mocks, while at a negligible
computational cost.
Unlike mocks, our analytic covariance is free of sampling noise, which often
leads to numerical problems and the need to inflate the errors.
In addition, our method can use the best-fit power spectrum as input, in
contrast to the standard procedure of using a fiducial model that may deviate
significantly from the truth.
We also show that a naive diagonal power spectrum covariance underestimates the
signal-to-noise ratio compared to our analytic covariance.
The code that accompanies this paper is available at
\url{https://github.com/eelregit/covdisc}.
}

\begin{document}
\maketitle

\section{Introduction}
\label{sec:intro}

Studies of the large-scale structure of the Universe have made rapid progress
over the last two decades especially with high precision measurements using
large galaxy surveys.
These surveys have provided sensitive tests on the cosmological models using
combinations of several independent probes including redshift space distortions
(RSD)~\cite{BlakeEtAl2011, BeutlerEtAl17, ReidEtAl16}, baryon acoustic
oscillations (BAO)~\cite{BeutlerEtAl2011, AndersonEtAl2014, AlamEtAl17},
weak gravitational lensing~\cite{TroxelEtAl2018, JoudakiEtAl2018,
SinghAlamEtAl2018, HikageOguriEtAl18}, supernovae~\cite{ReissEtAl2016} and
strong lensing~\cite{BonvinEtAl2017}.
The next generation of galaxy surveys, LSST~\cite{LSSTDESC2012},
Euclid~\cite{Euclid2013}, DESI~\cite{DESI2016}, WFIRST~\cite{WFIRST2018}, aim
to make even more precise measurements by observing even larger volume of the
universe.

With the increasing precision of the measurements, it is becoming increasingly
important to have accurate methods to infer and extract information from these
measurements.
In this work we focus on the probes of matter and galaxy two-point functions.
The two-point functions that measure clustering of galaxies or matter are still
the primary source to extract cosmological information from the observables.
One of the challenges in extracting this information is to accurately
model the covariance matrices of these probes in order to obtain unbiased
likelihoods of model parameters.
Inaccurate covariances make the inferences sub-optimal and in extreme cases
also introduce biases in the most likely values of the parameters.
Common methods account for the effects of noise in numerical covariances by
inflating the errors in the final constraints~\cite{HartlapSimonEtAl07,
DodelsonSchneider2013, SellentinHeavens2016}.

There are mainly three common approaches to quantify the covariance, including
estimations using mock simulations, internal estimations from data using
jackknife or bootstrapping methods, and analytic (or semi-analytic) modeling.
Covariance estimation using mock simulations has become the most popular
approach as it is in principle possible to include proper treatment of all the
observable effects as well as nonlinear and multiscale physics.
However, mock covariance require generating many independent realizations and
then computing the two point functions over all realizations.
This method becomes computationally very expensive, because the required number
of simulations scales with the number of data points and with the desired
accuracy of the parameter covariance~\cite{TaylorJoachimiEtAl13}.
To overcome these challenges, several fast methods to run approximate mocks
have been developed (see e.g.~\cite{LippichEtAl2018, BlotEtAl2018} for
comparison of covariances from different mocks).
Nonetheless, lack of correct physics in these mocks potentially leads to biased
covariances~\cite{BaumgartenChuang2018}.
In addition, it is difficult to guess a priori the right fiducial model at
which the mocks should be generated.
Any deviation of the model from the truth introduces errors in the covariance,
either underestimating or overestimating the covariance depending on the
relation of the fiducial model to the truth.

Computing covariance directly from the data has the potential to overcome these
challenges as it includes all possible observational effects and can be
estimated with relatively small computational overhead.
This approach requires splitting the data into smaller mutually exclusive
subsets, and the number of subsets required depends again on the number of data
points and the desired accuracy of parameter covariance.
However, splitting data into smaller subsamples limits the largest scales that
can be used in the analysis.
Furthermore, the subsamples are not totally independent (a requirement by the
jackknife and bootstrap methods), and as a result they bias the
covariance~\cite{NorbergBaughEtAl09}.
They are also biased by their subvolume windows that are different from the
survey window.

Compared to the mock and internal methods described above, calculating the
covariance matrices analytically has multiple advantages: its predictions are
noiseless; it can be evaluated at the best-fit cosmology with an iterative
procedure, and it is computationally efficient.
However, analytically modeling the covariance proves to be difficult due to the
following reasons.
The covariance of 2-point functions is related to 4-point functions that
consists of the disconnected and the connected contributions.
The disconnected covariance, including the Gaussian and Poisson errors, is a
full-rank matrix and dominant on large scales~\cite{FeldmanKaiserEtAl94}.
It is sensitive to the window of the survey, which is non-trivial to model
accurately.
On the other hand, the connected covariance is a low-rank matrix, becomes
important on small scales, and is hardly affected by the window function.
It receives many contributions, including nonlinear mode-coupling~\cite{MeiksinWhite99, ScoccimarroZaldarriagaEtAl99, CoorayHu01a, HarnoisPen12,
BertoliniSchutzEtAl16, MohammedSeljakVlah17, BarreiraSchmidt17cov, Lacasa18},
the super-sample covariance~\cite{HamiltonRimesEtAl06, HuKravtsov03,
TakadaJain09, TakadaHu13, LiHuEtAl14, LiHuEtAl14sss, AkitsuTakadaEtAl17,
LiSchmittfullEtAl18, BarreiraKrauseEtAl18}, shot noises~\cite{MeiksinWhite99,
Cohn06, Smith09}, and possible baryonic effects.
However, even as the subject of extensive studies, the connected part remains
intractable for accurate analytic prediction.
Due to these challenges, semi-analytic approaches have been developed, by
combining parametric covariance models with fewer number of mocks or the data~\cite{PopeSzapudi2008, OConnellEisenstein2016, PearsonSamushia2016,
HallTaylor2018, OConnellEisenstein18}.
These methods are able to relieve the computational burden by a factor of
$\order(10)$, but their accuracy is subject to the quality of the parametric
models.
They also inherit other drawbacks from the mock and internal methods, such as
fixation on fiducial cosmology, lack of correct physics, and missing
large-scale modes.

In this work, we propose a hybrid method to tackle the covariance challenge.
We develop a fully analytic method to compute the full-rank disconnected
covariance with a proper treatment of the window effect.
We demonstrate its accuracy and efficiency using galaxy mock simulations.
Given the difficulty in modeling the connected contributions, and the fact that
it is only important on small scales and not sensitive to the survey window,
one should be able to calibrate it internally from the data, e.g.\ by low-rank
approximation.
By treating the disconnected and connected components separately, our hybrid
approach enjoys the benefits of both the analytical and internal methods, and
at the same time avoid their disadvantages.
In this paper we focus on the disconnected covariance, and leave the internal
estimation of the connected part as a future project.

This paper is structured as follows.
In Sec.~\ref{sec:method} we develop the methodology to analytically model the
disconnected covariance matrix of the power spectrum or the correlation
function of various large-scale structure probes, taking into account the
survey window function.
Our analytic method involves oscillatory integrals with two Bessel functions,
which we solve in Sec.~\ref{sec:double_Bessel} with a novel quadrature
algorithm.
Using the mock simulations described in Sec.~\ref{sec:data}, we demonstrate the
accuracy and efficiency of the analytic method in Sec.~\ref{sec:results}.
App.~\ref{sec:disc} gives the detailed calculations that generalize our method
for different applications, and App.~\ref{sec:relations} provides the
mathematical tools useful in the derivations.

\section{Formalism and Methodology}
\label{sec:method}

In this section we develop the formalism of the analytic disconnected
covariance including the window effect.
We start with the power spectrum covariance, and take the case of the power
spectrum multipoles as the main example.
Our derivation assumes flat-sky approximation, and is applicable to wide
surveys by swapping in the curved-sky window functions.
We then generalize our method to the case of the correlation function, and
various other large-scale structure probes.

\subsection{3D Power Spectrum Covariance}
\label{sub:method_3D}

We begin by writing the number density of a discrete tracer field as a sum of Dirac
delta functions
\begin{equation}
    n_\data(\vx) = \sum_i \deltaD(\vx - \vx_i),
\end{equation}
whose underlying number density is a continuous field $\nbar(\vx) \equiv
\langle n_\data(\vx) \rangle$, determined by the tracer evolution and the
survey selection.
$\langle\,\rangle$ takes the ensemble average or the expected value of a
quantity.
The survey selection is most conveniently captured by a synthetic random
catalog, which in the simplest case is a Poisson process with mean $\langle
n_\rand(\vx) \rangle = \nbar(\vx) / \alpha$, where $\alpha$ is a constant
usually $\ll 1$ to reduce its shot noise.
We can write down the overdensity fields of those data and random
catalogs\footnote{
These overdensity fields of discrete catalogs are only formal, and they help to
simplify the calculation of shot noise terms, e.g.\ $\langle \delta_\data
\delta_\rand \rangle$ vanishes in \eqref{2pts}.}
\begin{align}
    \delta_\data &= - 1 + \frac1{\nbar(\vx)} \sum_{i \in\,\data} \deltaD(\vx - \vx_i),
    \nonumber\\
    \delta_\rand &= - 1 + \frac\alpha{\nbar(\vx)} \sum_{j \in\,\rand} \deltaD(\vx - \vx_j).
    \label{deltas}
\end{align}

Following~\cite{FeldmanKaiserEtAl94} (hereafter FKP), the galaxy overdensity is
estimated as the difference between the data and random catalogs weighted by a
weight function $w(\vx)$ (that maximizes the performance of an estimator like
the FKP weight, and/or minimizes some systematic effects)
\begin{align}
    \deltaW(\vx) &\equiv w(\vx) \bigl[ n_\data(\vx) - \alpha n_\rand(\vx) \bigr] \nonumber\\
    &\equiv W(\vx) \bigl[ \delta_\data(\vx) - \delta_\rand(\vx) \bigr]
    \equiv W(\vx) \delta(\vx).
    \label{deltaW}
\end{align}
On the second line we have rewritten the difference in number densities with
the difference in overdensities, and defined a combined overdensity field
$\delta \equiv \delta_\data - \delta_\rand$ that includes stochasticity from
both data and random catalogs.
We have also combined the survey selection and the weight function in the
definition of the window function
\begin{equation}
    W(\vx) \equiv \nbar(\vx) w(\vx).
\end{equation}
Throughout this paper $W$ denotes the windows on the fields, and is to be
distinguished from the other window factors introduced later for the 2-point
functions or their covariances.

Under the flat-sky approximation, the power spectrum can be simply estimated by
normalizing the squared Fourier-space overdensity before subtracting the shot
noise
\begin{equation}
    \hat{P}(\vk) = \frac{|\deltaW(\vk)|^2}{\W_0} - P_\shot,
    \label{Phat}
\end{equation}
where the shot noise power spectrum is
\begin{equation}
    P_\shot = \frac{\S_0}{\W_0},
    \label{Pshot}
\end{equation}
and the constant factors are given by\footnote{
In this paper we use the following shorthand notations for configuration-space
and Fourier-space integrals
\begin{equation*}
    \int_\vx \to \int \! \d^3\vx,
    \qquad
    \int_\vk \to \int \! \frac{\d^3\vk}{(2\pi)^3}.
\end{equation*}
}
\begin{align}
    \W_0 &= \int_\vx \nbar(\vx)^2 w(\vx)^2 = \int_\vx W(\vx)^2 = \int_\vk |W(\vk)|^2,
    \nonumber\\
    \S_0 &= (1 + \alpha) \int_\vx \nbar(\vx) w(\vx)^2.
    \label{W0S0}
\end{align}

Assuming flat sky and no redshift evolution, $\delta_\data(\vx)$ and
$\delta_\rand(\vx)$ are statistically homogeneous and satisfy
\begin{align}
    \ensavg{\delta_\data(\vk) \delta_\data(-\vk')}
    &= (2\pi)^3 \deltaD(\vk-\vk') P(\vk)
    + \int_\vx \frac1{\nbar(\vx)} e^{-i(\vk-\vk')\cdot\vx},
    \nonumber\\
    \ensavg{\delta_\rand(\vk) \delta_\rand(-\vk')}
    &= \int_\vx \frac\alpha{\nbar(\vx)} e^{-i(\vk-\vk')\cdot\vx},
    \nonumber\\
    \ensavg{\delta_\data(\vk) \delta_\rand(-\vk')} &= 0.
    \label{2pts}
\end{align}
Here the Dirac delta is a result of the translation invariance, $P(\vk)$ is the
3D power spectrum of the tracer, and the shot noise terms involving $1 / \nbar$
arise from the discrete nature of the data and random catalogs.
As independent samples from $\nbar(\vx)$, the data and the random overdensity
fields are not correlated, indicated by the third line in the above equations.

Plugging~\eqref{deltaW} and~\eqref{2pts} into the ensemble average of the
estimator in~\eqref{Phat} we get
\begin{equation}
    \ensavg{\hat{P}(\vk)}
    = \frac1{\W_0} \int_\vq P(\vk - \vq) |W(\vq)|^2
    \simeq \frac{P(\vk)}{\W_0} \int_\vq |W(\vq)|^2
    = P(\vk).
    \label{Phat_expect}
\end{equation}
The first equality shows that the expectation of $\hat{P}$ is a convolution of
the true power spectrum $P$ with a window.
To distinguish $\langle \hat{P} \rangle$ and $P$ we refer to them as the
convolved and the unconvolved power spectra, respectively.
In the second equality we have assumed that the scales of interest are much
smaller than the window size, i.e.\ $k \gg q$ for $\vq$ where $W(\vq)$ is
significant, so that a smooth power spectrum $P(\vk-\vq) \simeq P(\vk)$ can be
taken out of the integral.
Therefore the last equality proves that~\eqref{Phat} is an unbiased estimation
of the true power spectrum for modes much smaller than the survey scale.
In other words, the unconvolved and the convolved power spectra differ on large
scales due to the window.

The covariance function of the power spectrum estimator is define as
\begin{equation}
    \cov\bigl[\hat{P}(\vk), \hat{P}(\vk')\bigr]
    \equiv \ensavg{\hat{P}(\vk) \hat{P}(\vk')}
    - \ensavg{\hat{P}(\vk)} \ensavg{\hat{P}(\vk')}
\end{equation}
Substituting~\eqref{Phat} into the above equation, we split the covariance into
the disconnected and connected pieces in the multivariate cumulant expansion
\begin{align}
    \cov\bigl[\hat{P}(\vk), \hat{P}(\vk')\bigr]
    &= \cov^\disc\bigl[\hat{P}(\vk), \hat{P}(\vk')\bigr]
    + \cov^\conn\bigl[\hat{P}(\vk), \hat{P}(\vk')\bigr],
    \nonumber\\
    \cov^\disc\bigl[\hat{P}(\vk), \hat{P}(\vk')\bigr]
    &= \frac1{\W_0^2} \bigl| \ensavg{\deltaW(\vk) \deltaW(-\vk')} \bigr|^2
    + (\vk' \leftrightarrow -\vk'),
    \nonumber\\
    \cov^\conn\bigl[\hat{P}(\vk), \hat{P}(\vk')\bigr]
    &= \frac1{\W_0^2}
    \ensavg{\deltaW(\vk) \deltaW(-\vk) \deltaW(\vk') \deltaW(-\vk')}_\mathrm{c}.
    \label{cov_P}
\end{align}
The disconnected part $\cov^\disc$ captures the part of the 4-point correlation
arising from products of 2-point correlations, and the connected part
$\cov^\conn$ is from the excess correlation beyond $\cov^\disc$.
The subscript ``c'' on the ensemble average denotes the connected part or
the cumulant of the 4-point function.

In the conventional terminology, $\cov^\disc$ is referred to as the Gaussian
part and $\cov^\conn$ is named the non-Gaussian part.
This is true for a continuous random field, in which case the terms
``disconnected'' and ``Gaussian'' can be used interchangeably.
However, the conventional names are not accurate and can be confusing for point
processes like a galaxy catalog.
Since Gaussian and Poisson contributions enter both $\cov^\disc$ and
$\cov^\conn$, $\cov^\disc$ is not purely Gaussian and $\cov^\conn$ is not
completely free of Gaussian contribution.
Therefore in this paper we rename this covariance decomposition for
clarification.

Let's first look at the disconnected piece.
From~\eqref{deltaW} and~\eqref{2pts}
\begin{equation}
    \ensavg{\deltaW(\vk) \deltaW(-\vk')}
    = \int_{\vk''} P(\vk'') W(\vk-\vk'') W(\vk''-\vk')
    + (1 + \alpha) \int_\vx \nbar(\vx) w(\vx)^2 e^{-i(\vk-\vk')\cdot\vx},
\end{equation}
which includes both Gaussian and Poisson contributions.
For modes much smaller than the survey scale, the Gaussian term is only
important when $\vk''$ is close to both $\vk$ and $\vk'$, thus $P(\vk'')$
approximates $P(\vk)$ and $P(\vk')$.
So we can approximate the integral by taking the power spectrum out of the
convolution while preserving the $\vk \leftrightarrow \vk'$ exchange symmetry,
and then plug it into~\eqref{cov_P}
\begin{equation}
    \cov^\disc\bigl[\hat{P}(\vk), \hat{P}(\vk')\bigr]
    \approx \Bigl| \frac{P(\vk) + P(\vk')}2 \frac{\W(\vk-\vk')}{\W_0}
    + P_\shot \frac{\S(\vk-\vk')}{\S_0} \Bigr|^2
    + (\vk' \leftrightarrow -\vk'),
    \label{disc_P_}
\end{equation}
where we have introduced the following window factors
\begin{align}
    \W(\vq) = \int_\vx \W(\vx) e^{-i\vq\cdot\vx}
    &\equiv \int_\vx W(\vx)^2 e^{-i\vq\cdot\vx},
    \nonumber\\
    \S(\vq) = \int_\vx \S(\vx) e^{-i\vq\cdot\vx}
    &\equiv (1 + \alpha) \int_\vx \nbar(\vx) w(\vx)^2 e^{-i\vq\cdot\vx},
    \label{WS}
\end{align}
that modulate Gaussian and Poisson pieces, respectively.
$\W$ and $\S$ are to be distinguished from $W$, the window on the field.
Notice that the constants $\W_0$ and $\S_0$ introduced earlier in~\eqref{W0S0}
are special cases of $\W(\vq)$ and $\S(\vq)$ at $\vq = 0$.

The expansion of~\eqref{disc_P_} contains quadratic combinations $\W^2$,
$\W\S$, and $\S^2$, therefore we further define the window factor $\Q$'s as the
auto and cross 2-point correlation of $\W$ and $\S$.
In Fourier space
\begin{align}
    \QW(\vq) \equiv \W(\vq) \W(\vq)^* &= \int_\vs \QW(\vs) e^{-i\vq\cdot\vs},
    \nonumber\\
    \QS(\vq) \equiv \S(\vq) \S(\vq)^* &= \int_\vs \QS(\vs) e^{-i\vq\cdot\vs},
    \nonumber\\
    \QX(\vq) \equiv \W(\vq) \S(\vq)^* &= \int_\vs \QX(\vs) e^{-i\vq\cdot\vs},
    \label{Q}
\end{align}
with the configuration-space $\Q(\vs)$'s being the correlation functions of
$\W(\vx)$ and $\S(\vx)$, e.g.\
\begin{equation}
    \QX(\vs) = \int_\vx \W(\vx+\vs) \S(\vx).
    \label{Qs}
\end{equation}
So they can be measured as the correlation functions or power spectra of the
properly weighted random catalogs, which we describe in more details in
Sec.~\ref{sub:window}.

The final expression for the disconnected covariance given the power spectrum
$P$ and the window function $\Q$'s is
\begin{multline}
    \cov^\disc\bigl[\hat{P}(\vk), \hat{P}(\vk')\bigr]
    \approx \frac1{\W_0^2} \Bigl\{ P(\vk) P(\vk') \QW(\vk-\vk')
    \\
    + \bigl[P(\vk) + P(\vk')\bigr] \Re\bigl[\QX(\vk-\vk')\bigr]
    + \QS(\vk-\vk') \Bigr\}
    + (\vk' \leftrightarrow -\vk').
    \label{disc_P}
\end{multline}
Notice that under the same $P(\vk) \approx P(\vk')$ approximation we have
further simplified the expression by combining both $P(\vk) P(\vk)$ and
$P(\vk') P(\vk')$ terms into $P(\vk) P(\vk')$ while preserving the exchange
symmetry.
In general, the $\Q(\vk - \vk')$ windows have non-vanishing width determined by
the survey size, and its shape characterizes the correlation between
neighboring modes at $\vk$ and $\vk'$.
Also different $\Q$ windows generally have different shapes, and neglecting
this difference would lead to inaccurate $\cov^\disc$, which we show later in
Fig.~\ref{fig:shot_window}.

Now turning to the connected piece $\cov^\conn$, it is composed of a mixture of
the non-Gaussian, Poisson, and Gaussian contributions.
The non-Gaussian part arises from the gravitational mode-coupling, including
the trispectrum piece~\cite{MeiksinWhite99, ScoccimarroZaldarriagaEtAl99} and
the super-sample covariance (SSC)~\cite{HamiltonRimesEtAl06, HuKravtsov03,
TakadaHu13}.
The remaining parts of $\cov^\conn$ are various shot noise terms due to the
discrete nature of the tracer field~\cite{MeiksinWhite99}, and involves
Poisson, non-Gaussian (bispectrum), and Gaussian (power spectrum) components.
Overall $\cov^\conn$  is more difficult and complicated to model analytically
than $\cov^\disc$.
However, it is a smooth function of $\vk$ and $\vk'$ and can be well
approximated with a low-rank eigen-decomposition~\cite{HarnoisPen12,
MohammedSeljakVlah17}, which allows it to be measured from the data with the
internal covariance estimators.
Leaving the internal estimation of $\cov^\conn$ as a  future project, in this
paper we can obtain $\cov^\conn$ from the mock simulations by subtracting
$C^\disc$ from the mock sample covariance.
We find this empirical $\cov^\conn$ is indeed smooth and a low-rank component
with a principal component analysis.

\subsubsection{Diagonal Limit}
\label{ssub:diag}

We expect $\cov^\disc$ in~\eqref{disc_P} to reduce to the familiar diagonal
form when certain condition is met.
In the limit where $|\vk - \vk'|$ is much greater than the window scale, e.g.\
when they are from different bins with very wide bin width, $\Q(\vq)$
approaches $(2\pi)^3 \deltaD(\vq) \Q(\vs = 0)$, and $\cov^\disc$ reduces to a
diagonal covariance:
\begin{equation}
    \cov^\diag\bigl[\hat{P}(\vk), \hat{P}(\vk')\bigr]
    = (2\pi)^3 \deltaD(\vk - \vk') \biggl\{ \frac{P(\vk)^2}{V_\W}
    + \frac{2 P(\vk) P_\shot}{V_\times} + \frac{P_\shot^2}{V_\S} \biggr\}
    + (\vk' \leftrightarrow -\vk'),
    \label{diag}
\end{equation}
where $V$'s are effective volumes defined as follows
\begin{align}
    V_\W &\equiv \frac{\W_0^2}{\QW(\vs=0)}
    = \frac{\bigl[ \int_\vx \nbar(\vx)^2 w(\vx)^2 \bigr]^2}
        {\int_\vx \nbar(\vx)^4 w(\vx)^4},
    \nonumber\\
    V_\S &\equiv \frac{\S_0^2}{\QS(\vs=0)}
    = \frac{\bigl[ \int_\vx \nbar(\vx) w(\vx)^2 \bigr]^2}
        {\int_\vx \nbar(\vx)^2 w(\vx)^4},
    \nonumber\\
    V_\times &\equiv \frac{\W_0 \S_0}{\QX(\vs=0)}
    = \frac{\int_\vx \nbar(\vx)^2 w(\vx)^2 \int_{\vx'} \nbar(\vx') w(\vx')^2}
        {\int_\vx \nbar(\vx)^3 w(\vx)^4}.
\end{align}
We call~\eqref{diag} the diagonal limit, and take it as an ansatz for
$\cov^\disc$ for any $\vk$ and $\vk'$ even though the derivation only holds far
enough from the diagonal.
This diagonal covariance accounts for the size of the window which is captured
by $\Q(\vs = 0)$, but ignores the shape of $\Q(\vs)$, and as a result biases
the signal-to-noise ratio shown later in Sec.~\ref{ssub:SNR}.

In the diagonal limit, the $\cov^\disc$ matrix of the band-power, binned in
spherical $k$ shells following the later Sec.~\ref{sub:binning}, is
\begin{equation}
    \cov^\diag\bigl[\hat{P}(k_i), \hat{P}(k_j)\bigr] = 2 \deltaK_{ij}
    \int_{k_{i-\frac12}}^{k_{i+\frac12}} \frac{4\pi k^2\d k}{V_{k_i}}
    \biggl\{ \frac{P(k)^2}{N_\W}
    + \frac{2 P(k) P_\shot}{N_\times} + \frac{P_\shot^2}{N_\S} \biggr\},
    \label{diag_binned}
\end{equation}
where $\deltaK$ is the Kronecker delta, and $N$'s are the effective numbers of
modes associated with the effective volumes:
\begin{equation}
    N_\W = \frac{V_k V_\W}{(2 \pi)^3}, \quad
    N_\S = \frac{V_k V_\S}{(2 \pi)^3}, \quad
    N_\times = \frac{V_k V_\times}{(2 \pi)^3}.
    \label{N}
\end{equation}
Often only $N_\W$ is used to compute the diagonal covariance matrix, ignoring
the differences among $N$'s, that however could cause larger biases in the
signal-to-noise ratio (see Sec.~\ref{ssub:SNR}).
Eq.~\eqref{diag} and the constant effective volumes generalize the scale-dependent
effective volume commonly used in survey forecast~\cite{Tegmark97gal} for any
weights $w(\vx)$.

\subsection{Multipole Covariance}
\label{sub:method_pole}

In redshift space, because of the azimuthal symmetry about the line of sight
(LOS), the power spectrum does not have azimuthal dependence, i.e.\ $P(\vk) =
P(k, \uvk\cdot\uvn)$, where $\uvn$ is the LOS direction.
It is natural to decompose it into multipoles
\begin{equation}
    P(\vk) = \sum_\ell P_\ell(k) \L_\ell(\uvk\cdot\uvn),
    \label{pole}
\end{equation}
in which $\L_\ell$ is the Legendre polynomial of degree $\ell$.
The multipole moments can be obtained by\footnote{\label{fn:int_ang}
In this paper we use the following shorthand notation for averaging (instead of
integrating) over $4\pi$ solid angle of any vector $\vv$
\begin{equation*}
    \int_\uvv \to \int \frac{\d\Omega_\vv}{4\pi}.
\end{equation*}
}
\begin{equation}
    P_\ell(k) = (2\ell+1) \int_\uvk P(\vk) \L_\ell(\uvk\cdot\uvn).
    \label{Pl}
\end{equation}
In real space, all high-order multipoles usually vanish, leaving only the
monopole.

\eqref{pole} and~\eqref{Pl} also apply to their estimators $\hat{P}(\vk)$ and
$\hat{P}_\ell(k)$, so the disconnected covariance of multipoles follows
straightforwardly from~\eqref{disc_P}
\begin{multline}
    \cov^\disc\bigl[\hat{P}_\ell(k), \hat{P}_{\ell'}(k')\bigr]
    \approx \frac{2 (2\ell+1) (2\ell'+1)}{\W_0^2}
    \int_{\uvk, \uvk'} \L_\ell(\uvk\cdot\uvn) \L_{\ell'}(\uvk'\cdot\uvn)
    \Bigl\{
    \\
    P(\vk) P(\vk') \QW(\vk-\vk')
    + \bigl[P(\vk) + P(\vk')\bigr] \Re\bigl[\QX(\vk-\vk')\bigr]
    + \QS(\vk-\vk') \Bigr\}.
    \label{disc_pole_0}
\end{multline}
Since $P(\vk) = P(-\vk)$ for auto power spectrum, here we only consider the
even multipoles, for which the $\vk' \leftrightarrow -\vk'$ term in~\eqref{disc_P} simply doubles the first one giving the factor of 2 in the above
equation.

We compute the three terms in the curly brackets of~\eqref{disc_pole_0}
separately.
First we consider the $P^2$ term, expand the power spectra in multipoles, plug
in $\QW$ from~\eqref{Q}, and repeatedly couple Legendre polynomials with~\eqref{Legendre_product} and apply~\eqref{plane_wave_poles} to derive
\begin{align}
    & \int_{\uvk, \uvk'} \L_\ell(\uvk\cdot\uvn) \L_{\ell'}(\uvk'\cdot\uvn)
    P(\vk) P(\vk') \int_\vs \QW(\vs) e^{-i(\vk-\vk')\cdot\vs}
    \nonumber\\
    &= \sum_{\ell_1 \ell_2 \ell_3 \ell_4} P_{\ell_1}(k) P_{\ell_3}(k')
    (2\ell_2+1) \wj{\ell}{\ell_1}{\ell_2}000^2
    (2\ell_4+1) \wj{\ell'}{\ell_3}{\ell_4}000^2
    \nonumber\\
    &\hspace{20ex} \times (-i)^{\ell_2-\ell_4} \!
    \int \!4\pi s^2\d s\, \sum_{\ell''}
    \wj{\ell_2}{\ell_4}{\ell''}000^2
    \QW[\ell''](s) j_{\ell_2}(ks) j_{\ell_4}(k's).
    \label{disc_pole_1}
\end{align}
Similarly we obtain the cross term between power spectrum and shot noise ($P
\times P_\shot$)
\begin{align}
    & \int_{\uvk, \uvk'} \L_\ell(\uvk\cdot\uvn) \L_{\ell'}(\uvk'\cdot\uvn)
    \bigl[P(\vk) + P(\vk')\bigr]
    \Re\Bigl[ \int_\vs \QX(\vs) e^{-i(\vk-\vk')\cdot\vs} \Bigr]
    \nonumber\\
    &= \sum_{\ell_1 \ell_2} P_{\ell_1}(k)
    (2\ell_2+1) \wj{\ell}{\ell_1}{\ell_2}000^2
    (-i)^{\ell_2-\ell'} \! \int \!4\pi s^2\d s\, \sum_{\ell''}
    \wj{\ell_2}{\ell'}{\ell''}000^2
    \QX[\ell''](s) j_{\ell_2}(ks) j_{\ell'}(k's)
    \nonumber\\
    &\quad + (\ell, k \leftrightarrow \ell', k'),
    \label{disc_pole_2}
\end{align}
and the third ($P_\shot^2$) term
\begin{multline}
    \int_{\uvk, \uvk'} \L_\ell(\uvk\cdot\uvn) \L_{\ell'}(\uvk'\cdot\uvn)
    \int_\vs \QS(\vs) e^{-i(\vk-\vk')\cdot\vs}
    \\
    = (-i)^{\ell-\ell'} \!
    \int \!4\pi s^2\d s\, \sum_{\ell''}
    \wj{\ell}{\ell'}{\ell''}000^2
    \QS[\ell''](s) j_{\ell}(ks) j_{\ell'}(k's).
    \label{disc_pole_3}
\end{multline}

Note that in the above derivations the disconnected covariance only depends on
the multipole moments of the window $\Q$ due to the redshift-space symmetry
\begin{equation}
    \Q_\ell(s) = (2\ell+1) \int_\uvs \Q(\vs) \L_\ell(\uvs\cdot\uvn).
    \label{Ql}
\end{equation}
As shown in~\eqref{Q} and~\eqref{Qs}, $Q$ factors are really the 2-point
functions of the $\W$ and $\S$ windows, so their multipoles can be readily
measured from a random catalog of a survey.
Given $\Q_\ell(s)$ from the randoms and $P_\ell(k)$ from the data, we can
numerically evaluate~\eqref{disc_pole_1},~\eqref{disc_pole_2} and~\eqref{disc_pole_3} before summing them up in~\eqref{disc_pole_0} to obtain the
disconnected covariance for power spectrum multipoles.
The only remaining difficulty lies in the numerical integrals involving two
spherical Bessel functions, to which we provide a novel solution in
Sec.~\ref{sec:double_Bessel}.

\subsection{Window Functions}
\label{sub:window}

In deriving the disconnected covariance we have assumed the flat-sky
approximation.
However, many current and future surveys have wide sky coverages, making their
window functions fundamentally curved-sky entities.
Furthermore, the redshift-space clustering between two points depends on the
LOS direction that varies with the positions of the pair of points.
To account for this, the Yamamoto estimator proposed by
Ref.~\cite{YamamotoEtAl06} measures the multipoles of galaxy clustering with
respect to the midpoint direction of the two galaxies.
An alternative estimator also suggested by Ref.~\cite{YamamotoEtAl06} uses the
position of one galaxy of the pair as the reference direction, and can be
measured with a suite of efficient FFT-based algorithms~\cite{HandLiEtAl17,
SugiyamaShiraishiEtAl18, BianchiGil-MarinEtAl15, Scoccimarro15,
SlepianEisenstein16}.

Fortunately, it is easy to adapt our formalism to account for the curved-sky
effects both in the window function and the power spectrum.
For modes smaller than the window scale, our flat-sky equations remain a good
approximation to capture their local power-window coupling.
The LOS dependence can be simply incorporated in our window functions, e.g.\ in~\eqref{Ql} by letting the fixed line-of-sight direction $\uvn$ to depend on
$\vx$ and/or $\vx+\vs$.
Likewise, we can use the model fit to the power spectrum multipoles measured
with a LOS-dependent estimator, to compute $\cov^\disc$.

We use \texttt{nbodykit}, which calls \texttt{corrfunc}~\cite{corrfunc}, to
measure the window functions via fast pair counting.
With a random catalog that describes the survey geometry and weights, we count
the weighted pair in bins of both separation $s$ and the polar angle cosine
$\mu$, and then normalize the counts by the shell volume.
For example
\begin{equation}
    \QS(s,\mu) \approx \frac{3 \int_{\vx, \vs} \W(\vx+\vs) \S(\vs)}
        {2\pi (s_i^3-s_{i-1}^3) (\mu_j-\mu_{j-1})}
    \approx \frac{3 \alpha^2 \sum_{a b} \nbar(\vx_a) w(\vx_a)^2 w(\vx_b)^2}
        {2\pi (s_i^3-s_{i-1}^3) (\mu_j-\mu_{j-1})}.
\end{equation}
We take 441 logarithmic $s$ bins ranging from $1 \Mpch$ to $3.4 \Gpch$, and 100
$\mu$ bins from 0 to 1.
The \texttt{corrfunc} package measures the midpoint polar cosine
\begin{equation}
    \mu = \frac{\uvs\cdot(\vx+\vs/2)}{|\vx+\vs/2|}.
\end{equation}
To relate the double integral to the pair summation, we have simply replaced
$\int_\vx \nbar$ with $\alpha \sum_a$.
Given the $\Q$ windows in $s$ and $\mu$ bins, we then transform them into
multipoles to obtain an estimate of $\Q_\ell$ with~\eqref{Ql}.
All the odd order multipoles vanish because of the choice of midpoint for
measuring $\mu$.
With the same discretization trick, we can estimate the normalization constants
in~\eqref{W0S0} by replacing the integral with a summation over a random
catalog, e.g.\
\begin{equation}
    \W_0 \approx \alpha \sum_a \nbar(\vx_a) w(\vx_a)^2.
\end{equation}

We have seen in~\eqref{Phat_expect} that the expectation of the estimated power
spectrum is subject to a convolution with the window, and the unconvolved and
convolved power spectra differ significantly on large scales.
For accurate $\cov^\disc$, rather than directly using the estimated
$\hat{P}$, we should use an unconvolved model fit.
To achieve this we convolve the power spectrum model from~\cite{HandSeljakEtAl17} with the mask~\cite{WilsonPeacockEtAl15} before fitting
it to the data (more detailed description in Sec.~\ref{sub:results_pole} and
illustration in Fig.~\ref{fig:P}).
We then use the model before convolution to compute $\cov^\disc$.

\subsection{Corollary Covariances}
\label{sub:corollary}

We generalize our formalism below for more applications, and refer the readers
to App.~\ref{sec:disc} for detailed derivations.

\subsubsection{Cross Correlation and Cross Covariance}
\label{ssub:cross}

In Sec.~\ref{sub:method_3D} and~\ref{sub:method_pole} we have considered the
simplest case for 2-point function disconnected covariance: the covariance of
auto correlations of a single tracer.
In more general applications, one may need the cross covariance between
different sets of auto and cross correlations between different tracers.
For the most general case, we derive the equations for cross covariance of
cross 3D power spectra in App.~\ref{sub:disc_3D}.
The derivation parallels that of Sec.~\ref{sub:method_3D} and preserves the
Hermitian symmetry.
The equations can be easily generalized to all other cases including multipoles
and those in the rest of this subsections.

\subsubsection{Angular Power}

We can generalize the formalism developed in Sec~\ref{sub:method_pole} to the
angular power spectrum $C(\ell)$ which are generally used to measure the
correlations in observables from CMB, weak gravitational lensing as well as the
clustering of photometric samples of galaxies.
We derive the equations in App.~\ref{sub:disc_ang}, assuming flat sky and
Limber approximation.
Compared to the multipole case, the angular window effect is captured by some
$\Q$ windows similar to~\eqref{Ql} but defined on the sky, and the evaluation
involves double Bessel integral but with the zeroth-order Bessel function $J_0$
in place of the spherical Bessel functions.
The curved sky window can be incorporated simply by using the $\Q$ window
measured from the random catalog as a function of angular separation on the sky
similar to Sec.~\ref{sub:window}.
Our method can be a fast alternative to the calculation on the sphere as is
generally done for the CMB measurements~\cite{Planck15XI}.

\subsubsection{Projected Power and Projected Cross Multipoles}

When cross correlating a field with 3-dimensional information, such as
spectroscopic galaxies, with a field with poor radial information, e.g.\ weak
lensing shear or convergence, one can estimate their correlation projected at
fixed transverse separation using the radial information of the 3D field.
This is commonly done in galaxy-shear cross correlations measurements using
spectroscopic galaxies for which covariance was derived by~\cite{SinghEtAl2017}.
We derive the auto covariance for this case in App.~\ref{sub:disc_proj}, and
also its cross covariance with power spectrum multipoles, which is particularly
useful for their joint analysis aimed at testing theories of gravity~\cite{SimpsonEtAl2013, JoudakiEtAl2018}.
Because a projected power spectrum is indeed the transverse part of its 3D
counterpart, and the transverse power can be approximated by a series of
multipoles, we can adapt the multipole formalism to work for this cross
covariance.
We also derive alternative equations for the auto projected covariance similar
to the angular case in App.~\ref{sub:disc_ang}.

\subsubsection{Correlation Functions}
\label{ssub:corrfunc}

In configuration space, the Landy-Szalay estimator~\cite{LandySzalay93} is
typically employed to measure the correlation function, and it can be written
in our notation as
\begin{equation}
    \hat{\xi}(\vs) = \frac1{\Q_W(\vs)} \int_\vx \deltaW(\vx) \deltaW(\vx+\vs),
    \label{xihat}
\end{equation}
where we have defined the normalization factor analogous to~\eqref{Qs}
\begin{equation}
    \Q_W(\vs) \equiv \int_\vx W(\vx+\vs) W(\vx).
    \label{Q_W}
\end{equation}
\eqref{xihat} is free from the window effect, that can be shown easily in flat
sky by taking its expectation
\begin{align}
    \ensavg{\hat{\xi}(\vs)} &= \frac1{\Q_W(\vs)} \int_\vx W(\vx) W(\vx+\vs)
    \ensavg{\delta(\vx) \delta(\vx+\vs)}
    \nonumber\\
    &\equiv \frac{\xi(\vs)}{\Q_W(\vs)} \int_\vx W(\vx) W(\vx+\vs)
    = \xi(\vs),
\end{align}
where we have used the translation invariance of the correlation function
$\xi(\vs) \equiv \ensavg{\delta(\vx) \delta(\vx+\vs)}$.

The 3D correlation function and power spectrum are simply related by the
Fourier transform, again assuming translation invariance.
However, the relation between their estimators is more complex due to the
difference in normalizations:
\begin{align}
    \xi(\vs) &= \int_\vk P(\vk) e^{i\vk\cdot\vs},
    \nonumber\\
    \hat{\xi}(\vs)
    &= \frac{\W_0}{\Q_W(\vs)} \int_\vk \hat{P}(\vk) e^{i\vk\cdot\vs}.
\end{align}
Here $\W_0$ is equal to the value of $\Q_W(\vs=0)$ according to~\eqref{WS} and~\eqref{Q_W}.
So the covariance of the two 3D estimators are related by
\begin{equation}
    \cov\bigl[\hat{\xi}(\vs), \hat{\xi}(\vs')\bigr]
    = \frac{\W_0^2}{\Q_W(\vs) \Q_W(\vs')}
    \int_{\vk, \vk'} \cov\bigl[\hat{P}(\vk), \hat{P}(\vk')\bigr]
    e^{i\vk\cdot\vs} e^{i\vk'\cdot\vs'}.
\end{equation}

We are more interested in the correlation function covariances cast
in the multipole, angular, and projected form.
We develop their relations to the corresponding power spectrum covariances in
App.~\ref{sub:disc_corrfunc}.

\subsubsection{Wedges}
\label{ssub:wedge}

In this paper and especially Sec.~\ref{sub:method_pole}, we focus on using the
multipole moments to label the angular variation in 2-point functions.
Another popular choice is the so-called wedges, which are averages of the
2-point functions within bins of polar angle cosine, $\mu_i$, between bin edges
$(\mu_{i-\frac12}, \mu_{i+\frac12})$ for $i = 1, \ldots , N_\mathrm{wedge}$.
The power spectrum wedges are related to the multipoles by
\begin{equation}
    \hat{P}(k, \mu_i) = \int_{\mu_{i-\frac12}}^{\mu_{i+\frac12}}
    \frac{\d\mu}{\mu_{i+\frac12}-\mu_{i-\frac12}}
    \Bigl[ \sum_{\ell=0}^{\lmax} \hat{P}_\ell(k) \L_\ell(\mu) \Bigr]
    = \sum_{\ell=0}^{\lmax} \hat{P}_\ell(k) \bar{\L}_\ell(\mu_i),
\end{equation}
where we have truncated the multipoles at $\lmax$ and denoted the mean Legendre
polynomial across a wedge by $\bar{\L}$
\begin{equation}
    \bar{\L}_\ell(\mu_i) \equiv \int_{\mu_{i-\frac12}}^{\mu_{i+\frac12}}
    \frac{\L_\ell(\mu) \; \d\mu}{\mu_{i+\frac12}-\mu_{i-\frac12}}.
\end{equation}
Therefore
\begin{equation}
    \cov^\disc\bigl[\hat{P}(k, \mu_i), \hat{P}(k', \mu_j)\bigr]
    = \sum_{\ell=0}^{\lmax} \sum_{\ell'=0}^{\lmax}
    \cov^\disc\bigl[\hat{P}_\ell(k), \hat{P}_{\ell'}(k')\bigr]
    \bar{\L}_\ell(\mu_i) \bar{\L}_{\ell'}(\mu_j).
\end{equation}
The same relation applies to the correlation functions\footnote{
Note that the wedges are different from the $\mu$ bins used for estimating
$\hat{\xi}_\ell(s)$.
While one can directly estimate $\hat{P}_\ell(k)$, due to its normalization
$\hat{\xi}_\ell(s)$ is usually converted from $\hat{\xi}(s, \mu_i)$ which is
estimated in very fine $\mu$ bins.
Here the width of the wedges is assumed to be much wider than the width used
for such conversion.}
by replacing $P \to \xi$, $k \to s$.

\subsection{Binning}
\label{sub:binning}

All previous results in this section assume the power spectrum covariance as a
function of continuous $k$ and $k'$.
When numerically evaluating the equations we then discretize $k$ and $s$ on
logarithmic grids (more details in Sec.~\ref{sec:double_Bessel}).
In practice, the estimated power spectra are averaged in bins $k_i$ separated
by bin edges $(k_{i-\frac12}, k_{i+\frac12})$ for $i = 1, \ldots , N_\mathrm{bin}$.
In accordance, we integrate the analytic prediction within spherical $k$-shells
\begin{equation}
    \cov\bigl[\hat{P}(k_i), \hat{P}(k_j)\bigr]
    = \int_{k_{i-\frac12}}^{k_{i+\frac12}} \frac{4\pi k^2\d k}{V_{k_i}}
    \int_{k_{j-\frac12}}^{k_{j+\frac12}} \frac{4\pi k'^2\d k'}{V_{k_j}}
    \cov\bigl[\hat{P}(k), \hat{P}(k')\bigr],
\end{equation}
where $V_{k_i} = 4\pi (k_{i+\frac12}^3 - k_{i-\frac12}^3) / 3$ is the volume of
the $i$th $k$-shell.
The numerical integration is achieved by interpolating the covariance function
along each $k$ at a time with B-splines before integrating the piecewise
polynomials.\footnote{
We package and release this simple and general-purpose utility at
\url{https://github.com/eelregit/avgem}.}

In the case of correlation functions, because their denominators are scale
dependent, we need to average separately the numerator and denominator in $s$
bins before dividing them.
The binning integrals are similar to the above equation.

\section{Double Bessel Quadrature}
\label{sec:double_Bessel}

\begin{figure}[t]
    \centering
    \includegraphics[width=0.5\linewidth]{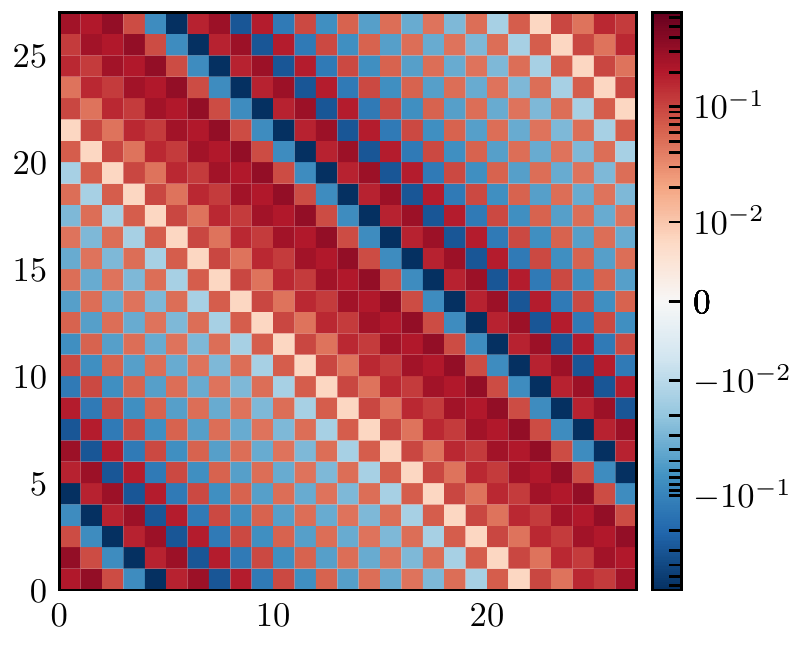}
    \includegraphics[width=0.49\linewidth]{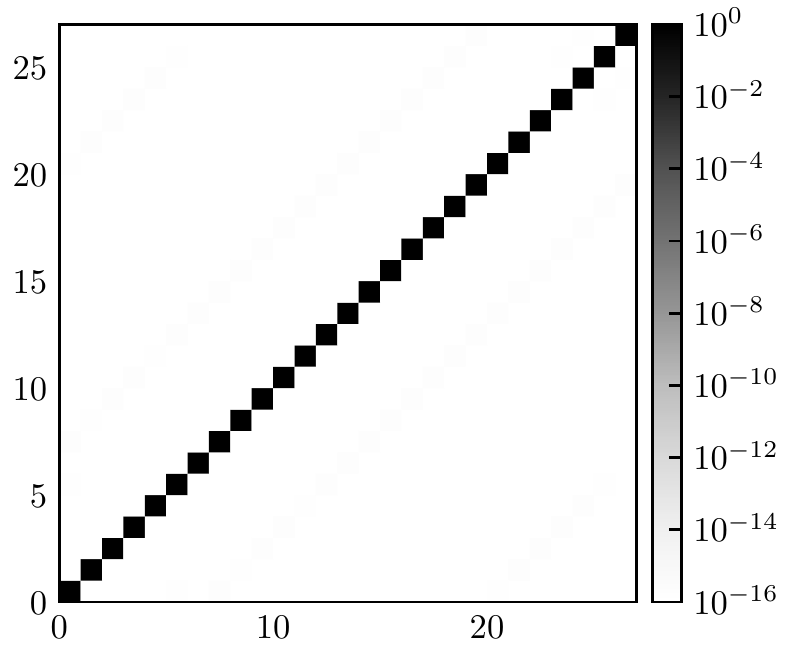}
    \caption{An example $27 \times 27$ Hankel matrix $H_0$, for the $J_0$
    Hankel transform.
    We visualize it in the left panel, and demonstrate in the right panel that
    the squared Hankel matrix is an identity matrix.
    As a circulant matrix, a Hankel matrix carries out the convolution
    operation in the FFTLog algorithm (Eq.~\ref{Hankel} and~\ref{Hankel_matrix}).
    It is also the building block of the double Bessel quadrature method
    (Eq.~\ref{double_Bessel} and~\ref{quadrature}).
    }
    \label{fig:Hankel}
\end{figure}

Eq.~\eqref{disc_pole_1},~\eqref{disc_pole_2}, and~\eqref{disc_pole_3} all have
integrals involving two spherical Bessel functions.
More generally, we are interested in evaluating the following integral with
Bessel functions
\begin{equation}
    G(y, y') = \int_0^\infty \!x\d x\; F(x) J_\nu(x y) J_{\nu'}(x y'),
    \label{double_Bessel}
\end{equation}
which proves useful also for the applications in Sec.\ref{sub:corollary}.
This applies to~\eqref{disc_pole_1},~\eqref{disc_pole_2}, and~\eqref{disc_pole_3} given
\begin{equation}
    j_n(t) = \sqrt\frac{\pi}{2t} J_{n+\frac12}(t).
\end{equation}
Usual quadrature methods struggle to converge for~\eqref{double_Bessel} because
the Bessel kernels are highly oscillatory and damp slowly.
Recent solutions proposed by Ref.~\cite{AssassiEtAl17, GebhardtJeong18}
generalized the FFTLog algorithm~\cite{Talman78, Hamilton00} to perform the
integral transform from $F(x)$ to $G(y, y')$ for each fixed ratio $y' / y$.

Here we present a fast and simple algorithm that potentially suits the
covariance calculation better than that in Ref.~\cite{AssassiEtAl17,
GebhardtJeong18}.
First let's look at a simpler integral with only one Bessel function
\begin{equation}
    B(x) = \int_0^\infty \!y\d y\; A(y) J_\nu(x y),
    \label{Hankel}
\end{equation}
known as the Hankel transform.
We can perform integral transforms like this one using the \texttt{mcfit}~\cite{mcfit} package, which implements and generalizes the FFTLog algorithm.
This method exploits the convolution theorem in terms of $\ln x$ and $\ln y$.
It approximates $A(e^{\ln y})$ with truncated Fourier series over one period of
the periodic approximant, and Fourier-transforms the kernel $J_\nu(e^{\ln x
+\ln y})$ analytically.
Because of the exact treatment of the kernel function, this algorithm is ideal
for oscillatory kernels like the Bessel functions.

Specifically, we compute an equivalent form of~\eqref{Hankel}
\begin{equation}
    x B(x) = \int_0^\infty \!\frac{\d y}y\; \bigl[ y A(y) \bigr]
    \bigl[ x y J_\nu(x y) \bigr],
    \label{Hankel_unitary}
\end{equation}
with the input, output, and kernel functions rescaled by corresponding linear
powers: $A(y) \to y A(y)$, $B(x) \to x B(x)$, and $J_\nu(t) \to t J_\nu(t)$.
\texttt{mcfit} evaluates~\eqref{Hankel_unitary} by discretizing $x$ and $y$
to grids of equal logarithmic intervals $\Delta \equiv \delta\ln x = \delta\ln
y$, on which the linear transform essentially takes the following matrix form
\begin{equation}
    B_i = x_i^{-1} \sum_{j=1}^N H_{\nu,ij} y_j A_j,
    \label{Hankel_matrix}
\end{equation}
where the $H_\nu$ is an $N \times N$ real matrix carrying out the convolution,
and $N$ is the number of grid points.
$H_\nu$ is a Hankel circulant matrix, i.e.\ $H_{\nu, ij} = h_{i+j}$ with $h$
being periodic ($h_{i+N} = h_i$).
The explicit form of $h$ can be found in Eq.(B26) of Ref.~\cite{Hamilton00}.
As an example we show in Fig.~\ref{fig:Hankel} a $27\times27$ $H_0$ matrix with
$\Delta = 0.34$.
The above rescaling in~\eqref{Hankel_unitary} is important because it (together
with the FFTLog low-ringing condition when $N$ is even) makes $H_\nu$ a unitary
matrix and involutory (being its own inverse), leading to the most numerically
stable results as demonstrated in the right panel of Fig.~\ref{fig:Hankel} by
the nearly perfect agreement between $H_0^2$ and the identity matrix to machine
precision.

Notice that formally~\eqref{Hankel_matrix} can be obtained by applying the
following replacement rule on~\eqref{Hankel_unitary} or~\eqref{Hankel}:
\begin{equation}
    x y J_\nu(x y) \longrightarrow \frac{H_{\nu, ij}}\Delta,
    \qquad
    \int \frac{\d y}y \longrightarrow \Delta \sum_j.
    \label{rule}
\end{equation}
While we emphasize that this rule is not rigorously true, i.e.\ $H_{\nu, ij}
\not\propto J_\nu(x_i y_j)$, coincidentally, our final recipe for the double
Bessel integral~\eqref{double_Bessel} also follows from this rule.

Now we are ready to derive our double Bessel quadrature algorithm.
Consider the following integral that further integrates~\eqref{double_Bessel}
with some arbitrary auxiliary functions $A(y)$ and $A'(y')$ over $y$ and $y'$
\begin{equation}
    I = \int_0^\infty \!y\d y \int_0^\infty \!y'\d y'\; A(y) G(y, y') A'(y').
    \label{I}
\end{equation}
We can change the order of integration to first integrate out $y$ and $y'$ to
get $B(x)$ and $B'(x)$, respectively, as shown from~\eqref{Hankel} to~\eqref{Hankel_matrix}, so that
\begin{equation}
    I = \int_0^\infty \!x\d x\; F(x) B(x) B'(x)
    \approx \Delta \sum_i F_i
    \sum_j H_{\nu, ij} y_j A_j \sum_{j'} H_{\nu', ij'} y_{j'} A'_{j'}.
\end{equation}
Alternatively, a direct discretization of~\eqref{I} reads
\begin{equation}
    I \approx \Delta^2 \sum_{j j'} y_j^2 A_j G_{j, j'} y_{j'}^2 A'_{j'}.
    \label{I_discretized}
\end{equation}
Since $A$ and $A'$ are arbitrary, by comparing the above two equations we
derive the formula to compute $G(y, y')$ on the logarithmic grid
\begin{equation}
    G_{j, j'} = \frac1\Delta \sum_i
    y_j^{-1} H_{\nu, ij} F_i H_{\nu', ij'} y_{j'}^{-1}.
    \label{quadrature}
\end{equation}
This result can also be obtained, formally, by applying the same replacement
rule~\eqref{rule} to~\eqref{double_Bessel}.

Evaluation of~\eqref{quadrature} is simple and fast, as the formula can be
optimized to $\order(N^2 \log N)$ time complexity:
constructing the $H_\nu$ matrix requires one Fast Fourier Transform per $\nu$;
the summation over $i$ is a convolution therefore can be done efficiently with
FFT by calling \texttt{mcfit};
and our algorithm does not require any additional special function
implementations as do Ref.~\cite{AssassiEtAl17, GebhardtJeong18}.

Furthermore,~\eqref{I_discretized} implies that $G_{j, j'}$ approximates $G(y,
y')$ as average values in the logarithmic intervals.
Generally $G(y, y')$ peaks sharply on the $y = y'$ diagonal and almost vanishes
elsewhere, with roughly constant peak width.
Our algorithm does not intend to compute the exact value of $G$ at point $(y,
y')$ but a smoothed version over the $(y_j, y_{j'})$ grid.
This turns out to be beneficial as we are only interested in its
coarse-grained values (see Sec.~\ref{sub:binning}) rather than the finer
structure, and~\eqref{quadrature} allows us to perform the numerical
integration without scanning a very fine grid.
On the other hand, the method in Ref.~\cite{AssassiEtAl17, GebhardtJeong18}
evaluates $G(y, y')$ itself, thus need to sample densely around the peak width
to arrive at the same coarse-grained result.
And because of the logarithmic grid and constant peak width, the sampling rate
is determined by the large $y$ end, and resolving the peak at large $y$ leads
to waste of computation at small $y$.

\begin{hide}
$G_{j, j'}$ conserves
\begin{equation}
    \int \d y \d y' G(y, y') \approx \Delta^2 \sum_{j j'} y_j G_{j, j'} y_j'
    = \Delta \sum_i F_i.
\end{equation}
How to use it for re-binning?
\end{hide}

\section{Simulations}
\label{sec:data}

\begin{figure}[t]
    \centering
    \includegraphics[width=0.7\linewidth]{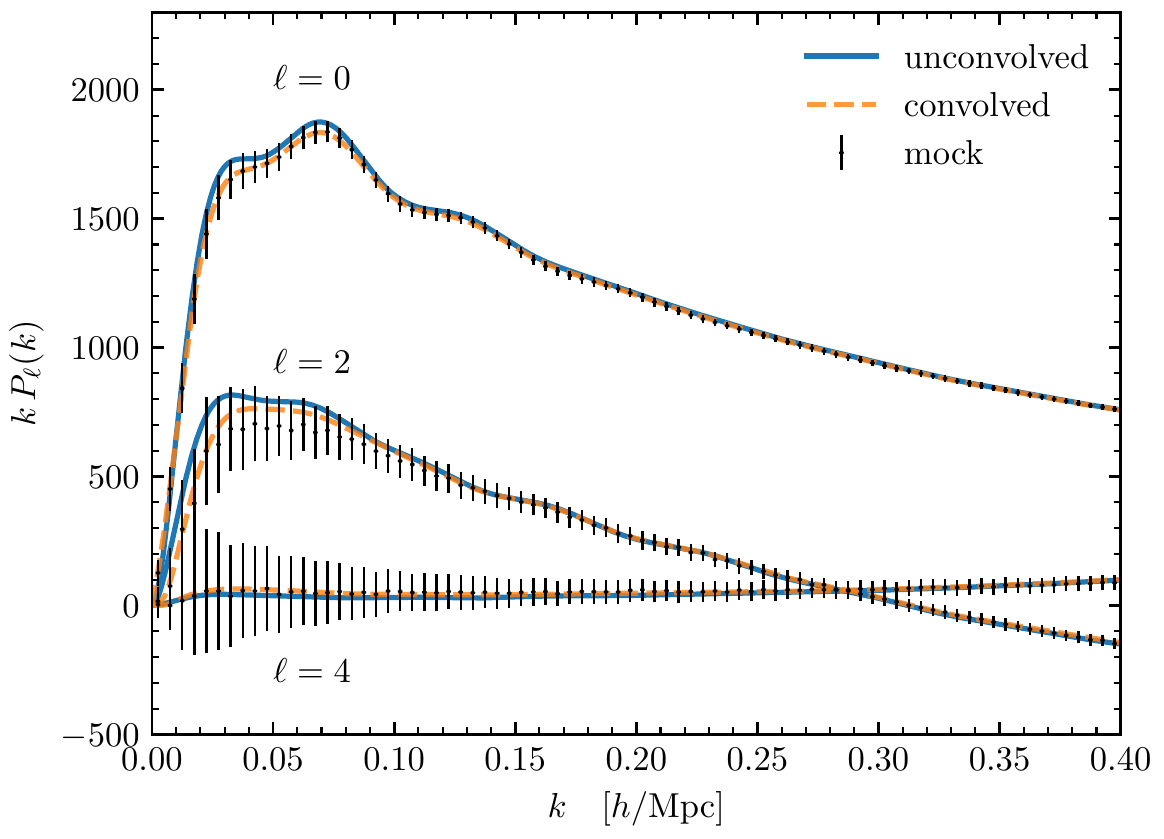}
    \caption{Power spectrum multipoles, of 1000 Patchy mocks for NGC of BOSS
    DR12 in $0.5 < z < 0.75$.
    The points with errobars show the sample mean and variance of the
    power spectrum multipoles of the galaxy mocks.
    The ``convolved'' curve shows the best-fit RSD model from~\cite{HandSeljakEtAl17}, including the convolution with the appropriate
    window function~\cite{WilsonPeacockEtAl15} to accounts for mixing of power
    by the window.
    We also show the ``unconvolved'' model prior to the convolution.
    We will use this ``unconvolved'' model along with the window factors $\Q$
    shown in Fig.~\ref{fig:Q} to compute the covariance matrices.
    }
    \label{fig:P}
\end{figure}

\begin{figure}[t]
    \centering
    \includegraphics[width=0.7\linewidth]{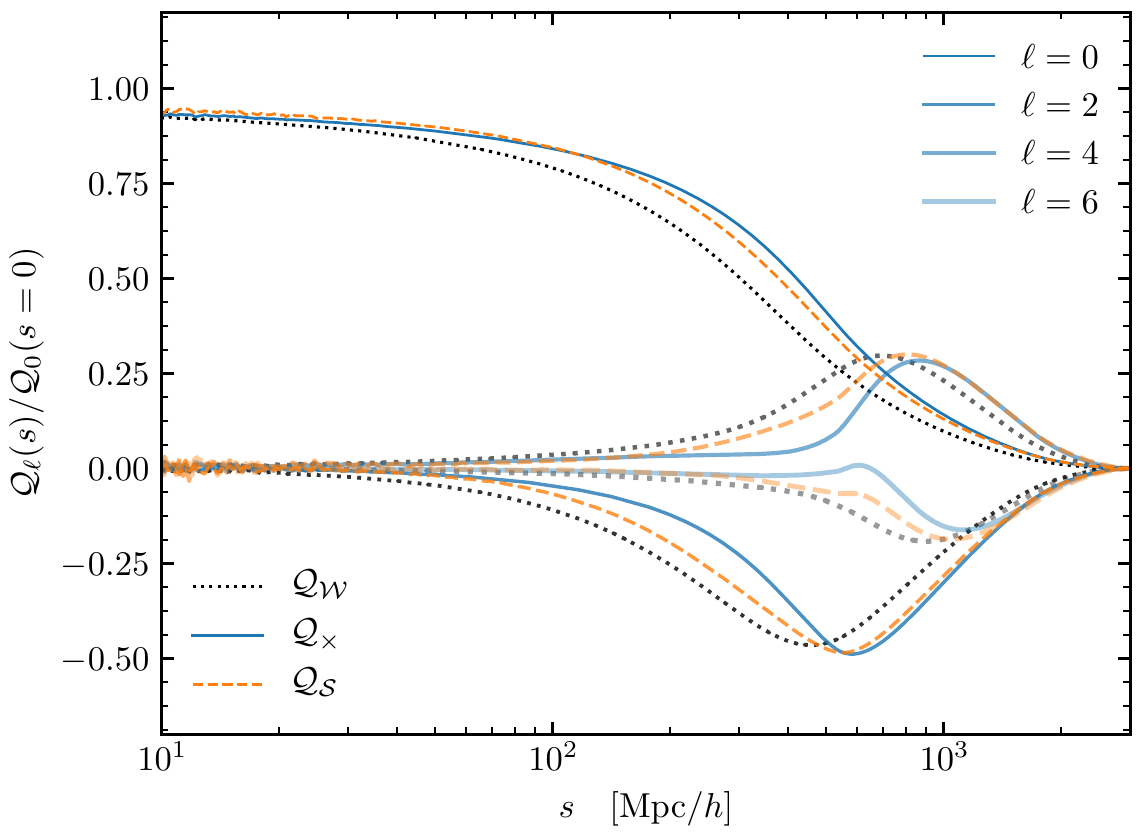}
    \caption{Window function multipoles, of 1000 Patchy mocks for NGC of BOSS
    DR12 in $0.5 < z < 0.75$.
    The $\Q$ window functions are directly related to the covariance, defined
    in~\eqref{Q},~\eqref{Qs} and~\eqref{Ql}.
    We normalize the window multipoles by their zero-lag monopoles to compare
    their shapes.
    As the covariance is $\sim (P + P_\shot)^2$, three $\Q$ factors in
    different colors describe the windows for different pieces: $\QW$ for
    $P^2$, $\QX$ for $P \times P_\shot$, and $\QS$ for $P_\shot^2$.
    Their normalized monopoles start from 1 on small scales and vanish on
    large scales beyond the size of the window.
    The higher order multipoles are only non-vanishing around the window scale
    and capture the anisotropy of the window.
    Combining the unconvolved $P_\ell$ in Fig.~\ref{fig:P} and the $\Q_\ell$
    here, we can compute the disconnected covariance of $\hat{P}_\ell$
    following Sec.~\ref{sub:method_pole}.
    }
    \label{fig:Q}
\end{figure}

To verify our analytic covariances, we compare them to the empirical results
measured from $N_\mock = 1000$ realizations of the Multidark-Patchy galaxy
mocks~\cite{KitauraEtAl16}, produced for BOSS data release 12 (DR12)~\cite{AlamEtAl17}.
The Patchy algorithm are based on the augmented Lagrangian perturbation theory
and a stochastic halo biasing scheme calibrated on high-resolution simulations.
It then uses halo occupation distribution to construct catalogs to match the
observed galaxy clustering and its redshift evolution.
In this work we use the North Galactic Cap (NGC) region covering 6800
$\mathrm{deg}^2$ of the sky, and two redshift ranges: the first ($0.2 < z <
0.5$) and the third ($0.5 < z < 0.75$) redshift bins of BOSS DR12.
The cosmology assumed by the Patchy mocks is $\Omega_\mathrm{m} = 0.307$,
$\Omega_\mathrm{b} = 0.048$, $h = 0.678$, $\sigma_8 = 0.829$.

We estimate the redshift-space power spectrum multipoles with the endpoint
Yamamoto estimator~\cite{YamamotoEtAl06} in the third redshift bin.
We employ the FFT-accelerated algorithm~\cite{HandLiEtAl17} enabled by
multipole decomposition~\cite{SlepianEisenstein15}, implemented in the
large-scale structure toolkit \texttt{nbodykit}~\cite{nbodykit}.
We use the standard Landy-Szalay~\cite{LandySzalay93, SinghEtAl2017} estimator
to estimate the correlation functions in the first redshift bin.
Correlation function multipoles are transformed from the correlation function
measured in 50 $\mu$ bins and the projected correlation function is computed
using the ratio of pair counts integrated over the line of sight.
Note that our projected correlation function estimator is dimensionless and
different from the projected correlation function which has dimensions of
length (see e.g.~\cite{SinghEtAl2017} for correlation function with
dimensions of length).

For a pair of 2-point observables, denoted by $O$ and $O'$, the unbiased sample
covariance matrix from the $N_\mock$ realizations is
\begin{equation}
    \widehat\cov(O, O') = \frac1{N_\mock-1} \sum_{r=1}^{N_\mock}
    (\hat{O}_r - \bar{O}) (\hat{O}'_r - \bar{O}'),
    \label{cov_sample}
\end{equation}
where $\bar{O}$ ($\bar{O}'$) is the sample mean of $\hat{O}_r$ ($\hat{O}'_r$)
for $r$ over all $N_\mock$ realizations.

For all the test cases in this paper, we use the following weight function
\begin{equation}
    w = w_\mathrm{FKP} \times w_\mathrm{veto},
\end{equation}
in which $w_\mathrm{FKP} \equiv 1 / (1 + \nbar \times 10^4 \, \Mpc^3 / h^3)$,
and the veto mask excludes certain types of unobservable regions, e.g.\ near
bright stars.
To estimate the power spectrum, we use synthetic random catalogs that have 50
times as many points as the mock galaxies, i.e.\ $\alpha = 0.02$.
For correlation functions we use 5 times as many randoms, i.e.\ $\alpha=0.2$.
And we estimate the window functions using pair counting with $\alpha =0.2$.

\section{Results}
\label{sec:results}

In this section, we validate our analytic disconnected covariance against the
sample covariance measured from the mock simulations.
We compare them in two test cases: the power spectrum multipole covariance, and
the joint covariance of projected correlation function and correlation function
multipoles.

\subsection{Power Spectrum Multipole Covariance Matrix}
\label{sub:results_pole}

\begin{figure}[t]
    \centering
    \includegraphics[width=0.48\linewidth]{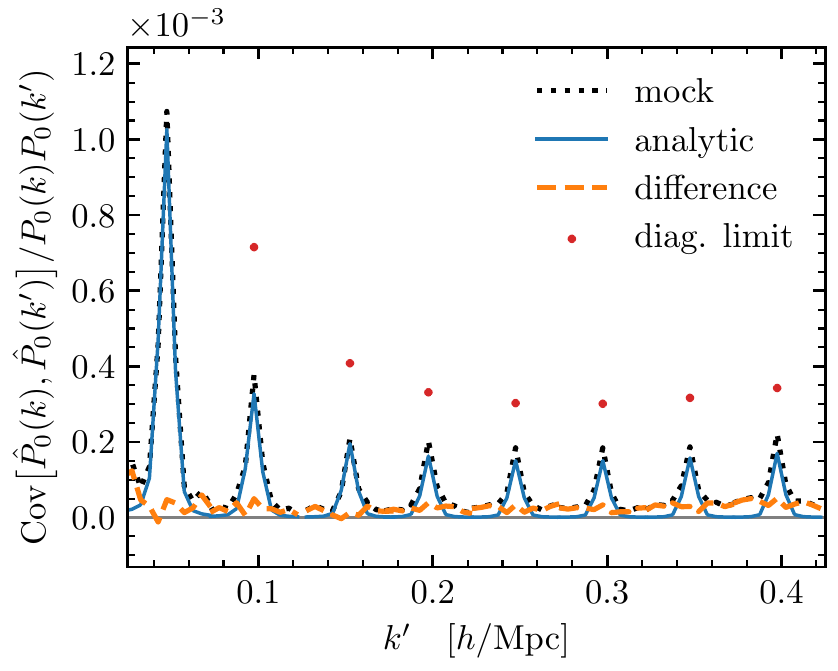}
    \includegraphics[width=0.51\linewidth]{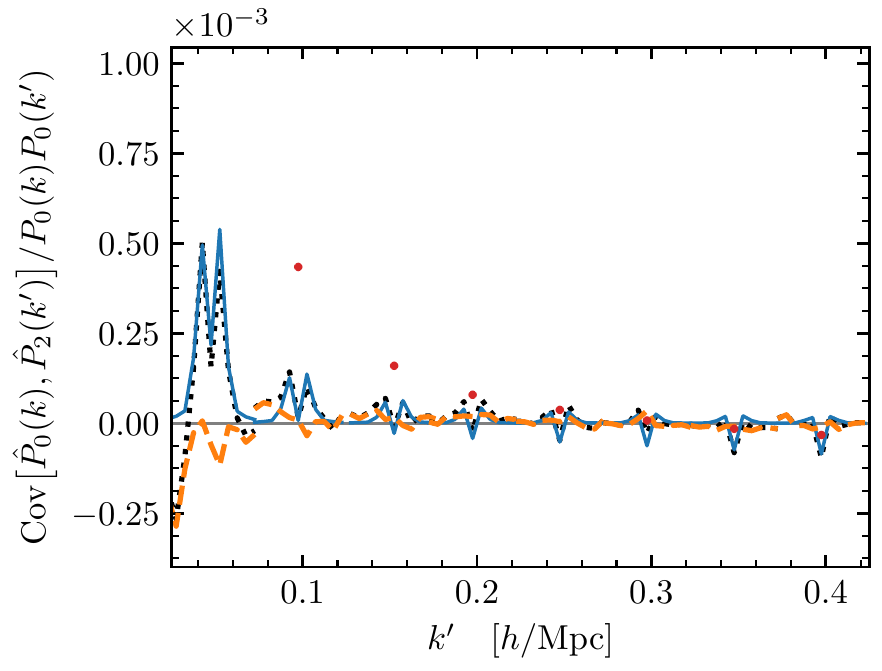}
    \includegraphics[width=0.475\linewidth]{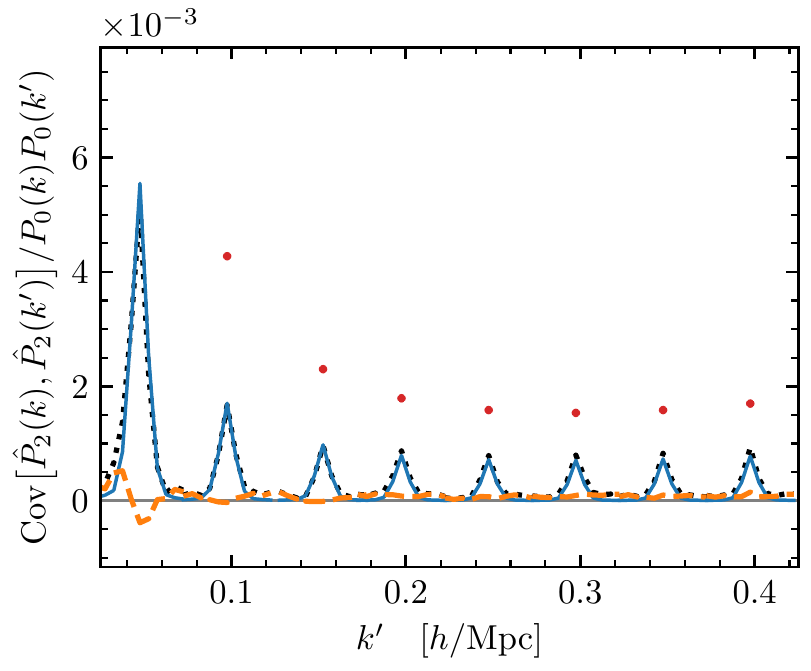}
    \includegraphics[width=0.5\linewidth]{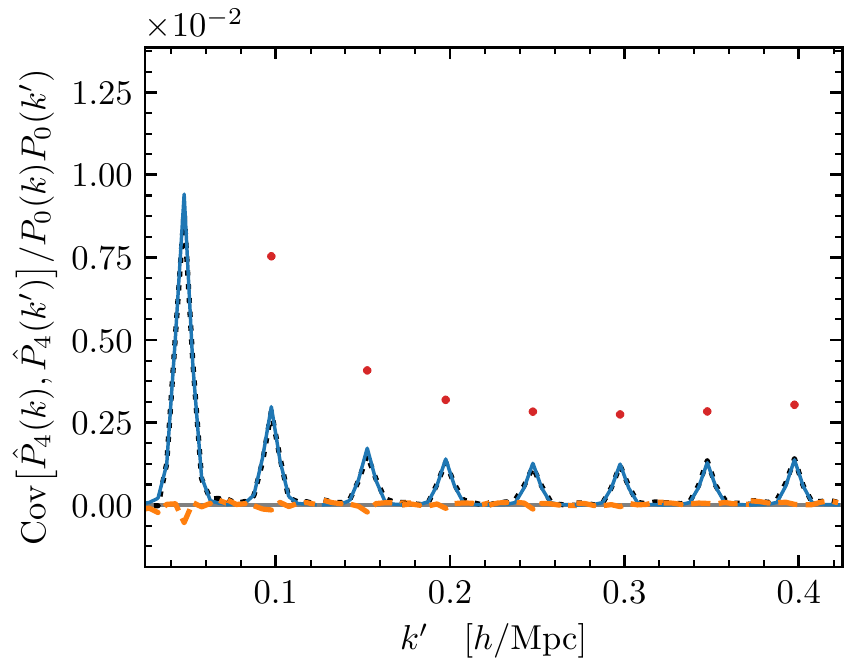}
    \caption{Slices of covariance matrices of power spectrum multipoles, of bin
    width $0.005 \hMpc$, normalized by the monopole.
    Different panels shows the auto and cross covariance matrices of different
    multipoles, as labeled next to the vertical axes.
    Every spike corresponds to a slice of covariance matrix near the diagonal
    ($k' \approx k$) at fixed $k$ (marked by the position of the peak).
    The difference (dashed orange) between the mock (dotted black) and analytic
    (solid blue) covariance is consistent with a smooth component, as expected
    from the connected covariance in the mocks.
    This demonstrates that the analytic result has accounted for most, if not
    all, of the disconnected covariance of the mocks.
    For comparison we also show the diagonal limit of the analytic covariances
    from~\eqref{diag_binned} in red dots, that ignores the inhomogeneity and
    anisotropy of the survey window and are only nonzero at the peak of each
    spike with very different values.
    }
    \label{fig:cov_slice}
\end{figure}

\begin{figure}[t]
    \centering
    \includegraphics[width=0.7\linewidth]{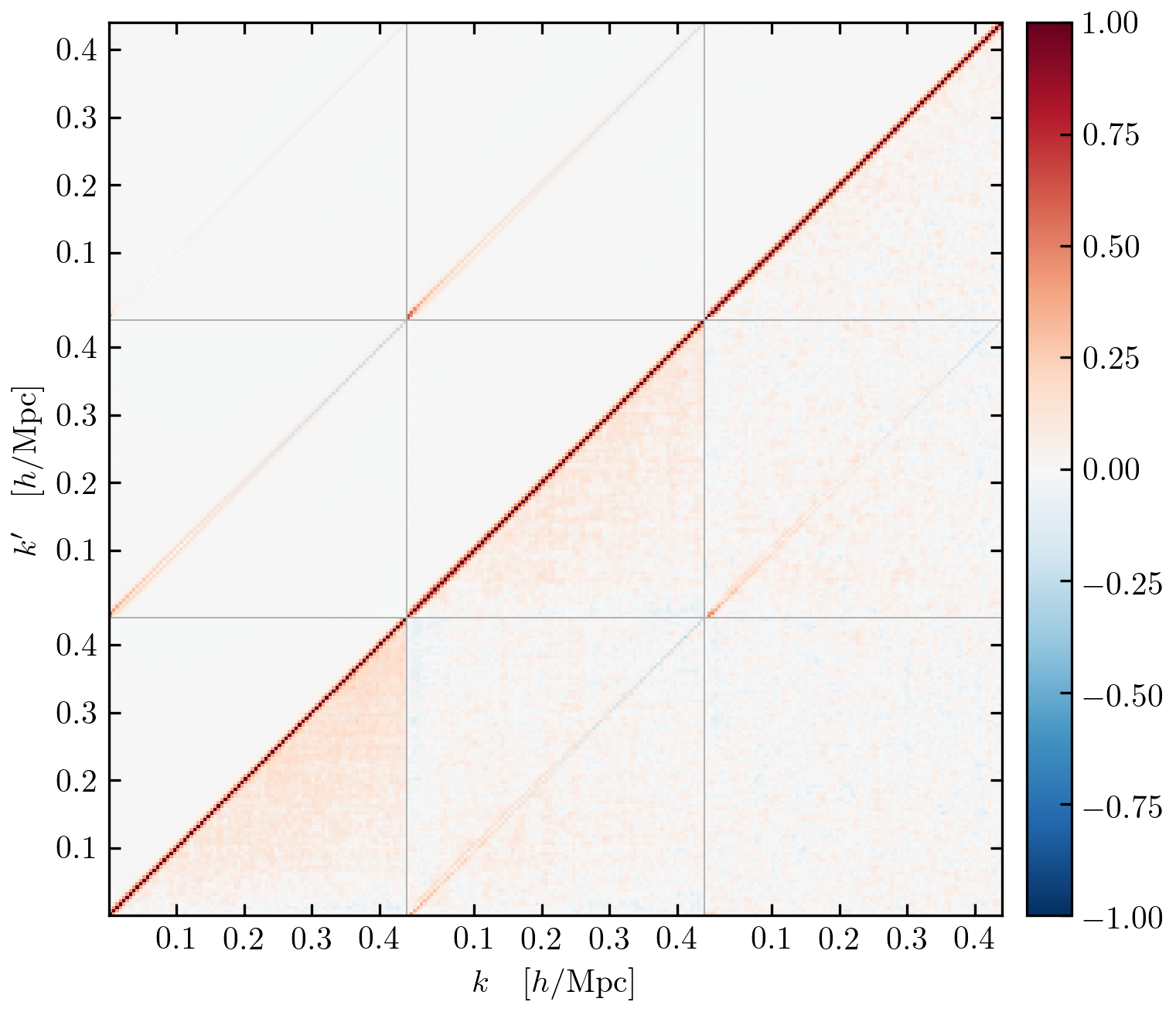}
    \includegraphics[width=0.7\linewidth]{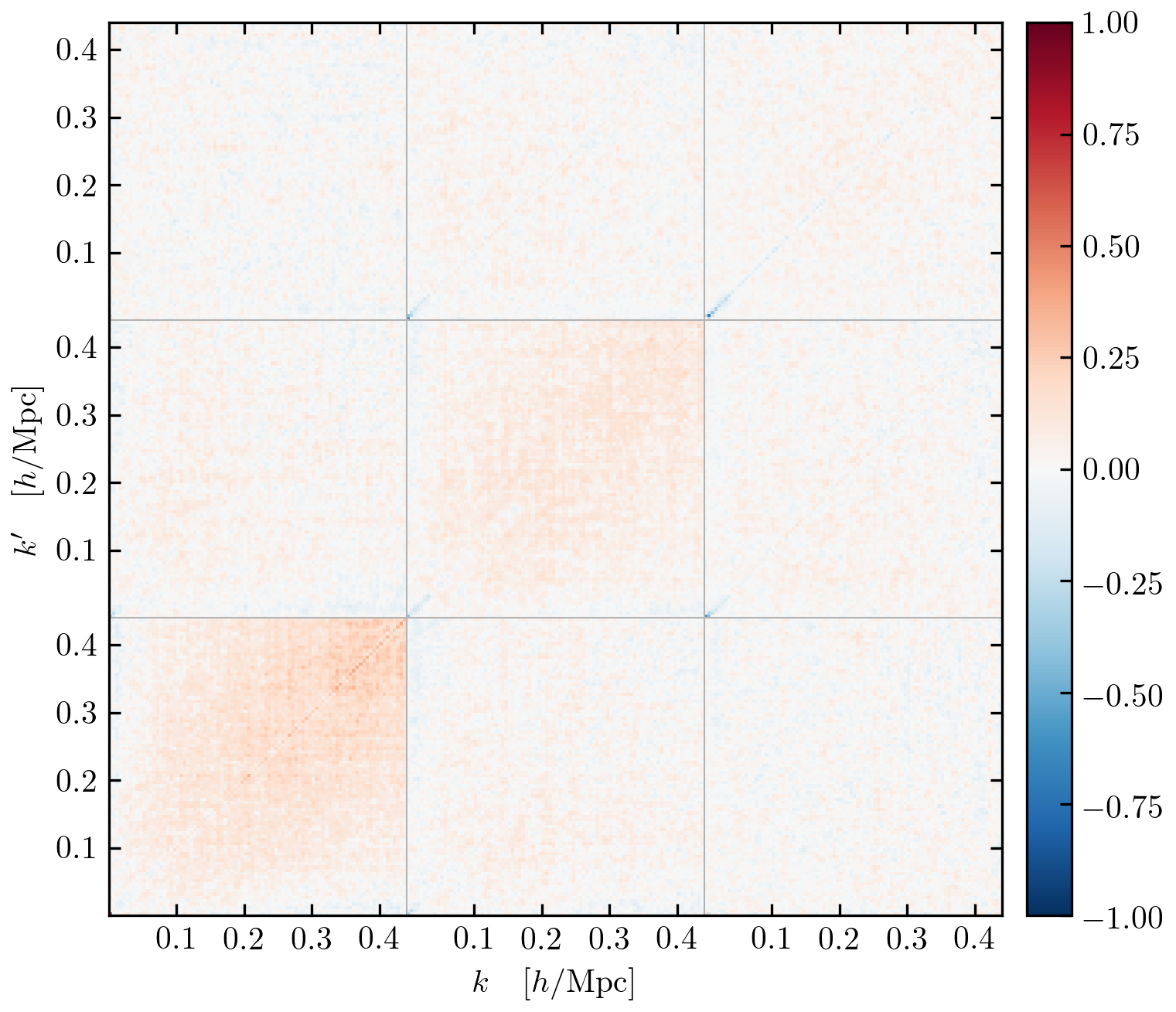}
    \caption{Correlation matrix of the power spectrum multipoles.
    In the $3 \times 3$ blocks, from bottom to top and from left to right, we
    visualize the auto and cross correlations of $\hat{P}_0$, $\hat{P}_2$,
    $\hat{P}_4$.
    The top panel compares the analytic result in its upper triangular block with
    the mocks covariance in the lower triangular block.
    The bottom panel shows the difference between the mock and analytic
    covariance matrices, normalized by the diagonal of the latter.
    As was shown in Fig.~\ref{fig:cov_slice}, this residual component is
    smooth, and captures the connected part of the covariance matrix.
    There are some remaining low $k$ diagonal features that are more prominent
    for larger $\ell$, and are likely due to the bias in mock $\hat{P}_\ell$
    for $\ell > 0$ caused by the sparse angular sampling of the Fourier grid.
    }
    \label{fig:corr_mat}
\end{figure}

\begin{figure}[t]
    \centering
    \includegraphics[width=0.49\linewidth]{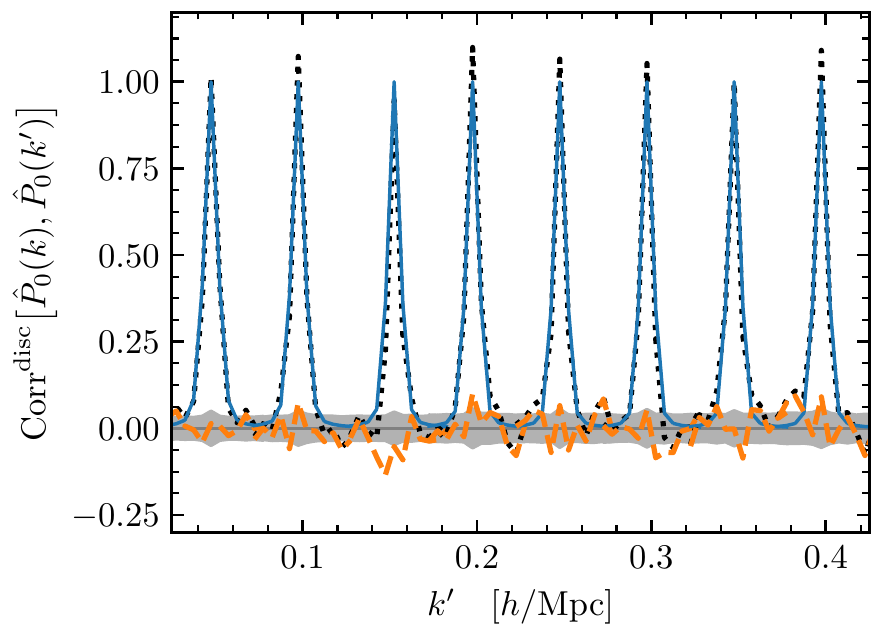}
    \includegraphics[width=0.49\linewidth]{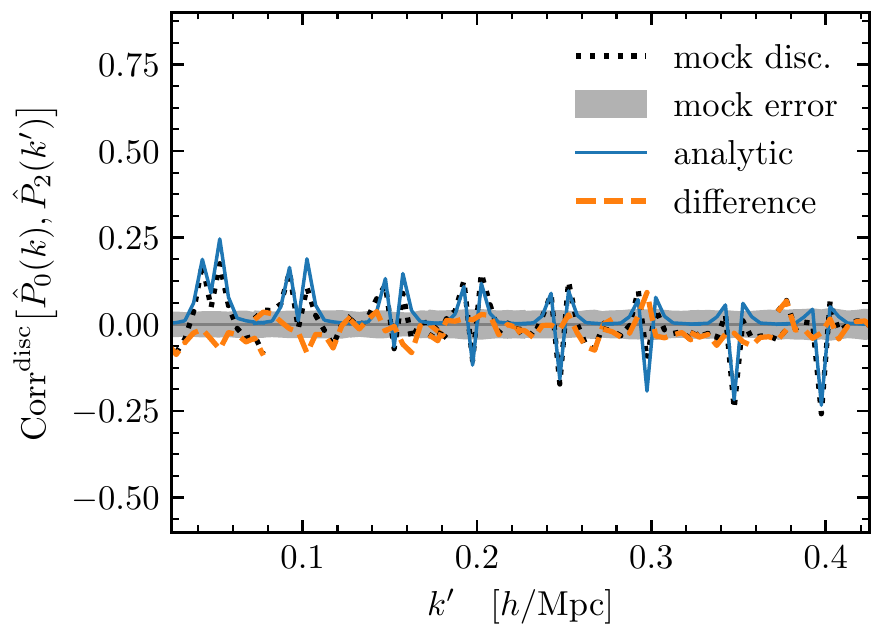}
    \includegraphics[width=0.49\linewidth]{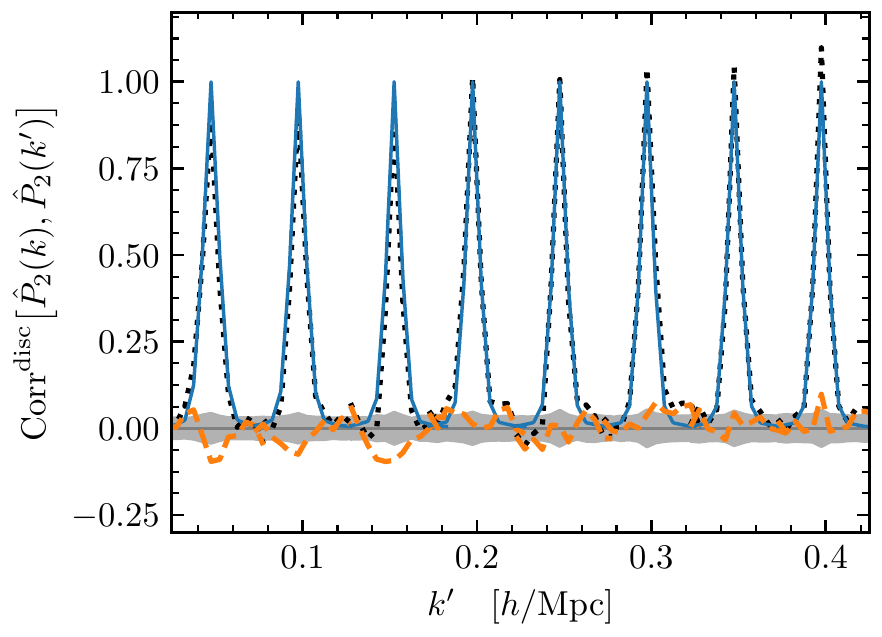}
    \includegraphics[width=0.49\linewidth]{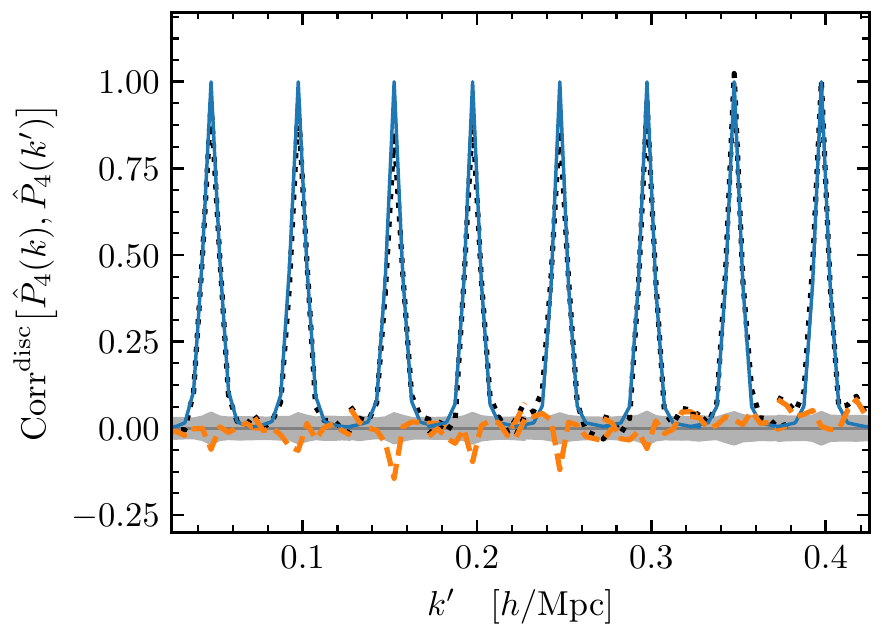}
    \caption{Slices of disconnected correlation matrix of power spectrum
    multipoles.
    Every spike corresponds to a slice of covariance matrix near the diagonal
    ($k' \approx k$) at fixed $k$ (marked by the position of the peak).
    For fair comparison, all covariance matrices are normalized by the diagonal
    of the analytic one.
    The mock disconnected covariance is obtained by subtracting the connected
    part, approximated with the first 4 principal components of the lower panel
    of Fig.~\ref{fig:corr_mat}, from its full covariance.
    Its difference from the analytic result reflects latter's accuracy and this
    residual is comparable to the bootstrapping errors on the mock sample
    covariance (grey band around zero).
    }
    \label{fig:corrdisc_slice}
\end{figure}

\begin{figure}[t]
    \centering
    \includegraphics[width=0.49\linewidth]{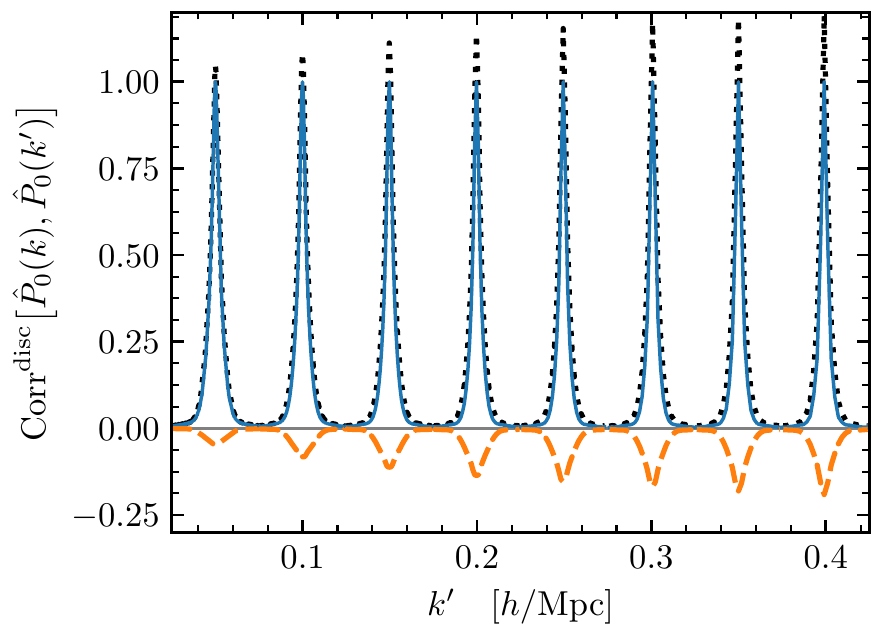}
    \includegraphics[width=0.49\linewidth]{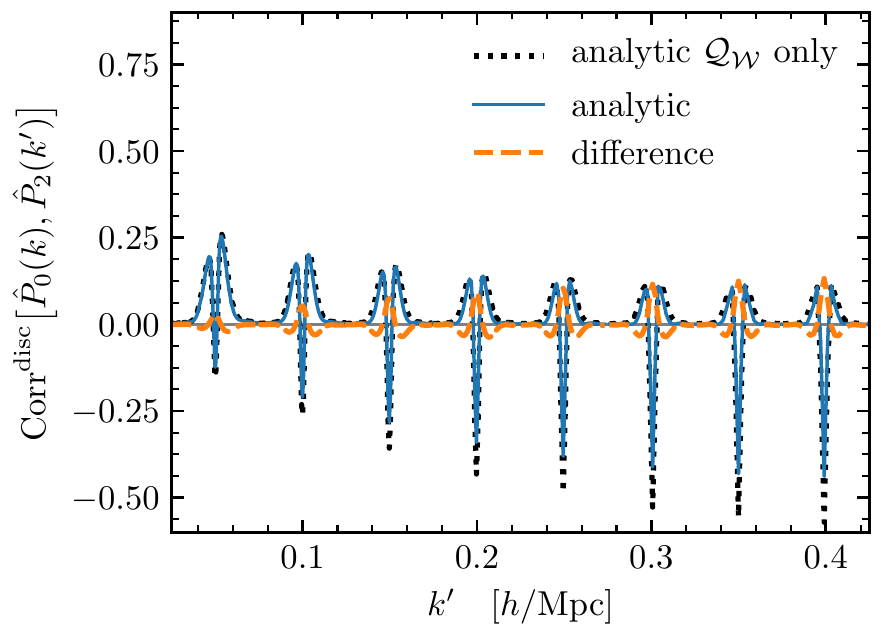}
    \includegraphics[width=0.49\linewidth]{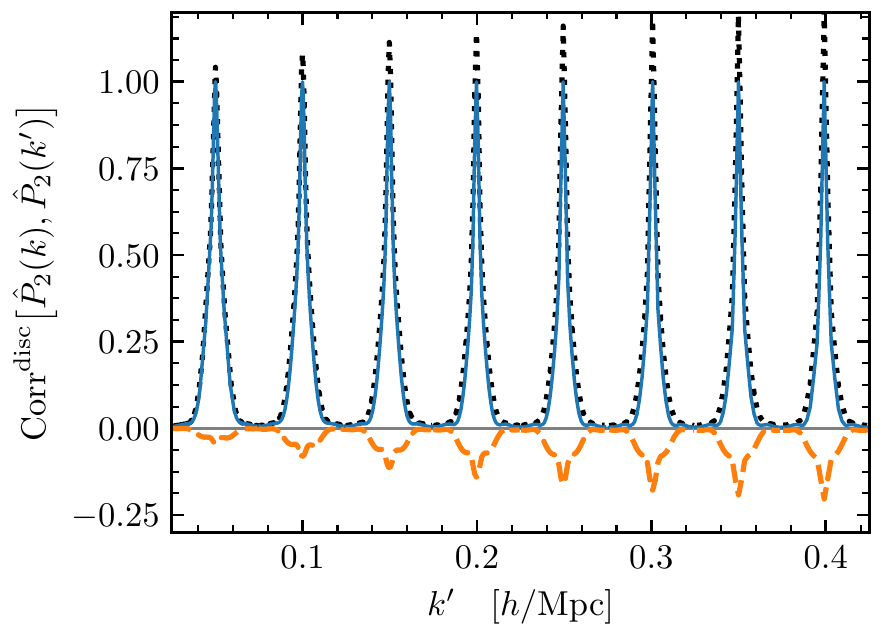}
    \includegraphics[width=0.49\linewidth]{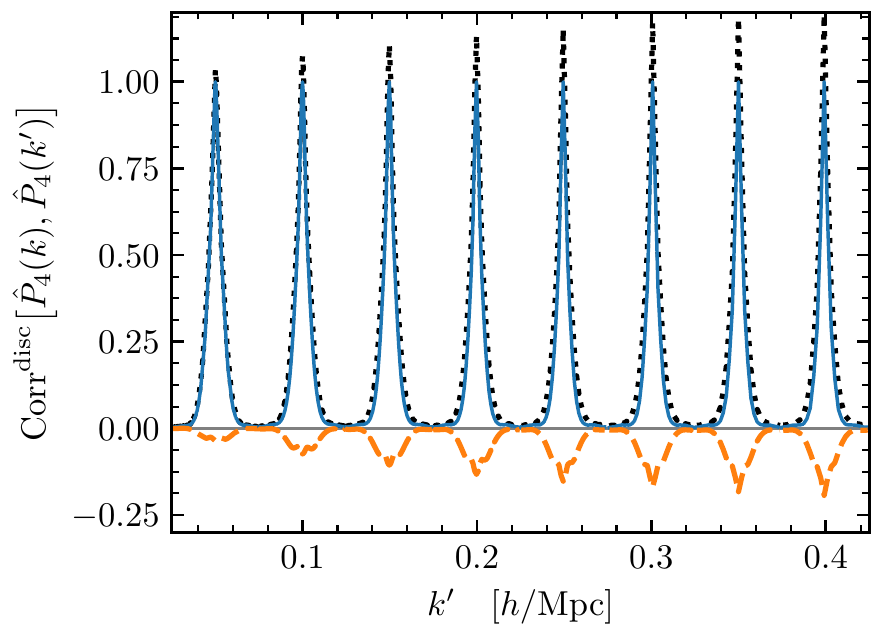}
    \caption{Slices of analytic disconnected correlation matrix of power
    spectrum multipoles, using different $\Q$ functions.
    The solid blue lines show the same but unbinned result as in previous
    figures, with which we normalize all the covariance in this figure for fair
    comparison.
    It was computed with the window functions $\QW$, $\QX$, and $\QS$, which
    are shown in Fig.~\ref{fig:Q}.
    To illustrate the necessity of modeling three separate windows, in dotted
    black line we show an approximation computed by assigning the same shape to
    all $\Q$'s: $\QX \leftarrow \QW$ and $\QS \leftarrow \QW$.
    The dashed orange lines give the difference between the two approaches.
    Although three $\Q$ functions have similar shapes, their differences can
    leads to 20\% difference in the disconnected covariance.
    This difference  depends on the shot noise magnitude, and is larger on
    small scales where shot noise is more important.
    }
    \label{fig:shot_window}
\end{figure}

We compute the analytic disconnected covariance of $P_\ell(k)$ using the
equations in Sec.~\ref{sub:method_pole}, that requires the unconvolved $P_\ell$
and the window function multipoles $\Q_\ell$ as inputs.
To obtain the former, we measure the sample mean and variance of the power
spectrum multipoles from the 1000 Patchy mocks, shown in Fig.~\ref{fig:P}.
With all model parameters initialized to their fiducial values, we use the RSD
model from~\cite{HandSeljakEtAl17} to compute a diagonal covariance
matrix~\cite{GriebEtAl16} and fit this model to the mock sample mean over the
$k$ range from $0.02 \hMpc$ to $0.4 \hMpc$.
We obtain the maximum a posteriori (MAP) estimate of the model parameters using
the L-BFGS algorithm, and subsequently the diagonal covariance is re-computed
using the best-fit model.
We then repeat the fitting process until convergence is reached.
This analysis is performed with the \texttt{pyrsd} package.
To account for the window convolution in~\eqref{Phat_expect}, we convolve the
model with the appropriate window~\cite{WilsonPeacockEtAl15} before fitting it
to the mocks.
We show both the unconvolved and convolved model curves in Fig.~\ref{fig:P}.

To measure the window functions represented by their multipole moments, as
defined in~\eqref{Q},~\eqref{Qs}, and~\eqref{Ql}, we use the random catalog of
Patchy mocks, and follow the procedure described in Sec.~\ref{sub:window}.
As shown from~\eqref{disc_P_} to~\eqref{disc_P}, because the covariance has a
quadratic form in power spectrum and its shot noise, i.e.\ $\sim (P +
P_\shot)^2$, the three different window factors describe the covariance shape
of different pieces: $\QW$ for $P^2$, $\QX$ for $P \times P_\shot$, and $\QS$
for $P_\shot^2$.
They capture the same survey geometry but differ in the weights, and therefore
have similar shapes and different normalizations.
Our analytic $\cov^\disc$ is derived in the flat-sky limit, so only depends on
the even order $\Q_\ell$'s.
We find that our results have converged when truncating the multipoles at
$\lmax = 10$.
In Fig.~\ref{fig:Q}, we show the shapes of the first 4 $\Q_\ell$'s after
normalizing them by their monopoles at zero lag.
Starting from the small scales, the window monopoles are $\lesssim 1$ while the
other multipoles are negligible, reflecting the fact that the windows are
homogeneous and isotropic in the small scale limit.
Moving to larger scales, the monopoles start to decrease, whereas the higher
order multipoles rise due to the anisotropy of the window on those scales.
And finally all the multipoles vanish beyond the size of the window.

With the unconvolved $P_\ell$ the $\Q_\ell$, we can evaluate the analytic
disconnected covariance of $\hat{P}_\ell$ using the equations in
Sec.~\ref{sub:method_pole}.
The equations involve double Bessel integrals that we compute using the
quadrature method introduced in Sec.~\ref{sec:double_Bessel}.
And finally we average the covariance function in linear $k$ bins of width
$0.005 \hMpc$ to obtain the covariance matrix, as described in
Sec.~\ref{sub:binning}.
The computation takes 2 minutes on a single CPU.

In Fig.~\ref{fig:cov_slice}, we compare our analytic covariance for power
spectrum multipoles, $P_0, P_2, P_4$, to the mock sample covariance computed
using eq.~\eqref{cov_sample}.
Out of the six combinations, we focus on the auto covariance of the $P_0$,
$P_2$, $P_4$, and the cross covariance between $P_0$ and $P_2$, in different
panels.
All panels are normalized by $P_0$ (without shot noise), to present the
relative error with respect to the monopole.
Because the covariance matrices peak sharply near the $k$-$k'$ diagonal where
the disconnected piece dominates, and are smooth otherwise (as better
illustrated by Fig.~\ref{fig:corr_mat}), for each 2D matrix we plot a few
slices across fixed $k$ near the diagonal ($|k' - k| \lesssim 0.0025 \hMpc$).

Recall that our analytic covariance matrix only has the disconnected
contribution, whereas the mock result also contains the connected piece.
By subtracting the analytic matrices from the mock ones, we can obtain an
estimate of the connected covariance matrices:
\begin{equation}
    \widehat\cov^\conn = \widehat\cov - \cov^\disc.
    \label{cov_conn}
\end{equation}
We also show the differences in mock and analytic covariance in
Fig.~\ref{fig:cov_slice} and find them smooth as expected, implying that the
analytic result has accounted for most, if not all, of the disconnected
covariance of the mocks.

Having shown the amplitude and shape of the covariance in a few slices near the
diagonal, in Fig.~\ref{fig:corr_mat} we present the shape of the full
covariance matrix by normalizing it as the linear correlation coefficients
given by
\begin{equation}
    \corr(O, O') = \frac{\cov(O, O')}{\sqrt{\cov(O, O) \cov(O', O')}},
\end{equation}
where $\cov \in \{\cov^\disc, \widehat\cov\}$ for analytic and mock
covariances, respectively.
The visualized matrix has $3 \times 3$ blocks, with the horizontal and vertical
blocks corresponding to $O \in \{P_0(k), P_2(k), P_4(k)\}$ and $O' \in
\{P_0(k'), P_2(k'), P_4(k')\}$, respectively.
Because the covariance is symmetric, we combine the analytic and mock
covariances in the top panel, with the upper triangular part showing the
analytic covariance and the lower triangular part showing the mock covariance.
The diagonals of corresponding blocks in the upper and lower triangles have
very similar shapes and scale dependence.
We again observe that far from the block diagonals the analytic blocks have
vanishing elements, while the mock blocks vary smoothly with some noise.

To visualize the full connected covariance, we normalize the~\eqref{cov_conn}
by the diagonal of the analytic disconnected covariance
\begin{equation}
    \corr^\conn(O, O')
    = \frac{\widehat\cov^\conn(O, O')}{\sqrt{\cov^\disc(O, O) \cov^\disc(O', O')}},
    \label{corr_conn}
\end{equation}
and show it in the lower panel of Fig.~\ref{fig:corr_mat}.
Note that it is not normalized properly as a correlation matrix, but correspond
to the ratio of connected to disconnected covariance.
As was shown in Fig.~\ref{fig:cov_slice}, this residual component is smooth,
suggesting that the disconnected part has been cleanly removed.
There are some notable remaining low $k$ diagonal features for $\ell = 4$ and
$\ell = 2$, and they are likely due to the bias in mock $\hat{P}_\ell$ caused
by the sparse angular sampling of the Fourier grid.

Note that the connected covariance estimated with~\eqref{cov_conn} also
includes possible error in our analytic $\cov^\disc$ as well as the noise in
the mock sample covariance.
From $\widehat\cov^\conn$ we can extract the connected part alone, by
exploiting the fact that it is smooth and has a low-rank approximation~\cite{MohammedSeljakVlah17, HarnoisPen12}.
We perform principal component analysis on $\corr^\conn$ in~\eqref{corr_conn},
and find that only the first 4 principal components have eigenvectors with
broadband features, while the rest are consistent with noise.
Therefore we can replace $\corr^\conn$ with this rank 4 substitute, and rescale
it by the denominator in~\eqref{corr_conn} to obtain a smooth estimate of
$\widehat\cov^\conn$, which we denote by $\widehat\cov^{\conn, \smooth}$.
Now if we subtract that from the mock sample covariance, the residual is an
estimate of its disconnected covariance from the mocks
\begin{equation}
    \widehat\cov^\disc = \widehat\cov - \widehat\cov^{\conn, \smooth}.
\end{equation}

This mock $\widehat\cov^\disc$ can be compared directly to the analytic
$\cov^\disc$, which we present in Fig.~\ref{fig:corrdisc_slice} as slices
across the correlation matrix.
The slices are arranged in the same way as in Fig.~\ref{fig:cov_slice}.
Also shown is the difference between the analytic and mock results, with
everything normalized by the analytic $\cov^\disc$ as in~\eqref{corr_conn} so
that the comparison is fair.
This residual contains possible flaw in the analytic $\cov^\disc$ and the error
on mock sample covariance.
We can quantify the level of the latter with the bootstrapping method, and find
it comparable to the residual.
Therefore our analytic disconnected covariance is accurate to the extent of the
errors on the sample covariance from 1000 mocks.

Before moving on, we examine the dependence of analytic $\cov^\disc$ on the
$\Q$ windows.
In this paper we have been modeling and measuring three distinct covariance
windows, namely $\QW$, $\QX$, and $\QS$.
Given their similarity in shapes (see Fig.~\ref{fig:Q}), one possible
simplification is to approximate all $\Q$ windows with the same shape.
In practice this can be achieved in~\eqref{disc_P_} by adding the shot noise to
the power spectrum and multiplying the squared sum by a single $\Q$ window, for
which we use $\QW$.
We compare this simplified treatment with our previous result in
Fig.~\ref{fig:shot_window}.
Since all results are analytic, the unbinned curves are shown for better
visualization.
We find that the approximation leads to 20\% error in the disconnected
covariance.
And this error depends on the shot noise magnitude since we have changed the
shot noise window, and is larger on small scales where shot noise is more
important.
Therefore, it is necessary to use three separate $\Q$ windows for accurate
evaluation of $\cov^\disc$, especially when shot noise is significant.

\FloatBarrier
\subsubsection{Signal-to-Noise Ratio}
\label{ssub:SNR}

\begin{figure}[t]
    \centering
    \includegraphics[width=0.485\linewidth]{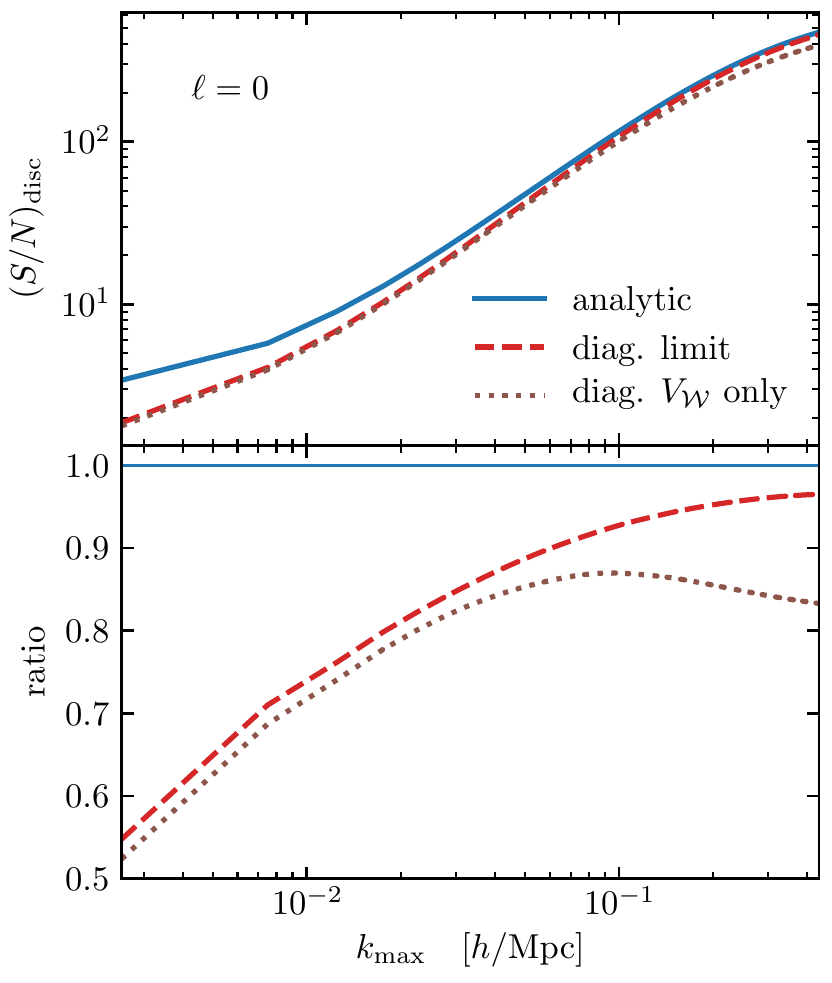}
    \includegraphics[width=0.50\linewidth]{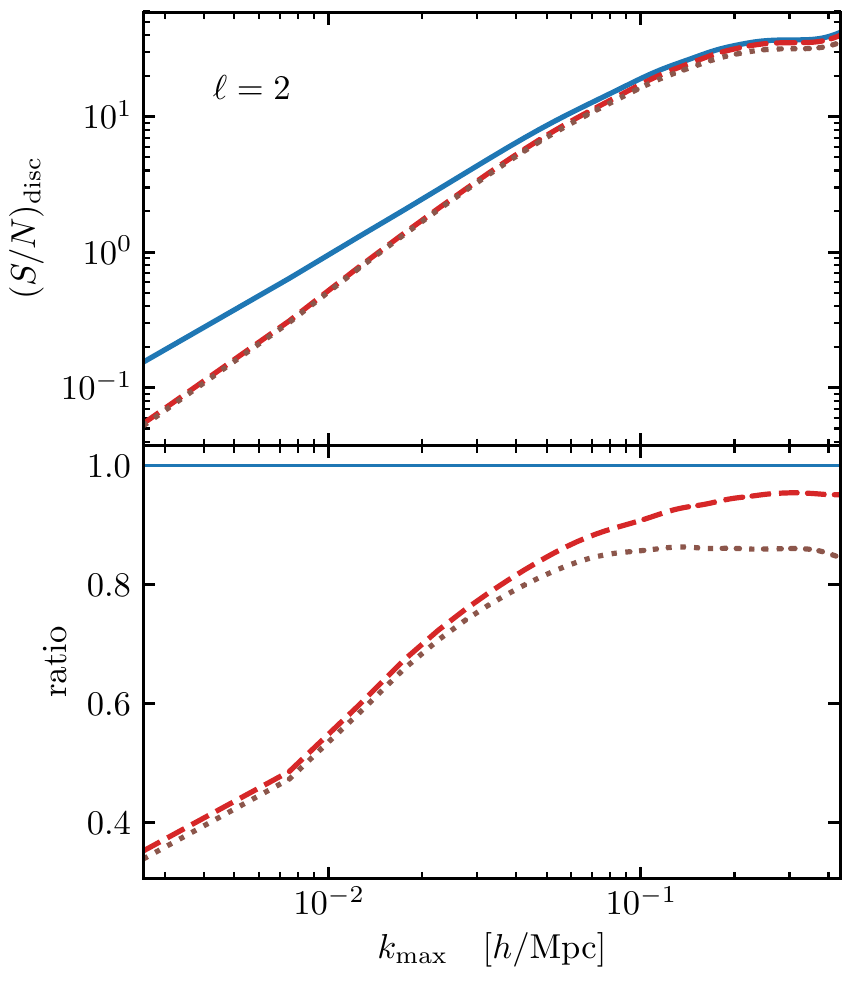}
    \caption{The signal-to-noise ratio of the power spectrum monopole and
    quadrupole with the disconnected covariance.
    We compare the signal-to-noise ratios using our analytic $\cov^\disc$ and
    its diagonal limit (Sec.~\ref{ssub:diag}).
    The latter captures the window size with three effective volumes in~\eqref{diag}, but ignores the shape of the window, resulting in very
    different covariance matrices (see Fig.~\ref{fig:cov_slice}).
    It underestimates the signal-to-noise ratio, and the difference is
    significant on large scales.
    This trend is expected because the window affects mostly the largest scales
    in the survey, and is roughly homogeneous and isotropic on small scales
    (see Fig.~\ref{fig:Q}).
    Also shown is the approximation that uses only $V_\W$ for all three
    effective volumes, which causes large bias on small scales where the shot
    noise terms are incorrectly normalized.
    }
    \label{fig:SNR}
\end{figure}

Our analytic disconnected covariance $\cov^\disc$ captures the correlation
between neighboring $k$ bins and multipoles due to the window effect.
In Sec.~\ref{ssub:diag} we have derived its diagonal limit $\cov^\diag$, that
only captures the survey size but neglect its inhomogeneity and anisotropy.
Fig.~\ref{fig:cov_slice} compares the two and shows they can have very
different shapes and values.
However such comparison is sensitive to binning: coarser bins would suppress
the off-diagonal elements of $\cov^\disc$ and reduce the difference between
$\cov^\disc$ and $\cov^\diag$ on their diagonals.
Here we compare their signal-to-noise ratios, which are independent of binning,
thereby study more carefully the impact of the survey window on the information
content.

The signal-to-noise ratio of each power spectrum multipole is defined as
\begin{equation}
    \Bigl(\frac{S}{N}\Bigr)_\disc^2 = \sum_{k_i, k_j < k_\mathrm{max}}
    P_\ell(k_i) \cov^\disc\bigl[\hat{P}(k_i), \hat{P}(k_j)\bigr]^{-1} P_\ell(k_j),
\end{equation}
as a function of the maximum wavenumber $k_\mathrm{max}$.
The covariance is limited to the disconnected part to compare $\cov^\disc$ and
$\cov^\diag$, and $P_\ell$ are the convolved power spectrum multipoles.
Fig.~\ref{fig:SNR} shows the results for $P_0$ and $P_2$.
We find that with the diagonal covariance $S / N$ is underestimated.
The difference is more significant on large scales: $\gtrsim 10 \%$ below $0.1
\hMpc$, and $\approx 5 \%$ even between $0.2 \hMpc$ and $0.4 \hMpc$.
This trend is expected because the window affect mostly the largest scales in
the survey, and is roughly homogeneous and isotropic on small scales as shown
in Fig.~\ref{fig:Q}.
We also show that it further biases $S / N$ on small scales if one ignores the
differences among the three effective volumes in~\eqref{diag} and uses only
$V_\W$, due to incorrect normalizations of the shot noise terms.

We warn however that this figure cannot be used for a quantitative estimate of
$S / N$ of cosmological parameters, and a more complete analysis that
marginalizes over all nuisance parameters, and includes also the connected
covariance, is needed.
This analysis will be presented in a future work.

\FloatBarrier
\subsection{Projected-Multipole Correlation Function Covariance Matrix}
\label{sub:results_proj_pole}

\begin{figure}[t]
    \centering
    \includegraphics[width=0.34\linewidth]{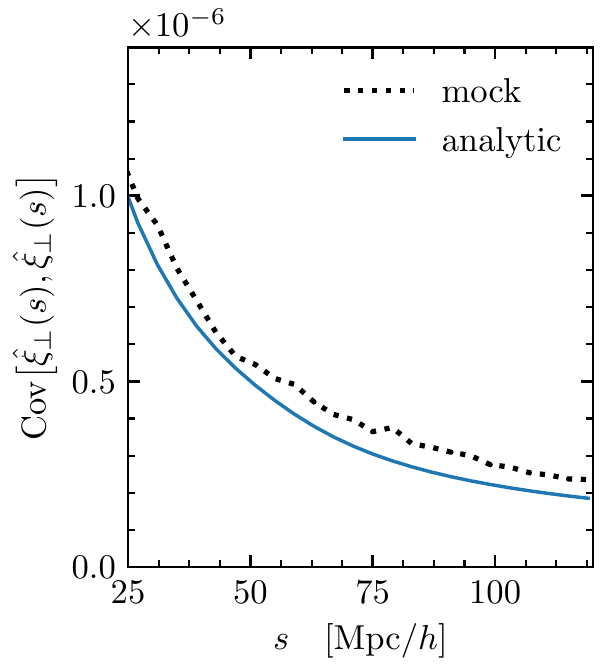}
    \includegraphics[width=0.32\linewidth]{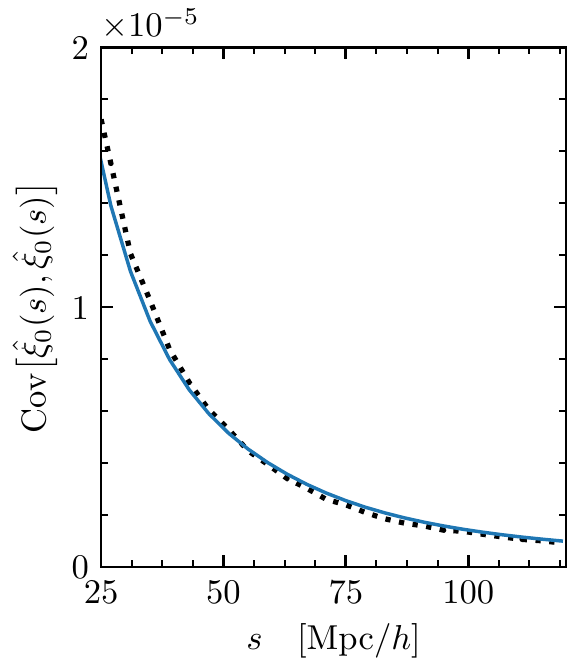}
    \includegraphics[width=0.32\linewidth]{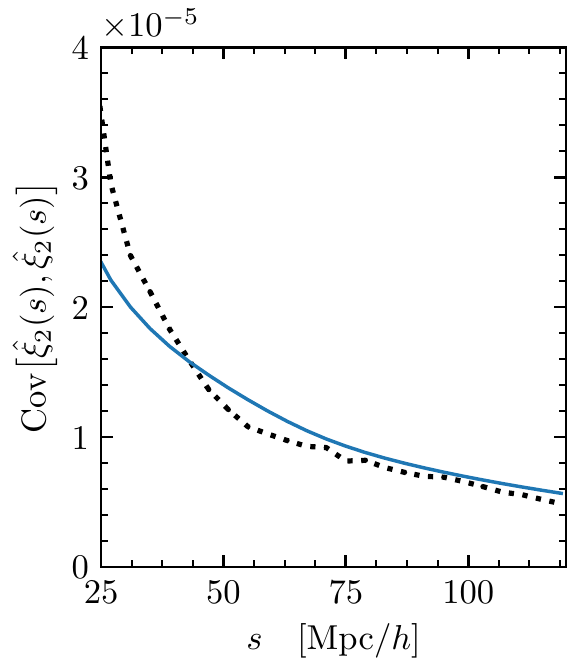}
    \includegraphics[width=0.33\linewidth]{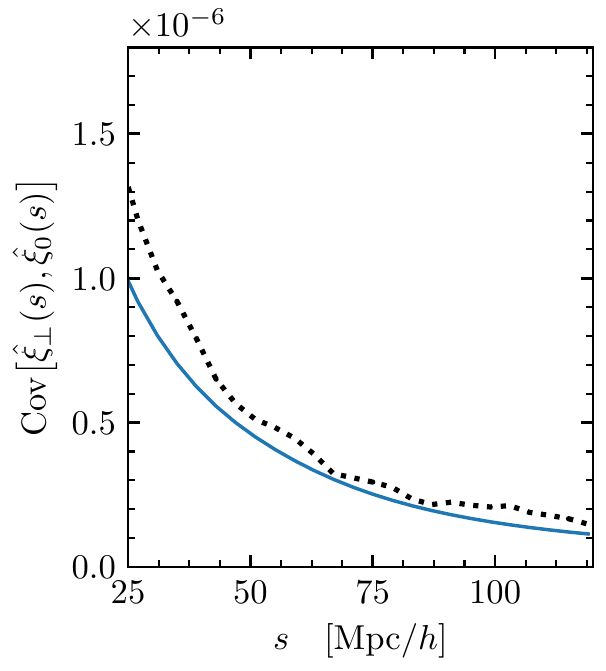}
    \includegraphics[width=0.325\linewidth]{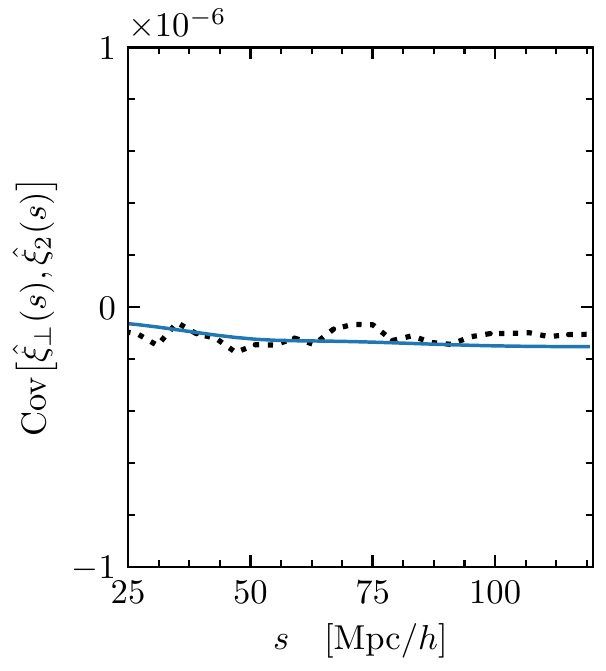}
    \includegraphics[width=0.325\linewidth]{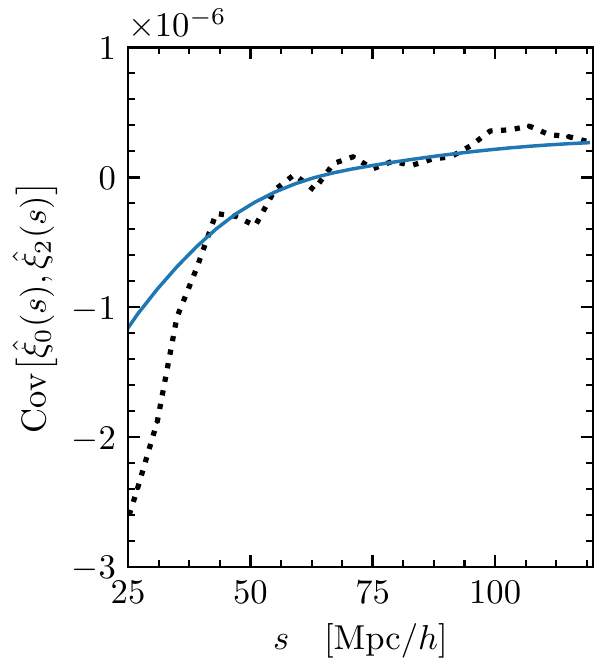}
    \caption{The diagonal of the joint covariance matrices of the projected
    correlation function and correlation function multipoles.
    We verify that the small-scale deviations in the $\xi_\ell$ covariances are
    due to the connected covariance.
    }
    \label{fig:cov_diag_corrfunc}
\end{figure}

\begin{figure}[t]
    \centering
    \includegraphics[width=0.7\linewidth]{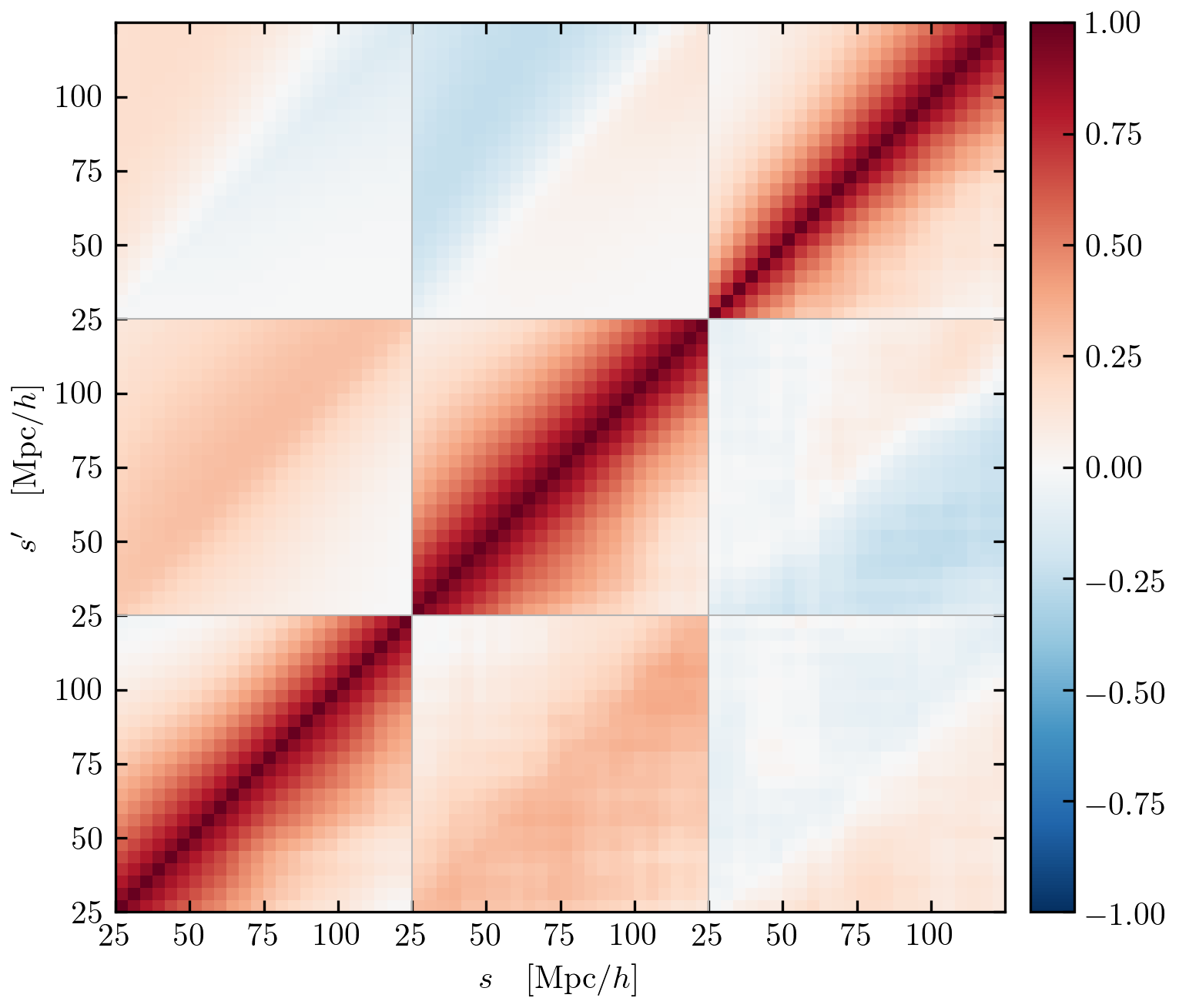}
    \includegraphics[width=0.7\linewidth]{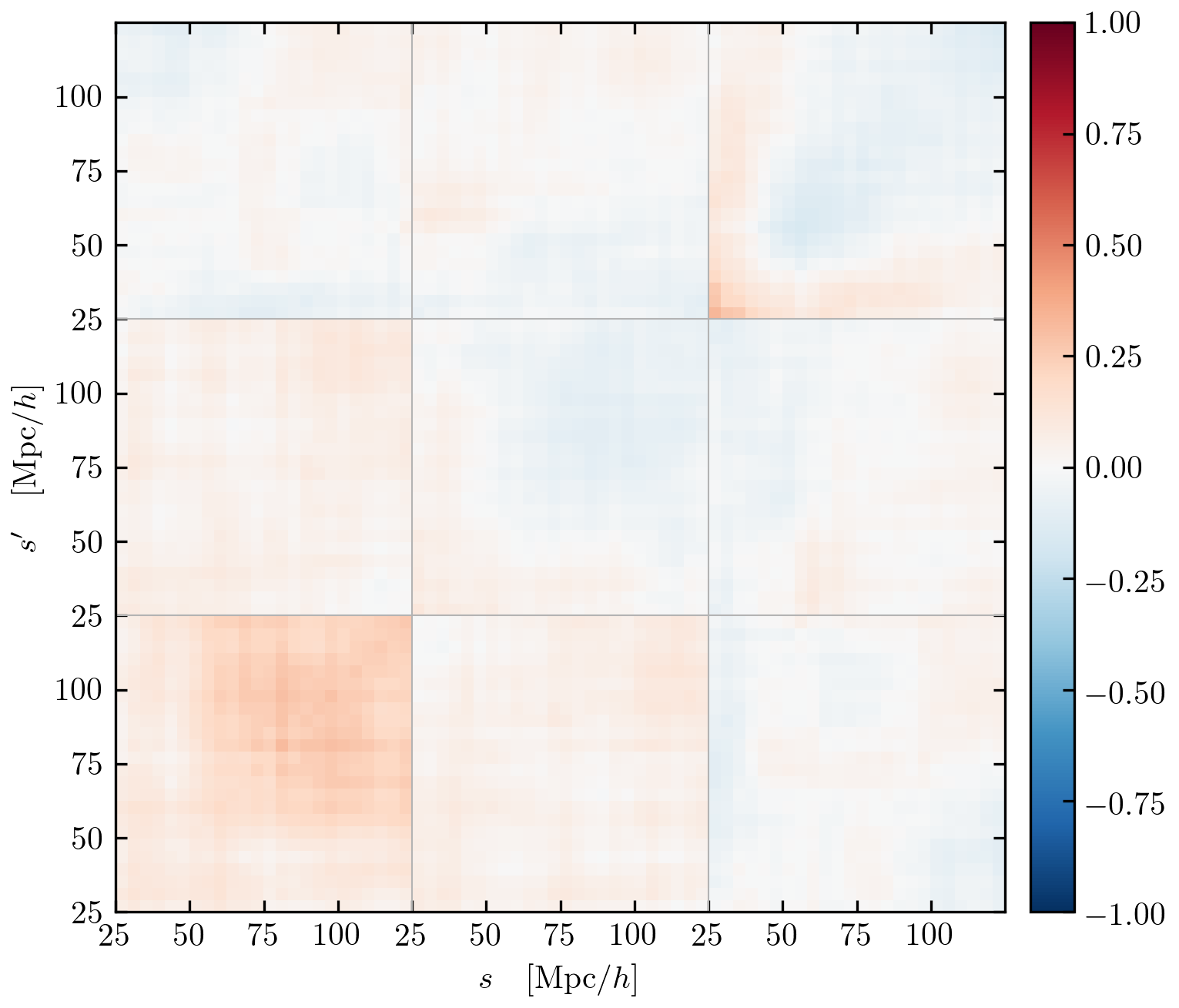}
    \caption{Correlation matrix of of the projected correlation function and
    correlation function multipoles.
    In the $3 \times 3$ blocks, from bottom to top and from left to right, we
    visualize the auto and cross correlations of $\hat\xi_\perp$, $\hat\xi_0$,
    $\hat\xi_2$.
    The upper figure compares the analytic result in its upper left corner with
    the mocks in the lower right corner.
    And the lower figure shows the difference between the mock and analytic
    covariance matrices, normalized by the diagonal of the latter.
    }
    \label{fig:corr_mat_corrfunc}
\end{figure}

Using the equations derived in Sec.~\ref{ssub:corrfunc} and
App.~\ref{sub:disc_corrfunc}, we can transform the power spectrum covariance to
that of the correlation functions.
Here we test this formalism on the joint covariance of the projected
correlation function $\xi_\perp$ and correlation function multipoles $\xi_\ell$.
Note that here $\xi_\perp$ is dimensionless unlike the $w_p$ estimator which is
more commonly used and has the dimension of length.
While it's possible to predict the covariance of $w_p$ with our formalism, we
choose $\xi_\perp$ as the test case as it is simpler to compute (see~\cite{SinghEtAl2017} for derivation of covariance for $w_p$).
$\xi_\perp$ and $w_p$ have similar signals and correlation matrices, as
$\xi_\perp$ is effectively $w_p$ with one very large bin of LOS separation,
although the amplitudes of the two estimators are different due to the factor
of length in $w_p$.
However, the correlation matrices for two estimators can be different for very
large LOS integration: the $w_p$ estimator gets larger contribution from
noisier bins from large LOS separation, while in $\hat\xi_\perp$ those bins are
damped by the LOS window.

First, we use the same power spectrum model $P_\ell$ and window function
multipoles $\Q_\ell$ as in Sec.~\ref{sub:results_pole}, to compute the auto
covariance of the projected power spectrum $P_\perp$, and the cross covariance
between $P_\perp$ and $P_\ell$'s, following App.~\ref{sub:disc_proj}.
Together with the auto covariance of $P_\ell$'s obtained in
Sec.~\ref{sub:results_pole}, we have the full joint covariance of $P_\perp(k)$
and $P_\ell(k)$'s.
Note that this is before the binning operation and $k$ is sampled on a
logarithmic grid.
Then we can perform Hankel transforms on this covariance twice with
$\texttt{mcfit}$ to replace $k$ with $s$, and $k'$ with $s'$.
The result is almost the correlation function covariance, except for the
difference in the normalization factors arising from the $\hat P$ and $\hat\xi$
estimators.
We account for this normalization difference following
App.~\ref{sub:disc_corrfunc}, before binning the covariance into matrices with
$s$ bins of width $4 \Mpch$.
The whole process takes less than 30 minutes on a single CPU, and is more than
ten times as long as the computation for the power spectrum.
Most of the time is spent on the $\xi_\perp$ cross $\xi_\ell$ blocks where we
approximate a Dirac delta function with multipole expansion truncated at $\lmax
= 16$, much higher than the power spectrum case.
The computation can be significantly faster when the cross covariance is not
needed.

Again we compare the analytic correlation function $\cov^\disc$ to the sample
covariance from mocks.
Fig.~\ref{fig:cov_diag_corrfunc} shows this comparison on the $s = s'$ diagonal
of the auto covariance matrices.
We find the analytic $\cov^\disc$ reproduce well the scale dependences.
On small scales $\lesssim 50 \Mpch$, the mock covariances of $\xi_0$ and
$\xi_2$ are larger than the analytic ones, and we find these excesses are due
to the connected covariance by transforming the power spectrum
$\widehat\cov^{\conn, \smooth}$ from Sec.~\ref{sub:results_pole}.
Apart from that, the analytic and mock covariance still differ by an offset for
$\xi_\perp$ and a 15\% deficit near $50 \Mpch$ for $\xi_2$.
This is probably due to the fact that the analytic $\cov^\disc$ is based on the
best-fit $P_\ell$ to the mock power spectrum, that in general can be somewhat different
if fitted against the correlation function measurement.

We also compare the shapes of the analytic and mock covariance by showing their
correlation matrices in Fig.~\ref{fig:corr_mat_corrfunc}.
The top panel shows the two have very similar shapes, and their difference
(normalized by the diagonal of the analytic $\cov^\disc$ similar to
Fig.~\ref{fig:corr_mat}) is $\lesssim 10\%$ except in the regions noted above.

\FloatBarrier
\section{Discussion}
\label{sec:discuss}

We presented an analytic method to compute the disconnected part of the covariance
matrix of 2-point functions in large-scale structure studies, accounting for
the survey window effect.
This method works for both power spectrum (Sec.~\ref{sub:method_3D},
App.~\ref{sub:disc_3D}) and correlation function (Sec.~\ref{ssub:corrfunc},
App.~\ref{sub:disc_corrfunc}), and applies to the covariances for various
probes including the multipoles (Sec.~\ref{sub:method_pole}) and the wedges
(Sec.~\ref{ssub:wedge}) of 3D clustering, the angular (App.~\ref{sub:disc_ang})
and the projected statistics of clustering and lensing
(App.~\ref{sub:disc_proj}), as well as their cross covariances
(App.~\ref{sub:disc_3D}, App.~\ref{sub:disc_proj}).

We verify our analytic covariance against the covariance from 1000 galaxy mock
simulations for BOSS DR12.
We compare the analytic and mock covariance matrices on power spectrum
multipoles, and demonstrate that we obtain an excellent agreement without the
sampling noise associated with numerical covariance matrices.
As a consequence, our method does not require various corrections and inflated
errors that have been developed for this purpose~\cite{HartlapSimonEtAl07,
SellentinHeavens2016}.

Another advantage of our method is that its predictions use the best-fit power
spectrum model to the data, and does not assume a fiducial model that may not
fit the data well.
The method is thus best viewed as part of a full data analysis pipeline, where
we determine both the power spectrum and the covariance matrix within the same
(iterative) procedure.
We also tested the joint covariance matrices of the projected correlation
function and correlation function multipoles, and find the accuracy of the
analytic prediction is satisfactory, given that our model is fitted to the
power spectrum instead of the correlation function.
The analytic computation is efficient and costs negligible CPU time compared to
the mocks.

In contrast to previous work~\cite{GriebEtAl16} our method includes the window
effect on the covariance.
This commonly adopted approximation to the disconnected covariance only
captures the size of the survey window but ignores its shape (inhomogeneity and
anisotropy), resulting in a diagonal power spectrum covariance matrix.
We show the proper diagonal limit arises as the homogeneous and isotropic limit
of our analytic covariance, and it generalizes the usual expression by
accounting for the inhomogeneous number density.
We show that such a diagonal covariance underestimates the signal-to-noise
ratio compared to our analytic covariance.
Other previous works modeled the window effect to different extents.
In Ref.~\cite{HowlettPercival17}, the authors accounted for the window function
in the diagonal elements of the power spectrum covariance matrix in the
flat-sky limit, by using FFT and pair summation over Fourier modes, but they
needed an ansatz to approximate the off-diagonal elements that again ignored
the window.
Ref.~\cite{BlakeCarterEtAl18} presented an analytic calculation of the
disconnected covariance of the galaxy power spectrum multipoles with variable
LOS, under some approximations that simplify the coupling between the power and
the window.
Our formalism assumes flat sky and similar simplified power-window coupling,
but uses the curved-sky window to account for the spherical geometry.
Compared to our method, the equations in ref.~\cite{BlakeCarterEtAl18} are more
complicated and need Monte-Carlo integration to evaluate.
Ref.~\cite{SinghEtAl2017} also derived the expression for the covariance and
cross covariance for 3D clustering and projected correlation functions, but
assumed uniform window when computing the covariance matrices.
Our method generalizes the disconnected covariance for arbitrary windows which
are normally encountered in the large-scale structure surveys.

Being able to compute the accurate disconnected covariance analytically opens the
possibility of calibrating the connected part using only small-volume mocks or
internal covariance estimators from the data, thereby substantially reduces the
computational cost required for estimating the full covariance matrices.
We will address this connected part of covariance matrix in the future work.
Having the complete noiseless covariance will allow for the optimal analyses of
the 2-point functions from the large-scale structure data.

\acknowledgments{
YL thanks Ue-Li Pen and Zachary Slepian for useful discussions.
YL acknowledges support from Fellowships at the Berkeley Center for
Cosmological Physics, and at the Kavli IPMU established by World Premier
International Research Center Initiative (WPI) of the MEXT, Japan.
We acknowledge grant support from NASA NNX15AL17G, 80NSSC18K1274, NSF 1814370,
and NSF 1839217.

This research used resources of the National Energy Research Scientific
Computing Center (NERSC), a U.S.\ Department of Energy Office of Science User
Facility operated under Contract No.\ DE-AC02-05CH11231.
}

\appendix

\section{Disconnected Covariance}
\label{sec:disc}

In this appendix we present calculations of the disconnected auto and cross
covariance of power spectra of various forms, including the general 3D power
spectra, the multipoles, the angular power spectra.

\subsection{Fully General 3D Case}
\label{sub:disc_3D}

As usual we model the tracer fields as a Poisson process that samples a
\emph{homogeneous} random field.
Under such assumption, the cross correlation between two fields $a$ and $b$ is
given by
\begin{align}
    \ensavg{\delta_\data^a(\vk) \delta_\data^b(-\vk')}
    &= (2\pi)^3 \deltaD(\vk-\vk') P^{ab}(\vk)
    + \int_\vx \frac{\nbar_{ab}(\vx)}{\nbar_a(\vx) \nbar_b(\vx)}
        e^{-i(\vk-\vk')\cdot\vx}
    \nonumber\\
    \ensavg{\delta_\rand^a(\vk) \delta_\rand^b(-\vk')}
    &= \int_\vx \frac{\alpha \deltaK_{ab}}{\nbar_a(\vx)} e^{-i(\vk-\vk')\cdot\vx}
    \nonumber\\
    \ensavg{\delta_\data^a(\vk) \delta_\rand^b(-\vk')}
    &= \ensavg{\delta_\data^b(\vk) \delta_\rand^a(-\vk')} = 0.
    \label{2pts_ab}
\end{align}
For each field the subscript `$\data$' and `$\rand$' labels the data and random
catalogs, respectively.
And the overdensity field for analyses is usually the difference between the
data and random overdensities
\begin{equation}
    \delta \equiv \delta_\data - \delta_\rand.
\end{equation}
The shot noise term is related to $\nbar_{ab}(\vx)$, the mean number density of
the overlap between the two samples.
Apparently $\nbar_{ab}$ reduces to $\nbar_a$ when $a=b$, and the cross shot
noise is at most as big as the minimum of the auto shot noises of the two
samples: $\nbar_{ab} / (\nbar_a \nbar_b) \leq \min(1/\nbar_a, 1/\nbar_b)$.

The windowed field in Fourier space is a convolution
\begin{equation}
    \deltaW^a(\vk)=\int_\vq \delta^a(\vk-\vq) W^a(\vq).
\end{equation}
The following estimator of 3D power spectrum measures the cross correlation
between $a$ and $b$ fields
\begin{equation}
    \hat{P}^{ab}(\vk) = \frac{\deltaW^a(\vk) \deltaW^b(-\vk)}{\W^{ab}(0)}
    - P_\shot^{ab}.
\end{equation}
This is a Hermitian function that satisfies $\hat{P}^{ab}(\vk)^* =
\hat{P}^{ab}(-\vk) = \hat{P}^{ba}(\vk)$, and a real function when $a = b$.
The auto and cross covariance functions of the estimators
\begin{equation}
    \cov\bigl[\hat{P}^{ab}(\vk), \hat{P}^{cd}(\vk')\bigr]
    \equiv \ensavg{\hat{P}^{ab}(\vk) \hat{P}^{cd}(\vk')^*}
        - \ensavg{\hat{P}^{ab}(\vk)} \ensavg{ \hat{P}^{cd}(\vk')^*}
\end{equation}
has the following disconnected part
\begin{equation}
    \cov^\disc\bigl[\hat{P}^{ab}(\vk), \hat{P}^{cd}(\vk')\bigr]
    = \frac{\ensavg{\deltaW^a(\vk) \deltaW^c(-\vk')}
            \ensavg{\deltaW^b(-\vk) \deltaW^d(\vk')}}
        {\W^{ab}(0) \W^{cd}(0)} + (c, \vk' \leftrightarrow d, -\vk').
\end{equation}
Both covariance are Hermitian forms that satisfy $\cov^{ab,cd}(\vk, \vk')^* =
\cov^{cd,ab}(\vk', \vk)$.
So the full Hermitian symmetry including those of the estimators are: $a, \vk
\leftrightarrow b, -\vk$; $c, \vk' \leftrightarrow d, -\vk'$; and $a, b, \vk
\leftrightarrow c, d, -\vk'$.

Plugging~\eqref{2pts_ab} into the following 2-point function
\begin{align}
    &\ensavg{\deltaW^a(\vk) \deltaW^b(-\vk')} \nonumber\\
    &= \int_\vq P^{ab}(\vk-\vq) W^a(\vq) W^b(\vk-\vk'-\vq)
    + (1 + \alpha\deltaK_{ab}) \int_\vx \nbar_{ab}(\vx) w_a(\vx) w_b(\vx)
        e^{-i(\vk-\vk')\cdot\vx}
    \nonumber\\
    &= \int_{\vk''} P^{ab}(\vk'') W^a(\vk-\vk'') W^b(\vk''-\vk')
    + (1 + \alpha\deltaK_{ab}) \int_\vx \nbar_{ab}(\vx) w_a(\vx) w_b(\vx)
        e^{-i(\vk-\vk')\cdot\vx}
    \nonumber\\
    &\approx \frac{P^{ab}(\vk)+P^{ab}(\vk')}2 \W^{ab}(\vk-\vk')
    + \S^{ab}(\vk-\vk'),
    \label{2pts_W_ab}
\end{align}
where the curly mode mixing windows $\W$ and $\S$ are define as
\begin{align}
    \W^{ab}(\vq) &= \int_\vx \W^{ab}(\vx) e^{-i\vq\cdot\vx}
    \equiv \int_\vx W^a(\vx) W^b(\vx) e^{-i\vq\cdot\vx},
    \nonumber\\
    \S^{ab}(\vq) &= \int_\vx \S^{ab}(\vx) e^{-i\vq\cdot\vx}
    \equiv (1 + \alpha\deltaK_{ab})
        \int_\vx \nbar_{ab}(\vx) w_a(\vx) w_b(\vx) e^{-i\vq\cdot\vx}.
    \label{WS_ab}
\end{align}
In the second equality of~\eqref{2pts_W_ab}, we have used change of variables
to rewrite it in the more symmetric way.
Under the assumption that the power spectrum is smooth, which is true at large
$k$ and $k'$, we can make the approximation $P(\vk) \approx P(\vk'') \approx
P(\vk')$ to obtain the last line.
This is due to the fact that the window functions are only significant when
$\vk \approx \vk'' \approx \vk'$ to the extent of the Fourier-space window
scale.
By replacing $P(\vk'')$ with its average at $\vk$ and $\vk'$, our approximation
preserves the Hermitian symmetry $a, \vk \leftrightarrow b, -\vk'$.

With~\eqref{2pts_W_ab} we can easily derive the following disconnected
covariance
\begin{align}
    \cov^\disc\bigl[\hat{P}^{ab}(\vk), \hat{P}^{cd}(\vk')\bigr]
    \approx \frac{1}{\W^{ab}(0) \W^{cd}(0)} \Biggl\{
    &\biggl[ \frac{P^{ac}(\vk) + P^{ac}(\vk')}2 \W^{ac}(\vk-\vk')
    + \S^{ac}(\vk-\vk') \biggr]
    \nonumber\\ \times
    &\biggl[ \frac{P^{bd}(\vk) + P^{bd}(\vk')}2 \W^{bd}(\vk-\vk')
    + \S^{bd}(\vk-\vk') \biggr]^*
    \nonumber\\ &\hspace{-20ex} +
    \biggl[ \frac{P^{ad}(\vk) + P^{ad}(\vk')^*}2 \W^{ad}(\vk+\vk')
    + \S^{ad}(\vk+\vk') \biggr]
    \nonumber\\ &\hspace{-20ex} \times
    \biggl[ \frac{P^{bc}(\vk) + P^{bc}(\vk')^*}2 \W^{bc}(\vk+\vk')
    + \S^{bc}(\vk+\vk') \biggr]^*
    \Biggr\}.
\end{align}
Furthermore because we have assumed $P(\vk) \approx P(\vk')$, in the very same
approximation we can combine certain terms by letting $P(\vk) P(\vk) \approx
P(\vk)P(\vk') \approx P(\vk')P(\vk')$ to simplify the expression while still
preserving its Hermitian symmetries
\begin{align}
    &\cov^\disc\bigl[\hat{P}^{ab}(\vk), \hat{P}^{cd}(\vk')\bigr]
    \approx \frac{1}{\W^{ab}(0) \W^{cd}(0)} \biggl\{
    \frac{P^{ac}(\vk) P^{bd}(\vk')^* + P^{ac}(\vk') P^{bd}(\vk)^*}2 \QW^{ac, bd}(\vk-\vk')
    \nonumber\\
    &+ \frac{P^{ac}(\vk) + P^{ac}(\vk')}2 \QX^{ac,bd}(\vk-\vk')
    + \frac{P^{bd}(\vk)^* + P^{bd}(\vk')^*}2 \QX^{bd,ac}(\vk-\vk')^*
    + \QS^{ac,bd}(\vk-\vk')
    \nonumber\\
    &+ \frac{P^{ad}(\vk) P^{bc}(\vk') + P^{ad}(\vk')^* P^{bc}(\vk)^*}2 \QW^{ad,bc}(\vk+\vk')
    \nonumber\\
    &+ \frac{P^{ad}(\vk) + P^{ad}(\vk')^*}2 \QX^{ad,bc}(\vk+\vk')
    + \frac{P^{bc}(\vk)^* + P^{bc}(\vk')}2 \QX^{bc,ad}(\vk+\vk')^*
    + \QS^{ad,bc}(\vk+\vk')
    \biggr\},
    \label{disc_3D}
\end{align}
where we have further defined the following auto and cross power spectra of
$\W$ and $\S$
\begin{align}
    \QW^{ab,cd}(\vq) \equiv \W^{ab}(\vq) \W^{cd}(\vq)^*
        = \int_\vs \QW^{ab,cd}(\vs) e^{-i\vq\cdot\vs},
    \nonumber\\
    \QS^{ab,cd}(\vq) \equiv \S^{ab}(\vq) \S^{cd}(\vq)^*
        = \int_\vs \QS^{ab,cd}(\vs) e^{-i\vq\cdot\vs},
    \nonumber\\
    \QX^{ab,cd}(\vq) \equiv \W^{ab}(\vq) \S^{cd}(\vq)^*
        = \int_\vs \QX^{ab,cd}(\vs) e^{-i\vq\cdot\vs},
    \label{Q_abcd}
\end{align}
with $\Q(\vs)$ being the auto and cross correlation functions of $\W$ and
$\S$, e.g.\
\begin{equation}
    \QX^{ab,cd}(\vs) = \int_\vx \W^{ab}(\vx+\vs) \S^{cd}(\vx).
\end{equation}
The different $\Q$ functions completely captures the mode mixing effect in the
disconnected covariance, therefore is the key to its evaluation.
Since the survey window is usually conveniently represented by a synthetic
random catalog, one can measure the $\Q$ window as either correlation functions
or power spectra of the random catalog.
We show how to quantify the $\Q$ window and then the disconnected covariance in
each specific case in the following subsections.

\subsection{Angular Case}
\label{sub:disc_ang}

We consider the angular power spectrum and its disconnected covariance in the
flat sky limit.
The angular field is related to the 3D field by an integral along the line of
sight
\begin{equation}
    \deltaW(\vtheta) = \int_\chi \delta(\vx) W(\vx),
    \label{deltaW_ang}
\end{equation}
where $\vx \equiv (\vx_\perp, x_\parallel) \equiv (\chi\vtheta, \chi)$.
The 3D window function is usually decomposable into angular and radial
distributions $W(\vx) \equiv W(\vtheta) W(\chi) = \nbar(\vtheta) w(\vtheta)
\nbar(\chi) w(\chi)$.
In Fourier space the angular field becomes
\begin{equation}
    \deltaW(\vl) \equiv \int_\vtheta \deltaW(\vtheta) e^{- i \vl \cdot \vtheta}
    = \int_{\chi, \vl'} \delta\Bigl(\frac{\vl-\vl'}\chi, \chi\Bigr)
    \frac{W(\vl') W(\chi)}{\chi^2}
    \label{deltaW_ang_l}
\end{equation}
Note that here $\delta(\vk_\perp, \chi)$ is only the 2D Fourier transform of a
transverse slice of $\delta(\vx)$, so that it has the dimension of area instead
of volume.

The flat-sky Limber approximation~\cite{Limber53, Kaiser92, LoverdeAfshordi08}
states that in the $k_\perp |\chi - \chi'| \gg 1$ limit, the correlation
between two density slices is given by
\begin{align}
    &\ensavg{\delta^a(\vk_\perp, \chi) \delta^b(-\vk'_\perp, \chi')}
    \equiv \int_{\vx_\perp, \vx'_\perp} \ensavg{\delta^a(\vx) \delta^b(\vx')}
    e^{- i (\vk_\perp \cdot \vx_\perp - \vk'_\perp \cdot \vx'_\perp)}
    \nonumber\\
    &\approx \deltaD(\chi - \chi') \Bigl[
    (2\pi)^2 \deltaD(\vk_\perp - \vk'_\perp) P^{ab}(\vk_\perp; z)
    + \int_{\vx_\perp} \frac{(1 + \alpha\deltaK_{ab}) \nbar_{ab}(\vx)}
                        {\nbar_a(\vx) \nbar_b(\vx)}
    e^{- i (\vk_\perp - \vk'_\perp) \cdot \vx_\perp} \Bigr],
    \label{Limber}
\end{align}
where $P(\vk; z)$ is the 3D power spectrum at redshift $z(\chi)$.

Using the Limber approximation one can derive the angular power spectrum of a
windowed field in a similar way that leads to~\eqref{2pts_W_ab}
\begin{align}
    &\ensavg{\deltaW(\vl) \deltaW(-\vl')}
    \nonumber\\
    &\approx \int_{\chi, \vl''} P\Bigl(\frac{\vl''}\chi; z\Bigr)
    \frac{W(\vl - \vl'') W(\vl'' - \vl') W(\chi)^2}{\chi^2}
    + (1 + \alpha) \int_{\chi, \vtheta} \frac{\nbar(\vx) w(\vx)^2}{\chi^2}
    e^{- i (\vl - \vl') \cdot \vtheta}
    \nonumber\\
    &= \int_{\vl''} C(\vl'') W(\vl - \vl'') W(\vl'' - \vl')
    + \S(\vl - \vl') \int_\chi \frac{\nbar(\chi) w(\chi)^2}{\chi^2}
    \nonumber\\
    &\approx \frac{C(\vl) + C(\vl')}2 \W(\vl - \vl')
    + \S(\vl - \vl') \int_\chi \frac{\nbar(\chi) w(\chi)^2}{\chi^2}
    \label{2pts_W_ang}
\end{align}
where $C(\vl)$ is the unmasked angular power spectrum in the flat-sky limit
\begin{equation}
    C(\vl) = \int_\chi P\Bigl(\frac\vl\chi; z\Bigr) \frac{W(\chi)^2}{\chi^2},
\end{equation}
and similar to~\eqref{WS_ab} the angular mode-mixing windows are
\begin{align}
    \W(\vl) &\equiv \int_\vtheta W(\vtheta)^2 e^{-i\vl\cdot\vtheta},
    \nonumber\\
    \S(\vl) &\equiv (1 + \alpha)
    \int_\vtheta \nbar(\vtheta) w(\vtheta)^2 e^{-i\vl\cdot\vtheta}.
    \label{WS_ang}
\end{align}
Likewise their correlation function $\QW$, $\QS$, and $\QX$ can be defined
analogous to~\eqref{Q_abcd}.
Notice that even though we have assigned the same symbols to both the 3D and
angular window factors, the specific one in use should be clear from the
context.
From~\eqref{2pts_W_ang} one can show that the following $C(\vl)$ estimator
\begin{equation}
    \hat{C}(\vl) = \frac{|\deltaW(\vl)|^2}{\W_0} - C_\shot
\end{equation}
is unbiased at high $\ell$ and the shot noise is
\begin{equation}
    C_\shot
    = \int_\chi \frac{\nbar(\chi) w(\chi)^2}{\chi^2} \frac{\S_0}{\W_0}
    = (1+\alpha) \int_\chi \frac{\nbar(\chi) w(\chi)^2}{\chi^2}
    \frac{\int_\vtheta \nbar(\vtheta) w(\vtheta)^2}
        {\int_\vtheta \nbar(\vtheta)^2 w(\vtheta)^2}.
\end{equation}

Because of the similarity of structure here to that of the 3D case, the
disconnected covariance of the angle-averaged angular power spectrum resembles
that of the power spectrum multipoles.
So analogous to~\eqref{disc_pole_0} we can write down immediately
\begin{align}
    &\cov^\disc\bigl[\hat{C}(\ell), \hat{C}(\ell')\bigr]
    = \int_{\uvl, \uvl'} \cov^\disc\bigl[\hat{C}(\vl), \hat{C}(\vl')\bigr]
    \approx 2 \int_{\uvl, \uvl'}
    \biggl\{ C(\ell) C(\ell') \frac{\bigl|\W(\vl-\vl')\bigr|^2}{\W_0^2}
    \nonumber\\ &\hspace{20ex}
    + \bigl[ C(\ell) + C(\ell') \bigr] C_\shot
    \frac{\Re\bigl[ \W(\vl-\vl') \S(\vl-\vl')^* \bigr]}{\W_0 \S_0}
    + C_\shot^2 \frac{\bigl|\S(\vl-\vl')\bigr|^2}{\S_0^2}
    \biggr\}
    \nonumber\\
    &= \frac2{\W_0^2} \biggl\{
    C(\ell) C(\ell') \int \!2\pi \theta\d\theta\,
    \QW(\theta) J_0(\ell\theta) J_0(\ell'\theta)
    \nonumber\\ &\quad
    + \bigl[C(\ell) + C(\ell')\bigr] \int \!2\pi \theta\d\theta\,
    \QX(\theta) J_0(\ell\theta) J_0(\ell'\theta)
    + \int \!2\pi \theta\d\theta\,
    \QS(\theta) J_0(\ell\theta) J_0(\ell'\theta)
    \biggr\},
    \label{disc_ang}
\end{align}
in deriving which we have used the Jacobi-Anger expansion~\eqref{Jacobi_Anger_expansion}.

As in the multipole case, given the $\Q$ measured from the survey random
catalog and the $C(\ell)$ constrained from the data, we can compute the
disconnected covariance of $\hat{C}(\ell)$ with this equation.
Its evaluation again includes three integrals involving two zeroth order Bessel
functions $J_0$.
We show their solution in Sec.~\ref{sec:double_Bessel}.

\begin{hide}
\subsection{Angular$\times$Multipole}
\label{sub:disc_ang_pole}

\HL{Unsure about this.}

We have already described how to compute the covariance of power spectrum
multipoles and of the angular power spectrum, and consider their
cross covariance here.
We use \eqref{deltaW_ang_l}, the following decomposition of the 3D field
\begin{equation}
    \deltaW(\vk) = \int_{\chi, \vl}
    \delta\Bigl(\vk_\perp - \frac\vl\chi, \chi\Bigr)
    W(\vl) W(\chi) e^{- i k_\parallel \chi},
\end{equation}
and the Limber approximation \eqref{Limber} to obtain the cross correlation
between an angular field and a 3D field
\begin{align}
    &\ensavg{\deltaW^a(\vl) \deltaW^b(-\vk')}
    \nonumber\\
    &\approx \int_\chi e^{i k'_\parallel \chi} \biggl\{
    \int_{\vl''} P^{ab}\Bigl(\frac{\vl''}\chi; z\Bigr)
    W^a(\vl - \vl'') W^b(\vl'' - \vk'_\perp\chi) W^a(\chi) W^b(\chi)
    \nonumber\\ &\hspace{30ex}
    + (1 + \alpha\deltaK_{ab}) \int_\vtheta \nbar_{ab}(\vx) w_a(\vx) w_b(\vx)
    e^{- i (\vl - \vk'_\perp\chi) \cdot \vtheta} \biggr\}
    \nonumber\\
    &\approx \int_\chi e^{i k'_\parallel \chi} \biggl\{
    P^{ab}(\vk'_\perp; z) \W^{ab}(\chi) \W^{ab}(\vl - \vk'_\perp\chi)
    + \S^{ab}(\chi) \S^{ab}(\vl - \vk'_\perp\chi) \biggr\},
    \label{2pts_W_ang_3D}
\end{align}
where $\W^{ab}(\chi) \equiv W^a(\chi) W^b(\chi)$, $\S^{ab}(\chi) \equiv
\nbar_{ab}(\chi) w_a(\chi) w_b(\chi)$, and $\W^{ab}(\vl)$ and $\S^{ab}(\vl)$
are defined as in \eqref{WS_ang} but for two different fields as in
\eqref{WS_ab}.
Notice that even here we denote either the radial or the angular window with
the same symbols $\W^{ab}$ and $\S^{ab}$, one can easily distinguish them by
their arguments, i.e.\ radius or angle, respectively.
Again because the angular window peaks sharply at high $\ell$, in the second
equality we have used the same approximation as before to take the power
spectrum out of the angular convolution integral.

The above cross correlation forms the basis to compute the cross disconnected
covariance between an angular power spectrum and multipoles
\begin{equation}
    \cov^\disc\bigl[\hat{C}^a(\ell), \hat{P}^b_{\ell'}(k')\bigr]
    = \frac{2 (2\ell'+1)}{\W^a(0) \W^b(0)} \int_{\uvl, \uvk'} \L_{\ell'}(\uvk'\cdot\uvn)
    \bigl|\ensavg{\deltaW^a(\vl) \deltaW^b(-\vk')}\bigr|^2
\end{equation}
Let's consider the first ($P^2$) term apart from the $\uvl$ and $\uvk$
integrals and the constant prefactor
\begin{align}
    &\int_{\chi, \chi'} e^{- i k'_\parallel (\chi'-\chi)}
    P^{ab}(\vk'_\perp; z) \W^{ab}(\chi) \W^{ab}(\vl - \vk'_\perp\chi)
    P^{ab}(\vk'_\perp; z') \W^{ab}(\chi') \W^{ab}(\vl - \vk'_\perp\chi')^*
    \nonumber\\
    &= \int_{\chi, \chi', \vtheta, \vtheta'} P^{ab}(\vk'_\perp; z) P^{ab}(\vk'_\perp; z')
    \W^{ab}(\chi) \W^{ab}(\chi') \W^{ab}(\vtheta) \W^{ab}(\vtheta')
    \nonumber\\ &\hspace{10ex}
    \times \exp\Bigl[i \vl \cdot (\vtheta'-\vtheta) - i k'_\parallel (\chi'-\chi)
    - i \vk'_\perp \frac{\chi+\chi'}2 \cdot (\vtheta'-\vtheta)
    - i \vk'_\perp (\chi'-\chi) \cdot \frac{\vtheta+\vtheta'}2\Bigr]
    \nonumber\\
    &\approx \left\{\begin{aligned}
    \int_{\chi, \chi', \vtheta} 2\pi \deltaD(\vk'_\perp\chi' - \vk'_\perp\chi)
    P^{ab}(\vk'_\perp; z)^2 \W^{ab}(\chi)^2
    \frac{\QW^{ab,ab}(\vtheta)}{\Qone^{ab,ab}(\vtheta)}
    e^{i (\vl - \vk'_\perp\chi) \cdot \vtheta},
    &\quad f_\mathrm{sky} \lesssim 1,
    \\
    2\pi \deltaD(k'_\parallel) \int_{\chi, \vtheta}
    P^{ab}(\vk'_\perp; z)^2 \W^{ab}(\chi)^2 \QW^{ab,ab}(\vtheta)
    e^{i (\vl - \vk'_\perp\chi) \cdot \vtheta},
    &\quad f_\mathrm{sky} \ll 1.
    \end{aligned}\right.
\end{align}
\HL{FIXME: missing $2\pi$ in the first case?}
One can compute the integrals more easily in two opposite regimes.
For a large sky coverage, we can integrate out $(\vtheta + \vtheta') / 2$ to
get a Dirac delta function in $\vk'_\perp\chi' - \vk'_\perp\chi$.
Note that while doing this we also replace the $\W$ correlation function $\QW$
with its sky average $\QW / \Qone$, where $\Qone$ is the unweighted correlation
function.
In the other limit where angle variation is small, we can drop the last phase
term in the second equality compared to the second one.
Assuming smooth radial window allows us to integrate out the second phase term
over $\chi' - \chi$ to get a different Dirac delta in $k'_\parallel$.

Now we can do the $\uvl$ and $\uvk$ integrals in either cases. In the wide sky
limit
\begin{align*}
    &\int_{\uvl, \uvk'} \L_{\ell'}(\uvk'\cdot\uvn)
    \int_{\chi, \chi', \vtheta} 2\pi \deltaD(\vk'_\perp\chi' - \vk'_\perp\chi)
    P^{ab}(\vk'_\perp; z)^2 \W^{ab}(\chi)^2
    \frac{\QW^{ab,ab}(\vtheta)}{\Qone^{ab,ab}(\vtheta)}
    e^{i (\vl - \vk'_\perp\chi) \cdot \vtheta}
    \\
    &= \int \d\mu' \L_{\ell'}(\mu') \frac\pi{k'^2_\perp}
    \int_\chi \frac1\chi P^{ab}(k'_\perp; z)^2 \W^{ab}(\chi)^2
    \int \!2\pi\theta\d\theta\, \frac{\QW^{ab,ab}(\theta)}{\Qone^{ab,ab}(\theta)}
    J_0(\ell\theta) J_0(k'_\perp\chi\theta),
\end{align*}
where $k'_\perp = k' \sqrt{1 - \mu'^2}$, whereas in the small angle limit
\begin{align*}
    &\int_{\uvl, \uvk'} \L_{\ell'}(\uvk'\cdot\uvn)
    2\pi \deltaD(k'_\parallel) \int_{\chi, \vtheta}
    P^{ab}(\vk'_\perp; z)^2 \W^{ab}(\chi)^2 \QW^{ab,ab}(\vtheta)
    e^{i (\vl - \vk'_\perp\chi) \cdot \vtheta}.
    \\
    &= \L_{\ell'}(0) \frac\pi{k'} \int_\chi
    P^{ab}_\perp(k'; z)^2 \W^{ab}(\chi)^2
    \int \!2\pi\theta\d\theta\, \QW^{ab,ab}(\theta) J_0(\ell\theta) J_0(k'\chi\theta)
    \\
    &\approx \L_{\ell'}(0) \frac{2\pi^2}{k'^2 \ell}
    P^{ab}_\perp(k'; z(\chi=\ell/k'))^2 \W^{ab}\Bigl(\frac\ell{k'}\Bigr)^2
    \QW^{ab,ab}(0)
\end{align*}
in which $P_\perp(k)$ is the transverse power.
It can be related to the multipoles by
\begin{align}
    P_\perp(k) \equiv P(k, \mu=0) &= \sum_\ell P_\ell(k) \L_\ell(0)
    = \sum_{\ell\in2\mathbb{Z}^*} \frac{(-1)^\frac\ell2 \ell!}{(\ell!!)^2} P_\ell(k)
    \nonumber\\
    &\approx P_0(k) - \frac12 P_2(k) + \frac38 P_4(k).
\end{align}

Likewise, for each case repeat the derivations for the remaining three terms
before summing up to arrive at the result.
For a large sky coverage
\begin{multline}
    \cov^\disc\bigl[\hat{C}^a(\ell), \hat{P}^b_{\ell'}(k')\bigr]
    \approx \frac{4\pi^2 (2\ell'+1)}{\W^a(0) \W^b(0)}
    \int \d\mu' \frac{\L_{\ell'}(\mu')}{k'^2_\perp} \biggl\{
    \\
    \int_\chi \frac1\chi P^{ab}(k'_\perp; z)^2 \W^{ab}(\chi)^2
    \int \!\theta\d\theta\, \frac{\QW^{ab,ab}(\theta)}{\Qone^{ab,ab}(\theta)}
    J_0(\ell\theta) J_0(k'_\perp\chi\theta)
    \\
    + 2 \int_\chi \frac1\chi P^{ab}(k'_\perp; z) \W^{ab}(\chi) \S^{ab}(\chi)
    \int \!\theta\d\theta\, \frac{\QX^{ab,ab}(\theta)}{\Qone^{ab,ab}(\theta)}
    J_0(\ell\theta) J_0(k'_\perp\chi\theta)
    \\
    + \int_\chi \frac1\chi \S^{ab}(\chi)^2
    \int \!\theta\d\theta\, \frac{\QS^{ab,ab}(\theta)}{\Qone^{ab,ab}(\theta)}
    J_0(\ell\theta) J_0(k'_\perp\chi\theta)
    \biggr\},
\end{multline}
and for a small patch of sky
\begin{multline}
    \cov^\disc\bigl[\hat{C}^a(\ell), \hat{P}^b_{\ell'}(k')\bigr]
    \approx \frac{4\pi^2 (2\ell'+1) \L_{\ell'}(0)}{\W^a(0) \W^b(0) k'} \biggl\{
    \\
    \int_\chi P^{ab}_\perp(k'; z)^2 \W^{ab}(\chi)^2
    \int \!\theta\d\theta\, \QW^{ab,ab}(\theta) J_0(\ell\theta) J_0(k'_\perp\chi\theta)
    \\
    + 2 \int_\chi P^{ab}_\perp(k'; z) \W^{ab}(\chi) \S^{ab}(\chi)
    \int \!\theta\d\theta\, \QX^{ab,ab}(\theta) J_0(\ell\theta) J_0(k'_\perp\chi\theta)
    \\
    + \int_\chi \S^{ab}(\chi)^2
    \int \!\theta\d\theta\, \QS^{ab,ab}(\theta) J_0(\ell\theta) J_0(k'_\perp\chi\theta)
    \biggr\}.
\end{multline}
\begin{multline}
    \cov^\disc\bigl[\hat{C}^a(\ell), \hat{P}^b_{\ell'}(k')\bigr]
    \approx \frac{4\pi^2 (2\ell'+1) \L_{\ell'}(0)}{\W^a(0) \W^b(0) k'^2 \ell} \biggl\{
    P^{ab}_\perp(k'; z)^2 \W^{ab}\Bigl(\frac\ell{k'}\Bigr)^2 \QW^{ab,ab}(0)
    \\
    + 2 P^{ab}_\perp(k'; z) \W^{ab}\Bigl(\frac\ell{k'}\Bigr) \S^{ab}\Bigl(\frac\ell{k'}\Bigr) \QX^{ab,ab}(0)
    + \S^{ab}\Bigl(\frac\ell{k'}\Bigr)^2 \QS^{ab,ab}(0)
    \biggr\}.
\end{multline}
These are in the form that we can compute numerically given the power spectrum,
sky masks and redshift distributions.
Again we use the algorithm in Sec.~\ref{sec:double_Bessel} to evaluate the
double Bessel integrals.
\end{hide}

\subsection{Projected and Projected$\times$Multipole Cases}
\label{sub:disc_proj}

When cross correlate a field with full 3-dimensional information, e.g.\
spectroscopic galaxies, with a field with poor radial information, e.g.\ weak
lensing shear or convergence, one can estimate their correlation projected at
fixed transverse separation $\vx_\perp$ using the radial information of the 3D
field.
Compared to the angular projection in~\ref{sub:disc_ang}, in such case each
field are effectively projected as
\begin{equation}
    \deltaW(\vx_\perp) = \int_\chi \delta(\vx) W(\vx),
\end{equation}
which in Fourier space reads
\begin{equation}
    \deltaW(\vk_\perp) = \int_{\vq} \delta(\vk_\perp-\vq) W(\vq)
    = \int_{\chi, \vl} \delta\Bigl(\vk_\perp - \frac{\vl}\chi, \chi\Bigr)
    W(\vl) W(\chi).
\end{equation}
The projected field is indeed the $k_\parallel = 0$ component of the
corresponding 3D field, therefore having the same equations for 2-point
estimator, shot noise, covariance, etc.
Therefore according to~\eqref{2pts_W_ab} the cross correlation between a
projected and a 3D field is
\begin{equation}
    \ensavg{\deltaW^a(\vk_\perp) \deltaW^b(-\vk')}
    \approx \frac{P^{ab}(\vk_\perp) + P^{ab}(\vk')}2
    \W^{ab}(\vk_\perp - \vk') + \S^{ab}(\vk_\perp - \vk'),
    \label{2pts_W_proj_3D}
\end{equation}
with $\W$ and $\S$ defined as in~\eqref{WS_ab}.
The auto correlation of two projected fields then reduces to a special case of
the above equation when $k'_\parallel = 0$.

As before, the above cross correlation forms the basis to compute the
disconnected cross covariance between a projected power spectrum and multipoles
\begin{align}
    &\cov^\disc\bigl[\hat{P}^a_\perp(k), \hat{P}^b_{\ell'}(k')\bigr]
    \nonumber\\
    &= \frac{2 (2\ell'+1)}{\W^a(0) \W^b(0)}
    \int_{\uvk, \uvk'} 2\deltaD(\uvk\cdot\uvn) \L_{\ell'}(\uvk'\cdot\uvn)
    \bigl|\ensavg{\deltaW^a(\vk) \deltaW^b(-\vk')}\bigr|^2
    \nonumber\\
    &\approx \sum_{\ell=0}^{\lmax}
    \frac{2 (2\ell+1)(2\ell'+1) \L_\ell(0)}{\W^a(0) \W^b(0)}
    \int_{\uvk, \uvk'} \L_\ell(\uvk\cdot\uvn) \L_{\ell'}(\uvk'\cdot\uvn)
    \bigl|\ensavg{\deltaW^a(\vk) \deltaW^b(-\vk')}\bigr|^2.
\end{align}
Since both power spectrum and the window are smooth functions in angular
direction, we can expand the Dirac delta above in multipoles using~\eqref{deltaD_pole} truncated at some $\lmax$.
This reduce the above expression to the multipole case that we have already
solved, as shown in~\eqref{disc_pole_0},~\eqref{disc_pole_1},
\eqref{disc_pole_2}, and~\eqref{disc_pole_3}.

One may want to compute the disconnected auto covariance of a projected power
spectrum in a similar way by expanding both Dirac deltas as follows
\begin{align*}
    &\cov^\disc\bigl[\hat{P}^a_\perp(k), \hat{P}^b_\perp(k')\bigr]
    \nonumber\\
    &\approx \sum_{\ell \ell'}
    \frac{2 (2\ell+1)(2\ell'+1) \L_\ell(0)\L_{\ell'}(0)}{\W^a(0) \W^b(0)}
    \int_{\uvk, \uvk'} \L_\ell(\uvk\cdot\uvn) \L_{\ell'}(\uvk'\cdot\uvn)
    \bigl|\ensavg{\deltaW^a(\vk) \deltaW^b(-\vk')}\bigr|^2.
\end{align*}
However, this turns out to converge much more slowly with $\lmax$ than the
equation before.
Instead, for the projected auto covariance we limit the computation only in the
transverse direction as in~\eqref{disc_ang}
\begin{align}
    &\cov^\disc\bigl[\hat{P}^a_\perp(k), \hat{P}^b_\perp(k')\bigr]
    = \frac2{\W^a(0) \W^b(0)} \int_{\uvk_\perp, \uvk'_\perp}
    \bigl|\ensavg{\deltaW^a(\vk_\perp) \deltaW^b(-\vk'_\perp)}\bigr|^2
    \nonumber\\
    &\approx \frac2{\W^a(0) \W^b(0)} \biggl\{
    P^{ab}_\perp(k) P^{ab}_\perp(k')
    \int \!2\pi s\d s\, \QW[\perp]^{ab,ab}(s) J_0(ks) J_0(k's)
    \nonumber\\ &\hspace{20ex}
    + \bigl[P^{ab}_\perp(k) + P^{ab}_\perp(k')\bigr]
    \int \!2\pi s\d s\, \QX[\perp]^{ab,ab}(s) J_0(ks) J_0(k's)
    \nonumber\\ &\hspace{20ex}
    + \int \!2\pi s\d s\, \QS[\perp]^{ab,ab}(s) J_0(ks) J_0(k's)
    \biggr\}.
\end{align}
Here $\vk \equiv \vk_\perp$ on the first line.
The projected power spectrum $P_\perp$ can be expressed as a linear combination
of multipoles $P_\perp(k) = \sum_\ell P_\ell(k) \L_\ell(0) \approx P_0(k) -
P_2(k) / 2 + 3 P_4(k) / 8$.
And the projected $\Q$ factors in the above equation have their radial
separation integrated and their direction of transverse separation averaged
\begin{equation}
    \Q_\perp^{ab,cd}(s_\perp) = \int_{s_\parallel, \uvs_\perp} \Q^{ab, cd}(\vs).
\end{equation}

\subsection{Correlation Function Case}
\label{sub:disc_corrfunc}

Sec.~\ref{ssub:corrfunc} shows that the Fourier transform relates the 3D
correlation function and power spectrum, and thus their covariances.
Here we derive similar relations for the more practical bases, including
multipole, angular and projected cases.
For multipoles, the Fourier transform connecting two representations becomes
the Hankel transform.
Apart from that, additional polar angle dependence in $\Q_W$ couples to that in
the power spectrum to give
\begin{align}
    \xi_\ell(s) &= i^{\ell} \int \frac{k^2\d k}{2\pi^2} P_\ell(k) j_\ell(k s),
    \nonumber\\
    \hat{\xi}_\ell(s)
    &= (2\ell+1) \W_0 \sum_{\ell_1 \ell_2} \wj{\ell}{\ell_1}{\ell_2}000^2
    \Q_{W \ell_2}^{-1}(s)
    \Bigl[ i^{\ell_1} \int \frac{k^2\d k}{2\pi^2} \hat{P}_{\ell_1}(k) j_{\ell_1}(k s) \Bigr],
\end{align}
where $\Q_{W \ell}^{-1}$ is the multipole moments of $1 / \Q_W(\vs)$, i.e.\
\begin{equation}
    \Q_{W \ell}^{-1}(s) \equiv (2\ell+1) \int_\uvs \frac1{\Q_W(\vs)} \L_\ell(\uvs\cdot\uvn).
\end{equation}
Assuming smooth angular variation, $\Q_{W \ell}^{-1}(s)$ can be computed given
$\Q_{W \ell}(s)$ and we show this procedure in App.~\ref{sub:inv_pole}.

Therefore the disconnected covariance of the correlation function multipoles is
\begin{multline}
    \cov^\disc\bigl[\hat{\xi}_\ell(s), \hat{\xi}_{\ell'}(s')\bigr]
    = (2\ell+1) (2\ell'+1) \W_0^2 \sum_{\ell_1 \ell_2 \ell_3 \ell_4}
    \wj{\ell}{\ell_1}{\ell_2}000^2 \wj{\ell'}{\ell_3}{\ell_4}000^2
    \\
    \Q_{W \ell_2}^{-1}(s) \Q_{W \ell_4}^{-1}(s) \,i^{\ell_1 - \ell_3}\!\!
    \int \frac{k^2\d k}{2\pi^2} \frac{k'^2\d k'}{2\pi^2}
    \cov^\disc\bigl[\hat{P}_{\ell_1}(k), \hat{P}_{\ell_3}(k')\bigr]
    j_{\ell_1}(k s) j_{\ell_3}(k' s').
\end{multline}
Once we have computed $\cov^\disc\bigl[\hat{P}_\ell(k),
\hat{P}_{\ell'}(k')\bigr]$ following Sec.~\ref{sub:method_pole}, we can
evaluate the above equation by calling \texttt{mcfit} to Hankel transform $k$
into $s$, and $k'$ into $s'$, respectively.

The flat-sky angular correlation function is related to the corresponding power
spectrum (App.~\ref{sub:disc_ang}) by the $J_0$ Hankel transform
\begin{align}
    w(\theta) &= \int \!\frac{\ell\d\ell}{2\pi}\; C(\ell) J_0(\ell \theta),
    \nonumber\\
    \hat{w}(\theta) &= \frac{\W_0}{\Q_W(\theta)}
    \int \!\frac{\ell\d\ell}{2\pi}\; \hat{C}(\ell) J_0(\ell \theta).
\end{align}
Thus the disconnected covariance of the angular correlation function is
\begin{equation}
    \hspace{-1ex}
    \cov^\disc\bigl[\hat{w}(\theta), \hat{w}(\theta')\bigr]
    = \frac{\W_0^2}{\Q_W(\theta) \Q_W(\theta')}
    \int \frac{\ell\d\ell}{2\pi} \frac{\ell'\d\ell'}{2\pi}
    \cov^\disc\bigl[\hat{C}(\ell), \hat{C}(\ell')\bigr]
    J_0(\ell \theta) J_0(\ell' \theta').
\end{equation}
Again this is straightforward to compute with \texttt{mcfit}, given
$\cov^\disc\bigl[\hat{C}(\ell), \hat{C}(\ell')\bigr]$ from
Sec.~\ref{sub:disc_ang}.

The same relations in the above two equations also connect the projected
correlation function and power spectrum (App.~\ref{sub:disc_proj}), if we
replace $w \to \xi_\perp$, $C \to P_\perp$, $\theta \to s$, and $\ell \to k$.
Here $\xi_\perp$ is the dimensionless projected correlation functions.
Combining the results on multipole and projected cases, we can write down the
disconnected covariance of projected$\times$multipole correlation functions
\begin{multline}
    \cov^\disc\bigl[\hat{\xi}_\perp(s), \hat{\xi}_{\ell'}(s')\bigr]
    = \frac{(2\ell'+1) \W_0^2}{\Q_{W \perp}(s)} \sum_{\ell_3 \ell_4}
    \wj{\ell'}{\ell_3}{\ell_4}000^2
    \\
    \Q_{W \ell_4}^{-1}(s') \,i^{- \ell_3}\!\!
    \int \frac{k\d k}{2\pi} \frac{k'^2\d k'}{2\pi^2}
    \cov^\disc\bigl[\hat{P}_\perp(k), \hat{P}_{\ell_3}(k')\bigr]
    J_0(k s) j_{\ell_3}(k' s').
\end{multline}
We verify this equation with galaxy mocks in Sec.~\ref{sub:results_proj_pole}.

\begin{hide}

\section{Connected Covariance}
\label{sec:conn}

The connected covariance includes the trispectrum term
\cite{MeiksinWhite99, ScoccimarroZaldarriagaEtAl99}, various shot noise
contributions \cite{MeiksinWhite99, Smith09}, and the super-sample effect
\cite{HamiltonRimesEtAl06, HuKravtsov03, TakadaHu13}
\begin{equation}
    \cov^\conn(\vk, \vk') = C^{T0}(\vk, \vk') + C^\SS(\vk, \vk') + C^\shot(\vk, \vk').
\end{equation}
We describe each component in this appendix.

It is clear that all terms are smooth functions of $k$ and $k'$, therefore so
is the whole $\cov^\conn$.
This implies that we can estimate $\cov^\conn$ of mocks as the difference
between their empirical full covariance and our analytic model for $C^\disc$,
and use its smoothness to test latter's performance.

\subsection{Trispectrum and Super-Sample Covariance}
\label{sub:T0_and_SS}

\subsection{Shot Noises}
\label{sub:shot}

Shot noises in the 2-point function and its covariance arise from the discrete
nature of the tracer field.
In \eqref{deltas}, we have constructed these compensated point processes with
zero mean to describe the overdensity fields of data and random catalogs.

These discrete catalogs give rise to self-correlation of a data or random point
with itself when we calculate the cumulants or connected correlation among $n$
points
\begin{align}
    \ensavg{\textstyle\prod_i \delta_{\data i}(\vx_i)}_\conn
    &= \sum_\setpart \xi(\setpart) \prod_{S \in \setpart}
        \nbar^{-|S|+1} \deltaD(S),
    \nonumber\\
    \ensavg{\textstyle\prod_i \delta_{\rand i}(\vx_i)}_\conn
    &= \alpha^{n-1} \nbar^{-n+1} \deltaD(\{1, \cdots, n\}),
    \nonumber\\
    \ensavg{\textstyle\prod_i \delta_{\data i}(\vx_i)
        \textstyle\prod_j \delta_{\rand j}(\vx_j)}_\conn &= 0,
    \label{shot}
\end{align}
where $|\;|$ denotes the cardinality of or the number of elements in a set.
$\setpart$ is a \emph{partition} of the set of $n$ points $\{1, \cdots, n\}$.
Each $\setpart$ is a set of nonempty subsets $\{S\}$ such that each point is in
exactly one of the subsets.
As an example, for $n=3$ there are 5 different $\setpart$'s: $\{\{1, 2, 3\}\}$,
$\{\{1, 2\}, \{3\}\}$, $\{\{2, 3\}, \{1\}\}$, $\{\{3, 1\}, \{2\}\}\}$,
$\{\{1\}, \{2\}, \{3\}\}$.
The number of all possible $\setpart$ is given by the Bell number $B_n$, e.g.\
$B_0 = 1$, $B_1 = 1$, $B_2 = 2$, $B_3 = 5$, $B_4 = 15$, $B_5 = 52$.
$\xi(\setpart)$ is a $|\setpart|$-point correlation function among these
subsets, with all points in each $S$ sharing the same position as guaranteed by
the Dirac delta $\deltaD(S)$.
$\xi(\setpart) = 1$ when $|\setpart| = 1$.
\eqref{shot} shows how to compute the connected piece, while the disconnected
pieces can be constructed as products of smaller connected pieces.

\end{hide}

\section{Useful Relations}
\label{sec:relations}

The product of two Legendre polynomials can be expanded in itself as
\begin{equation}
    \L_{\ell_1}(\mu) \L_{\ell_2}(\mu) =
    \sum_{\ell_3=|\ell_1-\ell_2|}^{\ell_1+\ell_2}
    (2\ell_3+1) \wj{\ell_1}{\ell_2}{\ell_3}000^2 \L_{\ell_3}(\mu),
    \label{Legendre_product}
\end{equation}
where the $2\times3$ matrix enclosed by parentheses denotes the Wigner 3$j$
symbol.
The completeness relation of the Legendre polynomials is
\begin{equation}
    \deltaD(\mu-\mu') = \sum_{\ell=0}^\infty
    \frac{2l+1}{2} \L_\ell(\mu) \L_\ell(\mu').
    \label{deltaD_pole}
\end{equation}

The plane wave expansion equates the kernel of Fourier transform in 3D with a
series in spherical Bessel function $j_\ell$ and Legendre polynomial
\begin{equation}
    e^{i \vk \cdot \vs} = \sum_{\ell=0}^\infty (2\ell+1) i^\ell j_\ell(k s)
    \L_\ell(\uvk\cdot\uvs),
    \label{plane_wave_expansion}
\end{equation}
and in 2D with a series in Bessel function $J_n$
\begin{equation}
    e^{i \vl \cdot \vtheta} = \sum_{n=-\infty}^\infty i^{|n|} J_{|n|}(\ell \theta)
    e^{i n \phi}
    = J_0(\ell \theta) + 2 \sum_{n=1}^\infty i^n J_n(\ell \theta)
    \cos(n \phi),
    \label{Jacobi_Anger_expansion}
\end{equation}
where $\cos\phi \equiv \uvl \cdot \uvtheta$.

The addition theorem for spherical harmonics relates the Legendre polynomial
with sum of products of spherical harmonics
\begin{equation}
    \L_\ell(\uvk\cdot\uvs)
    = \frac{4\pi}{2\ell+1} \sum_{m=-\ell}^\ell Y_\ell^m(\uvk) Y_\ell^m(\uvs)^*,
    \label{addition_theorem}
\end{equation}
with the spherical harmonics satisfying the following orthonormal relation
\begin{equation}
    \int_\uvv Y_\ell^m(\uvv) Y_{\ell'}^{m'}(\uvv)^*
    = \frac{\deltaK_{\ell\ell'} \deltaK_{m m'}}{4\pi},
    \label{Ylm_orthonormal}
\end{equation}
where an extra $4\pi$ factor appears due to our shorthand notation $\int_\uvv$
(footnote~\ref{fn:int_ang}) that averages rather than integrates over the solid
angle.

One can prove the following relation by using~\eqref{plane_wave_expansion},
\eqref{addition_theorem}, and~\eqref{Ylm_orthonormal}
\begin{equation}
    \int_\uvk \L_\ell(\uvk\cdot\uvn) e^{-i\vk\cdot\vs}
    = (-i)^\ell j_\ell(ks) \L_\ell(\uvs\cdot\uvn).
    \label{plane_wave_poles}
\end{equation}

\subsection{Inversion of Multipoles}
\label{sub:inv_pole}

The redshift-space correlation function has a $\mu$-dependent denominator.
To account for this in the covariance of $\xi_\ell(s)$, in one approach we can
compute the multipole moments of the inverse of the denominator, whose
multipoles are already known.
Let $a_\ell$ be the multipole moments of $a(\mu)$ and we want to compute
$b_\ell$ of $b(\mu)$ that satisfy
\begin{equation}
    1 \equiv a(\mu) b(\mu)
    = \sum_{\ell \ell'} a_\ell b_{\ell'} \L_\ell(\mu) \L_{\ell'}(\mu)
    = \sum_{\ell \ell' \ell''} a_\ell b_{\ell'}
    (2\ell''+1) \wj{\ell}{\ell'}{\ell''}000^2 \L_{\ell''}(\mu).
\end{equation}
The multipole coupling on the right hand side should result in only the
monopole on the left hand side, leading to the following linear system
\begin{equation}
    \sum_{\ell \ell'} a_\ell b_{\ell'}
    \wj{\ell}{\ell'}{\ell''}000^2 = \deltaK_{\ell''0},
\end{equation}
from which we can solve for $b_\ell$.

\begin{hide}

\section{Real space derivation}
\sukhdeep{Not tracking factors of $2\pi$, some signs}

We start with defining projected field as
\begin{align}
\delta(\vtheta)=\int_\chi\delta(\vtheta\chi,\chi)W(\chi)W(\vec \theta\chi)
\end{align}
The two point function with a 3-D field is defined as
\begin{align}
\mean{\delta_b^\perp\delta_a}(\vrp)=\int_\chi W_b(\chi)\int_{\vx'}\delta_a(\vx')
		\delta_b\left(\frac{\vx'_\perp+\vrp}{x_\parallel}\chi,\chi\right)W_a(\vx')W_{b,\perp}\left(\frac{\vx'_\perp+\vrp}{x_\parallel}\right)
\end{align}
where $\frac{\vx'_\perp+\vrp}{x_\parallel}\equiv\vec \theta$.

Moving into Fourier space
\begin{align}
&\int_\chi W_b(\chi)\int_{\vx',\vk_1,\vk_2,\vq_1,\vq_2^\perp}\delta_a(\vk_1)\delta_b(\vk_2)W_a(\vq_1)W_b(\vq_2)e^{i\vk_1\vx'}e^{i\vq_1\vx'}
e^{i\vk_2^{\perp}\vtheta\chi} e^{i\vq_2^{\perp}\vtheta\chi} e^{ik_2\chi}\\
&\int_\chi W_b(\chi)\int_{\vx',\vk,\vq_1,\vq_2^\perp}P(\vk)W_a(\vq_1)W_b(\vq_2)e^{i\vk\vx'}e^{i\vq_1\vx'}
e^{-i\vk^{\perp}\vtheta\chi} e^{i\vq_2^{\perp}\vtheta\chi} e^{-ik\chi}\\
&\int_\chi W_b(\chi)\int_{\vx',\vk,\vq_1,\vq_2^\perp}P(\vk)W_a(\vq_1)W_b(\vq_2)e^{i\vk\vx'}e^{i\vq_1\vx'}
e^{-i\vk^{\perp}\vtheta\chi} e^{i\vq_2^{\perp}\vtheta\chi} e^{-ik\chi}\\
&\int_\chi W_b(\chi)\int_{x'_\parallel,\vk,\vq_1}P(\vk)W_a(\vq_1)
W_b\left(\vk_\perp\left(1-\frac{\chi}{x_\parallel}\right)-\vq_\perp\frac{\chi}{x_\parallel}\right)
e^{ikx'}e^{iq_1x'}e^{-i(\vk^{\perp}+\vq_1^{\perp})\vrp} e^{-ik\chi}\\
&\int_\chi W_b(\chi)\int_{x'_\parallel,\vk,\vq_1}P(\vk)W_a(\vq_1)
W_b(-\vq_1^\perp)
e^{ikx'}e^{iq_1x'}e^{-i(\vk^{\perp}+\vq_1^{\perp})\vrp} e^{-ik\chi}\\
&\int_\chi W_b(\chi)\int_{\vk,\vq_1^{\perp}}P(\vk)W_a(\vq_1^{\perp})W_a(k)
W_b(-\vq_1^\perp)
e^{-i(\vk^{\perp}+\vq_1^{\perp})\vrp} e^{-ik\chi}\\
&\int_{\chi,\chi'} W_b(\chi)W_a(\chi')\int_{\vk,\vq_1^{\perp}}P(\vk)W_a(\vq_1^{\perp})
W_b(-\vq_1^\perp)
e^{-i(\vk^{\perp}+\vq_1^{\perp})\vrp} e^{-ik(\chi-\chi')}\\
\end{align}
In the first step we take Dirac delta in $\vk_1,\vk_2$. Next we integrate over $\vx_\perp$, convert $\vtheta$ to $\vrp$ and get rid of
$\vq_2^\perp$. Next we argue that since the correlations only work on small separations, $\frac{\chi}{x_\parallel}\sim 1$, so we ignore \
these terms in the $W_b$. Next we integrate over $x_\parallel$ and get rid of $q_\parallel$. Finally we convert the line of sight window $W_b$
back to real space.

Now we want to get rid of $k_\parallel$. We consider two limiting cases. We always assume $W_b(\chi)$ is  wide in $\chi$. In first case we assume $W_a(\chi)$ is wide, in which case $W_{a}(k_\parallel)$ forces $k_\parallel$ to be small and we can ignore the dependence of power spectra on $k_\parallel$, assuming $k_\perp\gg k_\parallel$. This allows us to integrate over $k_\parallel$ to get $\delta_D(\chi-\chi')$. In the second case, we assume $W_a(\chi')$ to be narrow, such that $W_{a}(\chi')\approx\delta_D(\chi')$, in which case we can then integrate over $\chi$ to get $\delta_D(k_\parallel)$. In both case we get expressions that can be written as
\begin{align}
&\int_{\chi} W_b(\chi)W_a(\chi)\int_{\vk_\perp,\vq_1^{\perp}}P(\vk)W_a(\vq_1^{\perp})
W_b(-\vq_1^\perp)
e^{-i(\vk^{\perp}+\vq_1^{\perp})\vrp}\\
\end{align}

\subsection{2D-3D covariance}
\begin{align}
\delta(\vtheta)=\int_\chi\delta(\vtheta\chi,\chi)W(\chi)W(\vec \theta\chi)
\end{align}

\begin{align}
\mean{\delta_1^\perp\delta_2}(\vtheta)=\int_{\chi,\chi'}
W_1(\chi)W_2(\chi')\int_{\vx_\perp}
\delta_1\left({\vx_\perp},\chi'\right)
\delta_2\left(\vx_\perp+\vtheta\chi,\chi\right)
        W_1(\vx_\perp)W_{2}\left(\vx_\perp+\vtheta\chi\right)
\end{align}
where $\frac{\vx'_\perp+\vrp}{x_\parallel}\equiv\vec \theta$.

Covariance
\begin{align}
C({\delta_1\delta_2(\vtheta)\delta_3\delta_4(\vs)})&=
\frac{1}{A_W(\vtheta)V_W(\vs)}
\int_{\chi,\chi'}W_1(\chi')W_2(\chi)\int_{\vx_\perp,\vx'}
\delta_1\left({\vx_\perp},\chi'\right)
\delta_2\left(\vx_\perp+\vtheta\chi,\chi\right)
\delta_3(\vx')\delta_4(\vx'+\vs)\nonumber\\
&W_1(\vx_\perp)W_{2}\left(\vx_\perp+\vtheta\chi\right)
W_3(\vx')W_{4}\left(\vx'+\vs\right)\\
&=
\frac{1}{A_W(\vtheta)V_W(\vs)}
\int_{\chi,\chi'}W_1(\chi')W_2(\chi)\int_{\vx_\perp,\vx'}
\int_{\vk_{1-4},\vq_{\perp,1-2},\vq_{3-4}}
\delta_1(\vk_1)\delta_2(\vk_2)\delta_3(\vk_3)\delta_4(\vk_4)
\nonumber\\
&W_1(\vq_{1,\perp})W_2(\vq_{2,\perp})
W_3(\vq_3)W_{4}(\vq_4)
e^{i\vk_1(\vx_\perp,\chi')}e^{i\vk_2(\vx_\perp+\theta\chi,\chi)}
e^{i\vk_3(\vx')}e^{i\vk_4(\vx'+\vs)}
e^{i\vq_{1,\perp}(\vx_\perp)}e^{i\vq_{2\perp}(\vx_\perp+\theta\chi)}
e^{i\vq_3(\vx')}e^{i\vq_4(\vx'+\vs)}
\\
&=
\frac{1}{A_W(\vtheta)V_W(\vs)}
\int_{\chi,\chi'}W_1(\chi')W_2(\chi)
\int_{\vk_{1-4},\vq_{\perp,1-2},\vq_{3-4}}
\delta_1(\vk_1)\delta_2(\vk_2)\delta_3(\vk_3)\delta_4(\vk_4)
\nonumber\\
&W_1(\vq_{1,\perp})W_2(\vq_{2,\perp})
W_3(\vq_3)W_{4}(\vq_4)
e^{ik_1(\chi')}e^{i\vk_2(\theta\chi,\chi)}
e^{i\vk_4(\vs)}e^{i\vq_{2\perp}(\theta\chi)}e^{i\vq_4(\vs)}
\nonumber\\
&\delta_D(\vk_{1,\perp}+\vk_{2,\perp}+\vq_{1,\perp}+\vq_{2,\perp})
\delta_D(\vk_{3}+\vk_{4}+\vq_{3}+\vq_{4})
\end{align}

1-3,2-4 term
\begin{align}
&=\frac{1}{A_W(\vtheta)V_W(\vs)}
\int_{\chi,\chi'}W_1(\chi')W_2(\chi)
\int_{\vk_{1-4},\vq_{\perp,1-2},\vq_{3-4}}
P_{13}(\vk_1)P_{24}(\vk_2)
\nonumber\\
&W_1(\vq_{1,\perp})W_2(\vq_{2,\perp})
W_3(\vq_3)W_{4}(\vq_4)
e^{ik_1(\chi')}e^{i\vk_2(\theta\chi,\chi)}
e^{i\vk_4(\vs)}e^{i\vq_{2\perp}(\theta\chi)}e^{i\vq_4(\vs)}
\nonumber\\
&\delta_D(\vk_{1,\perp}+\vk_{2,\perp}+\vq_{1,\perp}+\vq_{2,\perp})
\delta_D(\vk_{3}+\vk_{4}+\vq_{3}+\vq_{4})
\delta_D(\vk_1+\vk_3)\delta_D(\vk_2+\vk_4)\\
&=\frac{1}{A_W(\vtheta)V_W(\vs)}
\int_{\chi,\chi'}W_1(\chi')W_2(\chi)
\int_{\vk_{1-2},\vq_{\perp,1-2},\vq_{3-4}}
P_{13}(\vk_1)P_{24}(\vk_2)
\nonumber\\
&W_1(\vq_{1,\perp})W_2(\vq_{2,\perp})
W_3(\vq_3)W_{4}(\vq_4)
e^{ik_1(\chi')}e^{i\vk_2(\theta\chi,\chi)}
e^{-i\vk_2(\vs)}e^{i\vq_{2\perp}(\theta\chi)}e^{i\vq_4(\vs)}
\nonumber\\
&\delta_D(\vk_{1,\perp}+\vk_{2,\perp}+\vq_{1,\perp}+\vq_{2,\perp})
\delta_D(-\vk_{1}-\vk_{2}+\vq_{3}+\vq_{4})\\
&=\frac{1}{A_W(\vtheta)V_W(\vs)}
\int_{\chi,\chi'}W_1(\chi')W_2(\chi)
\int_{\vk_{1},k_2,\vq_{\perp,1-2},\vq_{3-4}}
P_{13}(\vk_1)P_{24}(-\vk_1,k_2)
\nonumber\\
&W_1(\vq_{1,\perp})W_2(\vq_{2,\perp})
W_3(\vq_3)W_{4}(\vq_4)
e^{ik_1(\chi')}e^{-i\vk_{1,\perp}(\theta\chi)}e^{ik_2\chi}
e^{i\vk_1(\vs)}e^{i\vq_{2\perp}(\theta\chi)}e^{i\vq_4(\vs)}
\nonumber\\
&\delta_D(\vq_{1,\perp}+\vq_{2,\perp}+\vq_{3,\perp}+\vq_{4,\perp})
\delta_D(-k_{1}-k_{2}+q_{3}+q_{4})\\
&=\frac{1}{A_W(\vtheta)V_W(\vs)}
\int_{\chi,\chi'}W_1(\chi')W_2(\chi)
\int_{\vk_{1},\vq_{\perp,1-2},\vq_{3-4}}
P_{13}(\vk_1)P_{24}(-\vk_1,k_1-q_3-q_4)
\nonumber\\
&W_1(\vq_{1,\perp})W_2(\vq_{2,\perp})
W_3(\vq_3)W_{4}(\vq_4)
e^{ik_1(\chi'-\chi)}e^{-i\vk_{1,\perp}(\theta\chi)}e^{i(q_3+q_4)\chi}
e^{i\vk_1(\vs)}e^{i\vq_{2\perp}(\theta\chi)}e^{i\vq_4(\vs)}
\nonumber\\
&\delta_D(\vq_{1,\perp}+\vq_{2,\perp}+\vq_{3,\perp}+\vq_{4,\perp})
\\
&\approx\frac{1}{A_W(\vtheta)V_W(\vs)}
\int_{\chi,\chi'}W_1(\chi')W_2(\chi)W_3(\chi)W_4(\chi+s_\parallel)
\int_{\vk_{1},\vq_{\perp,1-4}}
P_{13}(\vk_1)P_{24}(-\vk_1)
\nonumber\\
&W_1(\vq_{1,\perp})W_2(\vq_{2,\perp})
W_3(\vq_3)W_{4}(\vq_4)
e^{ik_1(\chi'-\chi)}e^{-i\vk_{1,\perp}(\theta\chi)}
e^{i\vk_1(\vs)}e^{i\vq_{2\perp}(\theta\chi)}e^{i\vq_4(\vs_\perp)}
\nonumber\\
&\delta_D(\vq_{1,\perp}+\vq_{2,\perp}+\vq_{3,\perp}+\vq_{4,\perp})\\
&=\frac{1}{A_W(\vtheta)V_W(\vs)}
\int_{\chi}W_1(\chi-s_\parallel)W_2(\chi)W_3(\chi)W_4(\chi+s_\parallel)
\int_{\vk_{1\perp},\vq_{\perp,1-4}}
P_{13}(\vk_1)P_{24}(-\vk_1)
\nonumber\\
&W_1(\vq_{1,\perp})W_2(\vq_{2,\perp})
W_3(\vq_3)W_{4}(\vq_4)
e^{-i\vk_{1,\perp}(\theta\chi)}
e^{i\vk_1(\vs)}e^{i\vq_{2\perp}(\theta\chi)}e^{i\vq_4(\vs_\perp)}
\nonumber\\
&\delta_D(\vq_{1,\perp}+\vq_{2,\perp}+\vq_{3,\perp}+\vq_{4,\perp})\\
\end{align}

\end{hide}

\bibliographystyle{JHEP}
\bibliography{cosmo,cosmo_preprints}
\end{document}